\documentclass[12pt,epsf,showkeys]{article}

\usepackage{graphicx,float}
\usepackage{mathrsfs,array,multirow}
\usepackage{amstext}
\usepackage{subfigure}
\usepackage{bm,latexsym,amsmath,amsfonts,amssymb}
\usepackage[usenames,dvipsnames]{color}
\usepackage{color}
\usepackage[colorlinks=true,linkcolor=blue]{hyperref}
\usepackage{soul}
\usepackage{epsfig}

\newcommand{\be}{\begin{equation}}
\newcommand{\ee}{\end{equation}}
\newcommand{\ba}{\begin{eqnarray}}
\newcommand{\ea}{\end{eqnarray}}

\graphicspath{{./Figs/}}

\begin{document}
\begin{titlepage}
\vspace{.3in}
\begin{flushright}
\end{flushright}
\vspace{0.3cm}

\begin{center}
{\Large\bf Chaotic behaviors of particles around the black hole with an anisotropic matter immersed in a magnetic field}\\
\vspace{.3in}

  {$\mbox{Khusan \,\, Alibekov}^{a}$},\,\,
  {$\mbox{Hocheol \,\, Lee}^{b}$},\,\,
  {$\mbox{Pakhlavon \,\, Yovkochev}^{c j}$},\,\,
  {$\mbox{Bobomurat J.\,\, Ahmedov}^{d e h}$},\,\,
  {$\mbox{Bum-Hoon \,\, Lee}^{f g i}$},\,\,
  {$\mbox{Ahmadjon \,\, Abdujabbarov}^{d e h}$},\,\,
  {$\mbox{Wonwoo \,\, Lee}^{f}$},\,\,  \\

\vspace{.1in}

{\small a \it Department of Physics and Astronomy, Faculty of Physical Sciences and Engineering,
University of Southampton, Southampton, SO17 1BJ, United Kingdom } \\
{\small b \it Department of Physics, Dongguk University, Seoul 04620, Korea}\\
{\small c \it Institute of Fundamental and Applied Research, National Research University TIIAME, Kori Niyoziy 39, Tashkent 100000, Uzbekistan}\\
{\small d \it Ulugh Beg Astronomical Institute of the Uzbekistan Academy of Sciences,
Astronomy Street 33, Tashkent 100052, Uzbekistan}\\
{\small e \it Institute of Theoretical Physics, National University of Uzbekistan, Tashkent 100174, Uzbekistan}\\
{\small f \it Center for Quantum Spacetime, Sogang University, Seoul 04107, Korea}\\
{\small g \it Department of Physics, Sogang University, Seoul 04107, Korea}\\
{\small h \it School of Physics, Harbin Institute of Technology, Harbin 150001, People's Republic of China}\\
{\small i \it Department of Physics, Shanghai University, 99 Shangda Road, Shanghai, 200444, China} \\
{\small j \it Samarkand State Pedagogical Institute, Spitamen Shokh Street 166, Samarkand 140100, Uzbekistan}
\vspace{.2in}

\end{center}
\begin{center}
{\large\bf Abstract}
\end{center}
\begin{center}
\begin{minipage}{5.0in}

{\small \,\,\,\,

We present an exact solution to the Einstein-Maxwell equations that describes a static black hole coexisting with anisotropic matter immersed in an external magnetic field, obtained via the Harrison transformation. Our findings reveal that an increase in the anisotropic matter parameter systematically suppresses the local chaotic behavior, as indicated by a reduction in the Lyapunov exponent. Conversely, variations in the external magnetic field lead to qualitative changes in global chaotic behavior. This is analyzed through Poincar\'e sections, which demonstrate transitions between regular and chaotic trajectories resulting from the nonlinear gravitational-magnetic interactions. These factors play distinct yet complementary roles in shaping chaotic particle dynamics around black holes. This study would offer a new theoretical framework for exploring nonintegrable particle motion within magnetized black hole spacetimes.
}

\end{minipage}
\end{center}

\let\thefootnote\relax\NoHyper\footnotetext{email: $^1$K.A.U.Alibekov@soton.ac.uk, $^2$insaying@dongguk.edu, $^3$yovqochevp@gmail.com, $^4$ahmedov@astrin.uz, $^5$bhl@sogang.ac.kr, \, $^6$ahmadjon@astrin.uz, \, $^7$warrior@sogang.ac.kr }
\end{titlepage}

\newpage

\section{Introduction \label{sec1}}

\quad

The environments surrounding astrophysical black holes are far more complex than the idealized vacuum solutions of general relativity.
Observations of active galactic nuclei, x-ray binaries, and the Galactic center indicate that black holes are embedded in magnetized plasmas and interact with surrounding matter fields, both of which play essential roles in accretion, jet formation, particle acceleration, and high-energy radiation~\cite{Ghez:2008ms, Gillessen:2008qv, Blandford:1977ds, Johnson:2015iwg, Blandford:2018iot, LIGOScientific:2020iuh, EventHorizonTelescope:2022wkp, EventHorizonTelescope:2025vum, Fernandes:2025osu}.
Although the Schwarzschild~\cite{Schwarzschild:1916uq} and Kerr~\cite{Kerr:1963ud, Kerr:2007dk} geometries
successfully describe isolated vacuum black holes,
realistic astrophysical black holes would be expected to interact with surrounding matter and magnetic fields.
Understanding how these environmental effects modify the dynamics of particles in strong gravitational fields is therefore an important problem in relativistic astrophysics.

A remarkable feature of several black hole spacetimes is the existence of hidden symmetries.
In addition to the conserved quantities associated with stationarity and axisymmetry,
the Kerr geometry admits a nontrivial second-rank Killing tensor that generates Carter's constant.
The presence of this hidden symmetry renders the Hamilton-Jacobi equation completely separable and guaranties the complete integrability of geodesic motion~\cite{Carter:1968rr, Carter:1968ks, Walker:1970un, Benenti:1979erw}.
Consequently, particle trajectories can be described by four independent first-order differential equations~\cite{Bardeen:1973tla, Dymnikova1986}. In contrast, once the hidden symmetry is broken, the Hamilton-Jacobi equation generally becomes nonseparable,
the equations of motion remain coupled, and the resulting dynamics become intrinsically nonintegrable.

The loss of integrability provides one of the fundamental mechanisms responsible for the global chaotic motion in general relativity.
Unlike Newton's theory of gravitation with only Keplerian orbits, Einstein's theory of gravitation allows homoclinic orbits as unstable circular orbits~\cite{Levin:2008yp, Pugliese:2010ps, Li:2023bgn}, which organize the phase-space structure of geodesic motion.
Chaotic behavior reflecting from these structures has been investigated in various black hole spacetimes using Lyapunov exponents for the local chaotic behavior in the integrable system~\cite{Cardoso:2008bp, Gwak:2022xje, Jeong:2023hom, Lei:2023jqv, Lee:2025vih, Targema:2026anu}, Poincar\'e sections for the global chaotic behavior in the nonintegrable system~~\cite{Karas:1992, Wang:2016wcj, Li:2018wtz}, and other nonlinear dynamical methods. As shown in these extensive studies, the geometric origin of chaos remains closely connected to the existence--or absence--of hidden spacetime symmetries.

Magnetic fields~\cite{Melvin:1963qx} constitute one of the most important environmental ingredients surrounding astrophysical black holes.
Einstein-Maxwell solutions describing black holes immersed in external magnetic fields, including Ernst geometries~\cite{Ernst:1976mzr, Ernst:1976bsr}, have long provided valuable models for studying electromagnetic processes in strong gravitational fields and astrophysics~\cite{Harrison:1968wue, Wald:1974np, Hiscock:1980zf, Thorne:1986iy, Rezzolla:2000dk, Podolsky:2025tle, Astorino:2025lih, Ma:2026ima}.
Independently, black holes coexisting with anisotropic matter have attracted increasing interest as phenomenological descriptions of effective matter distributions and possible dark-sector environments~\cite{Kiselev:2002dx, Toshmatov:2015npp, Cho:2017nhx, Visser:2019brz, Kim:2019hfp, Kim:2021vlk, Kim:2025sdj, Lee:2026cit, Xu:2026jjg}.
Although these two classes of solutions have been studied extensively, considerably less attention has been devoted to understanding how anisotropic matter and external magnetic fields jointly modify spacetime symmetries and consequently influence the transition from integrable to chaotic particle dynamics.

In this paper, we construct an exact solution of the Einstein-Maxwell equations describing a static black hole coexisting with anisotropic matter and immersed in an external magnetic field through a Harrison transformation~\cite{Harrison:1968wue}.
We show that the resulting spacetime possesses only the Killing symmetries associated with stationarity and axisymmetry and admits no second-rank Killing tensor analogous to that responsible for Carter's constant.
As a consequence, the Hamilton-Jacobi equation cannot be separated, leading to fundamentally nonintegrable geodesic motion. We investigate the resulting particle trajectories through homoclinic orbits, Lyapunov exponents, and Poincar\'e sections.

The remainder of this paper is organized as follows. In Sec.~\ref{sec2} we construct the exact Einstein-Maxwell solution
and discuss its geometrical and physical properties.
We analyze the separability structure and derive the equations that govern particle motion.
Section \ref{sec3} investigates homoclinic orbits and chaotic dynamics using Lyapunov exponents and Poincar\'e sections.
Finally, Sec.~\ref{sec4} summarizes our results and discusses their implications for relativistic dynamics in the geometry of the black hole with anisotropic matter immersed in a magnetic field.

\section{A black hole immersed in a magnetic field \label{sec2}}
\quad

In this section, we construct the geometry of a black hole that coexists with anisotropic matter immersed in a magnetic field. We explore the structure of the event horizon in this configuration and demonstrate the magnetic field configuration surrounding the black hole. Additionally, we clearly outline the separability structure of the spacetime and derive the Hamilton-Jacobi equation.

\subsection{Construction of the background geometry \label{sec2-1}}

\quad

We want to consider the geometry of a black hole coexisting with anisotropic matter immersed in the magnetic field.
For this analysis, we first consider a black hole coexisting with the matter field outside the event horizon~\cite{Kiselev:2002dx, Cho:2017nhx}
\begin{eqnarray}
\label{metric01}
   ds^2 &&= -f(r)dt^2 + \frac{dr^2}{f(r)} + r^2 (d\theta^2 +  \sin^2\theta d\phi^2 ) \nonumber \\
        && = \left(-\frac{\Delta}{r^2}dt^2 + \frac{r^2}{\Delta}dr^2 + r^2d\theta^2\right) +  r^2\sin^2\theta d\phi^2 \,,
\end{eqnarray}
where
\begin{equation}
\label{Delta}
    f(r)= 1 -\frac{2M}{r} + \frac{K}{r^{2w}}\,, ~~ \Delta = r^2 - 2 M r +K r^{2\left(1-\omega\right)}  \,,
\end{equation}
where $M$ represents the Arnowitt-Deser-Misner mass and $K$ is a constant that represents the anisotropic fluid matter. This has the relation $(2w-1)K=r^{2w}_o$, in which $r_o$ is a chargelike quantity of dimension of length. When $(2w-1)K \geq 0$ the energy density is always non-negative.  The radial null energy condition is satisfied. In this study, our analysis is limited to asymptotically flat cases, which restricts our consideration to systems with $w > 1/2$.

To obtain the geometry of such a black hole immersed in the magnetic field, let us consider the following  action~\footnote{Due to the absence of a recognized action for such fluid matter, it is difficult to show the corresponding explicit action; as an alternative, one may begin with an action that includes a nonlinear electrodynamics term~\cite{Lee:2026cit}.}:
\begin{equation}
\label{action}
    I=\int_{\mathcal{M}} d^4x\sqrt{-g} \left[ \frac{1}{16 \pi} (R-F_{\mu\nu}F^{\mu\nu})  + J^{\mu}A_{\mu} + \mathcal{L}_{\rm am} \right] + I_b  \,,
\end{equation}
where $R$ is the Ricci scalar of spacetime, $F^{\mu\nu}$ is the  electromagnetic field tensor, we take $G=1$ for simplicity, $\mathcal{L}_{\rm am} $ denotes the effective anisotropic matter fields, and $I_b$ corresponds to
the boundary term~\cite{Gibbons:1976ue, Hawking:1995ap}. One obtains the Einstein equations
\begin{equation}
\label{Eineq}
    R_{\mu\nu} - \frac{1}{2}R g_{\mu\nu} = 8\pi T_{\mu\nu} \,,
\end{equation}
where $T_{\mu\nu}=\frac{1}{4\pi}\left( F_{\mu\alpha}F^{~\alpha}_{\nu} -\frac{1}{4} g_{\mu\nu} F_{\alpha\beta} F^{\alpha\beta} \right)+T^{\rm am}_{\mu\nu}$, and   the Maxwell equations
\begin{equation}
\label{Maxweq}
   \nabla_{\nu}F^{\mu\nu}= \frac{1}{\sqrt{-g}} \left[ \partial_{\nu} ({\sqrt{-g}}F^{\mu\nu}) \right] = 4\pi J^{\mu}  \,,
\end{equation}
where the source term will be related to the anisotropic matter, and we show this in the following.

We now apply the above black hole geometry according to Refs.~\cite{Harrison:1968wue, Ernst:1976mzr, Hiscock:1980zf} and take the form:
\begin{equation}
\label{metric02}
    ds^2 = \zeta^{-1}[-\rho^2 dt^2 + 2P^{-2} d\xi^*d\xi] +\zeta(d\phi -\omega dt)^2 \,,
\end{equation}
where $\omega=0$, $\rho=\Delta^{1/2} \sin\theta$,
$P=(r^2\sin\theta)^{-1}$, $d\xi=\frac{1}{\sqrt 2}\left( \frac{dr}{\Delta^{1/2}} + i d\theta \right)$, $\zeta=r^2\sin^2\theta$ gives a   gravitational complex potential $\varepsilon=-r^2\sin^2\theta$.

We employ the Harrison transformation~\cite{Harrison:1968wue} as
\begin{equation}
\label{Harrisonmap}
   \Phi'=\frac{\alpha  \varepsilon}{1+|\alpha|^2 r^2 \sin^2\theta} \ , ~~ \varepsilon'= \frac{\varepsilon}{1+|\alpha|^2 r^2 \sin^2\theta} \,,
\end{equation}
where $\alpha$ could be complex. However, we choose $\alpha=-B_o/2$~\cite{Vigano:2022hrg} as the real value in the absence of the external electric field. $\Phi'$ is the electromagnetic potential that gives $A_{\phi}$.

Then we obtain the black hole geometry with the anisotropic matter field immersed in the magnetic field
\begin{equation}
\label{metric}
    ds^2 = \Lambda^2 \left(-\frac{\Delta}{r^2}dt^2 + \frac{r^2}{\Delta}dr^2 + r^2d\theta^2\right) + \frac{1}{\Lambda^2} r^2\sin^2\theta d\phi^2 \,,
\end{equation}
where
\begin{equation}
   \Lambda = 1+\frac{1}{4}B_o^2 r^2\sin^2\theta \,.
\end{equation}
This is not asymptotically flat due to the magnetic field.
It is the static axisymmetric black hole solution of the Einstein-Maxwell equations.

Then the source term of the Maxwell equations is nonzero, and $J^{\phi}=\frac{2 B_o \rho_{\rm am} }{\Lambda^2}$ \cite{Lungu:2024iob}
\footnote{During the writing of the manuscript, it was observed that the paper was present.}
with $\rho_{\rm am} =\frac{r^{2w}_o}{8\pi r^{2(w+1)}}$ \cite{Cho:2017nhx, Jeong:2023hom}, in which $A_{\phi}= \frac{B_o r^2 \sin^2\theta}{2\Lambda}$. This one satisfies the Bianchi identity.

The components of the Maxwell tensor are given by
\begin{equation}
\label{compb}
 F^{\theta\phi} =\frac{B_o \cos\theta}{r^2 \sin\theta \Lambda^2} \,,~~ F^{\phi r} = - \frac{B_o \Delta }{r^3 \Lambda^2} \,.
\end{equation}
Equation \eqref{compb} satisfies Eq.~\eqref{Maxweq},
in which $\sqrt{-g} = \Lambda^2 r^2 \sin\theta $. The value of $F^{\phi r}$ is zero at the horizon of a black hole, and its sign changes inside the horizon.

From now on, we present a new solution and the results for the horizon structure. The orthonormal tetrad is employed to measure physical quantities, such as magnetic fields. We adopt the orthonormal frame shown as
\begin{eqnarray}
\label{orthonor}
   e^{\hat t}_{\mu} = \left( \frac{\Lambda \sqrt{\Delta}}{r}, 0, 0, 0 \right) \,,~ e^{\hat r}_{\mu} = \left( 0, \frac{\Lambda r}{ \sqrt{ \Delta}}, 0, 0 \right) \,,
   e^{\hat \theta}_{\mu} = \left(0, 0, \Lambda  r, 0 \right) \,,~ e^{\hat \phi}_{\mu} = \left(0, 0, 0, \frac{r \sin\theta}{\Lambda} \right)\,,
\end{eqnarray}
and
\begin{eqnarray}
\label{orthonoru}
   e_{\hat t}^{\mu} = \left( \frac{r}{\Lambda \sqrt{\Delta}}, 0, 0, 0 \right) \,,~ e_{\hat r}^{\mu} = \left( 0, \frac{ \sqrt{\Delta}}{\Lambda r}, 0, 0 \right) \,,
   e_{\hat \theta}^{\mu} = \left(0, 0, \frac{1}{\Lambda  r}, 0 \right) \,,~ e_{\hat \phi}^{\mu} = \left(0, 0, 0, \frac{\Lambda}{r \sin\theta} \right)\,.
\end{eqnarray}

Magnetic fields take the form of $F^{\hat{a}\hat{b}}=e^{\hat a}_{\mu}e^{\hat b}_{\nu}F^{\mu\nu}$, which gives
\begin{eqnarray}
\label{bfieldortho}
  B^{\hat r} =\frac{B_o \cos\theta}{\Lambda^2} \,,~~ B^{\hat \theta}= -\frac{B_o\sin\theta \sqrt{\Delta} }{r \Lambda^2} \,.
\end{eqnarray}
In the asymptotic region, $r \gg M$ and $Kr^{1-2w}$, $B^{\hat r}\simeq B_o \cos\theta \,, B^{\hat \theta} \simeq - B_o \sin\theta$, thus, $\vec B \simeq  B_o \hat z$.

The Einstein equations and the nonvanishing components of the corresponding stress-energy tensor are given by
\begin{eqnarray}
\label{setensor}
-G^t_t=8\pi{\mathbf{\epsilon}} =&&\frac{B^4_o \sin^4\theta(\Delta + r(r-\Delta'))}{16 \Lambda^4}
   -\frac{B^2_o r^2 \sin^2\theta(-3\Delta+ r(r-\Delta')}{2 r^4 \Lambda^4} \nonumber \\
   &&+\frac{(\Delta+ r(r+B^2_o r^3-\Delta'))}{ r^4 \Lambda^4} \nonumber \ , \\
G^r_r=8\pi {\mathbf{p}}_{\hat{r}}=&&-\frac{B^4_o r^4\sin^4\theta(\Delta + r(r-\Delta'))}{16 r^4 \Lambda^4}
   +\frac{B^2_o r^2 \sin^2\theta(\Delta+ r(r-\Delta')}{2 r^4 \Lambda^4} \nonumber \\
   &&-\frac{(\Delta+ r(r+B^2_o r^3-\Delta'))}{ r^4 \Lambda^4} \nonumber \ , \\
G^{\theta}_{\theta}=8\pi {\mathbf{p}}_{\hat{\theta}}=&&\frac{2B^2_o r^2\cos^2\theta+ 2(1-\frac{1}{4}B^2_o r^2 \sin^2\theta)^2\Delta }{2 r^4 \Lambda^4}
   +\frac{-2\Delta' + r\Delta''}{2 r^3\Lambda^2} \nonumber \ , \\
G^{\phi}_{\phi} = 8\pi {\mathbf{p}}_{\hat{\phi}}=&&\frac{8B^2_o r^2\sin^2\theta(2\Delta + r(2\Delta' + r(-8+ \Delta'') )) }{32 r^4 \Lambda^4}
   + \frac{16(2\Delta + r(2B^2_or^3 - 2\Delta' + r \Delta'' )) }{32 r^4 \Lambda^4}  \nonumber \\
   &&+ \frac{B^4_o r^4\sin^4\theta(-6\Delta + r(6\Delta' + r(-8+ \Delta'') )) }{32 r^4 \Lambda^4}   \ , \nonumber     \\
G^r_{~\theta}= 8\pi T^{r}_{~\theta}=&&\frac{2B^2_o \cos\theta\sin\theta\Delta  }{r \Lambda^4} \ , \nonumber \\
G^{\theta}_{~r}= 8\pi T^{\theta}_{~r}=&&\frac{2B^2_o \cos\theta\sin\theta}{r \Lambda^4} \ ,
\end{eqnarray}
where ${\mathbf{\epsilon}}$ is the energy density, ${\mathbf{p}}_{\hat{r}}$ is the radial pressure, ${\mathbf{p}}_{\hat{\theta}}$ and ${\mathbf{p}}_{\hat{\phi}}$ are tangential pressures, $\Delta$ is shown in Eq.~(\ref{Delta}), $\Delta'=2r-2M + 2(1-w)Kr^{1-2w}$, $\Delta''=2+2(1-w)(1-2w)Kr^{-2w}$,  $T^{r}_{~\theta}$ represents the shear stress or the $r$-component of momentum flux density in the $\theta$ direction, and $T^{\theta}_{~r}$ represents the $\theta$-component of momentum flux density in the $r$ direction.
Regarding the current geometry: $8\pi{\mathbf{\epsilon}}=e^{\mu}_{\hat{t}}e^{\nu}_{\hat{t}}G_{\mu\nu} = g^{tt}G_{tt}=-G^t_t$, $8\pi {\mathbf{p}}_{\hat{r}}=e^{\mu}_{\hat{r}}e^{\nu}_{\hat{r}}G_{\mu\nu} = g^{rr}G_{rr}=G^r_r$, $8\pi {\mathbf{p}}_{\hat{\theta}}=e^{\mu}_{\hat{\theta}}e^{\nu}_{\hat{\theta}}G_{\mu\nu}=g^{\theta\theta}G_{\theta\theta}=G^{\theta}_{\theta}$, $8\pi {\mathbf{p}}_{\hat{\phi}}=e^{\mu}_{\hat{\phi}}e^{\nu}_{\hat{\phi}}G_{\mu\nu}=g^{\phi\phi}G_{\phi\phi}=G^{\phi}_{\phi}$.
The physical shear-stress components in the orthonormal frame are given by
\begin{eqnarray}
\label{orthoshear}
8\pi T_{\hat r \hat \theta} =8\pi T_{\hat \theta \hat r} = \frac{2B^2_o \cos\theta \sin\theta \sqrt{\Delta}}{r \Lambda^4 } \,.
\end{eqnarray}
This satisfies the null-energy condition ${\mathbf{\epsilon}}+{\mathbf{p}}_{\hat{r}}= \frac{2B^2_o \sin^2\theta \Delta}{r^2\Lambda^4} \geq 0$, in which the equality holds at the horizon, $\Delta=0$, on the axis $\sin\theta=0$, and for $B_o=0$.
In this study, the black hole geometry with $w > 1/2$ asymptotically resembles that of the Melvin magnetic universe.

We examine how the anisotropic matter and the equation of state parameters affect the black hole's event horizon structure~\cite{Carter:1969zz}.
The radius of the horizon $r_H$ is determined by solving $g^{rr}=0$ or $\Delta(r_H) = 0$.
It does not depend on whether there is a magnetic field in this geometry.

All physical quantities have dimensions. The radial distance $r$ has the same dimensions as the mass $M$, while the magnetic field $B_o$ has dimensions of $1/r$. The parameter $K$ is a dimensional quantity determined by the relation $(2w-1)K = r^{2w}_o$ for a given value of $w$. In the numerical analysis that follows, we adopt dimensionless variables for all quantities. The equation-of-state parameter $w$ remains a dimensionless constant throughout.
\begin{figure}[H]
    \centering
    \includegraphics[width=7.2cm]{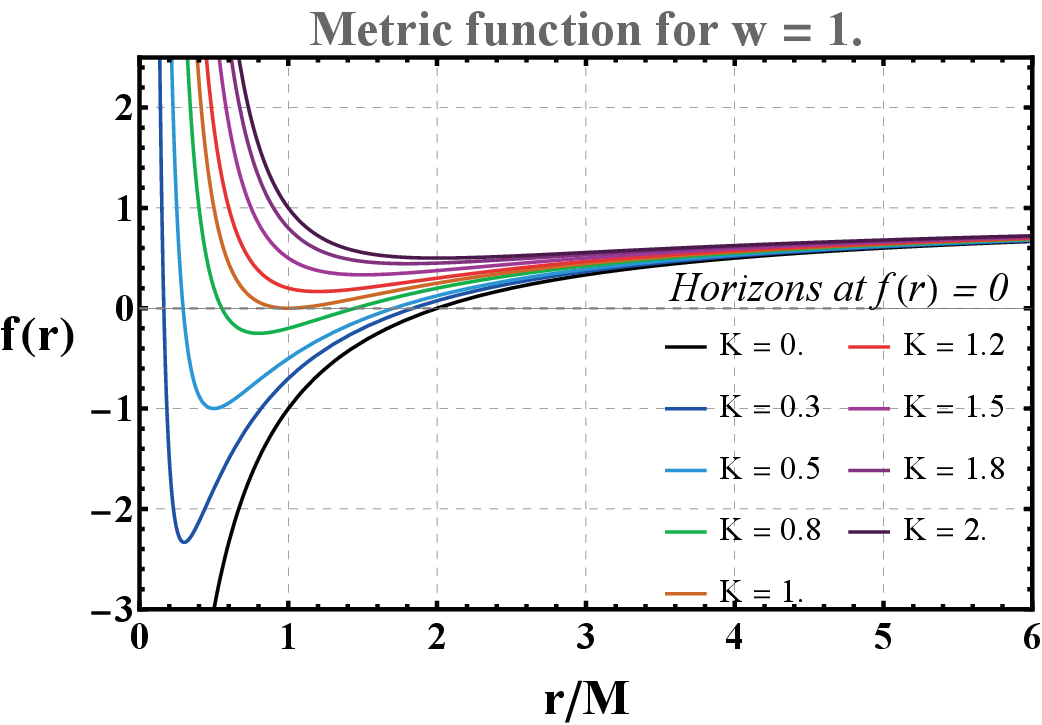}
    \includegraphics[width=7.2cm]{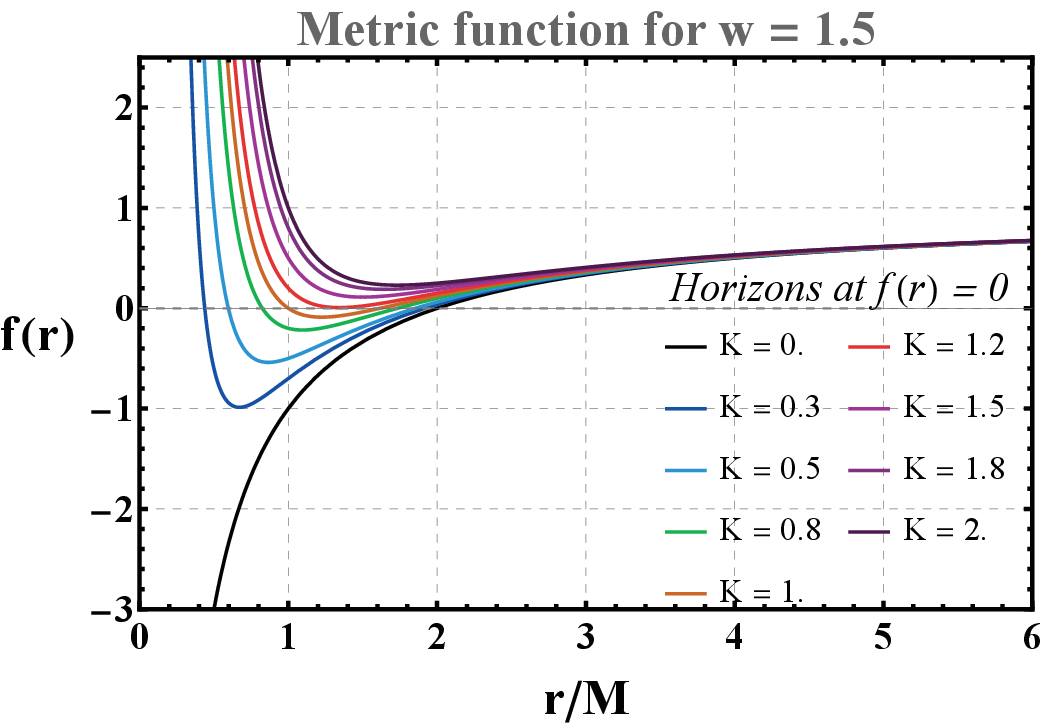}
    \includegraphics[width=7.2cm]{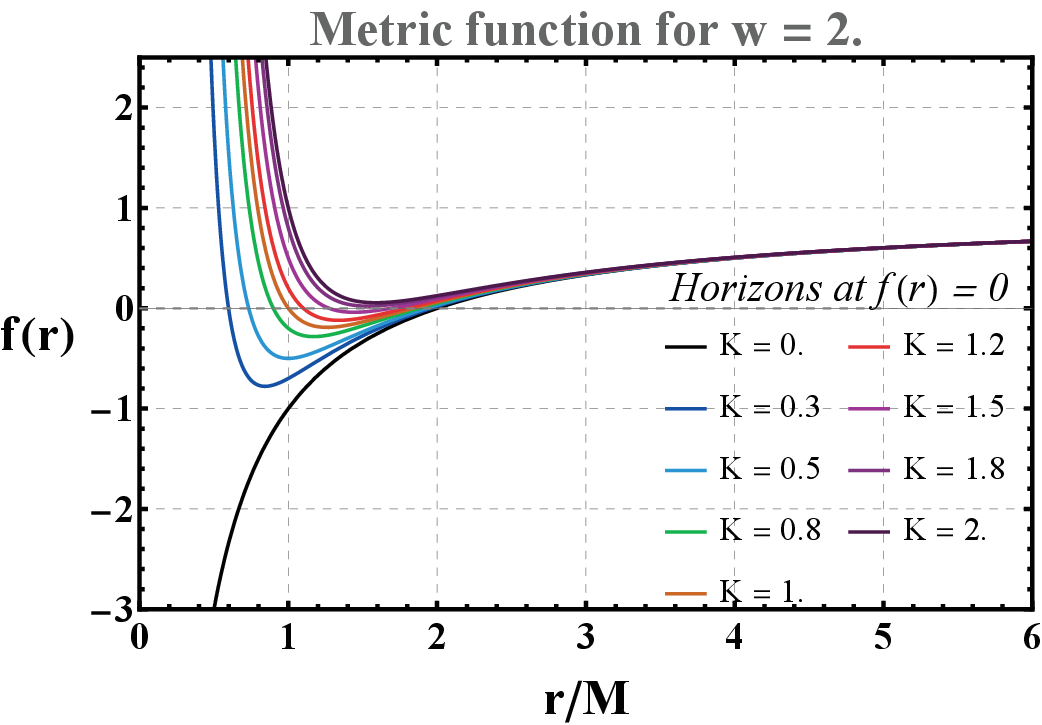}
    \includegraphics[width=7.2cm]{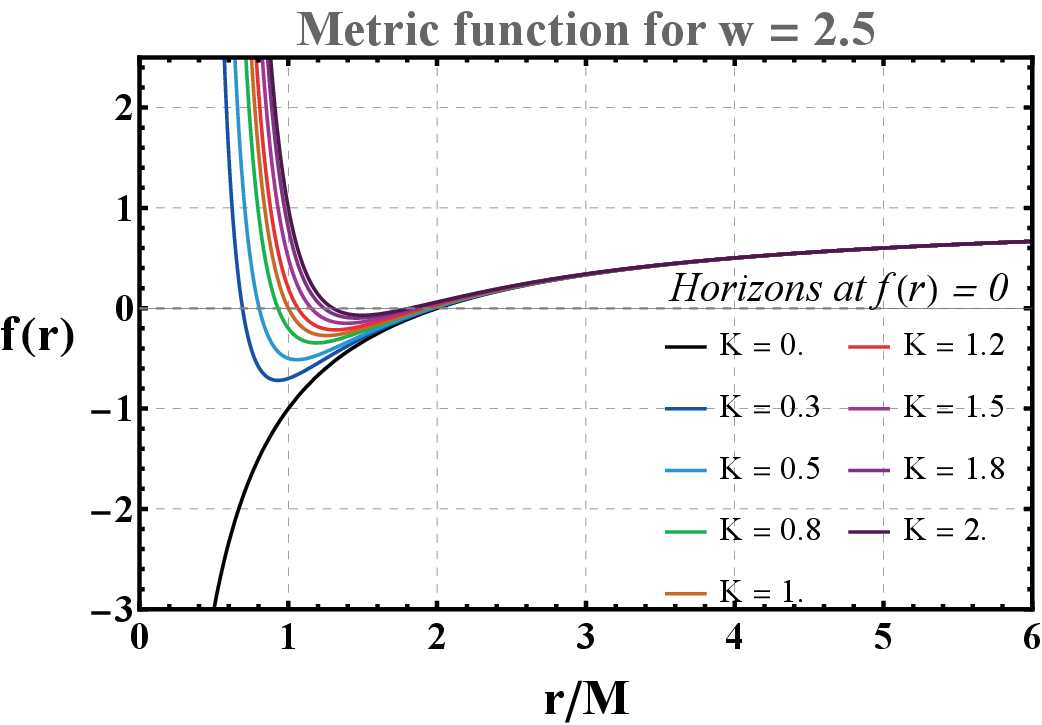}
    \includegraphics[width=7.2cm]{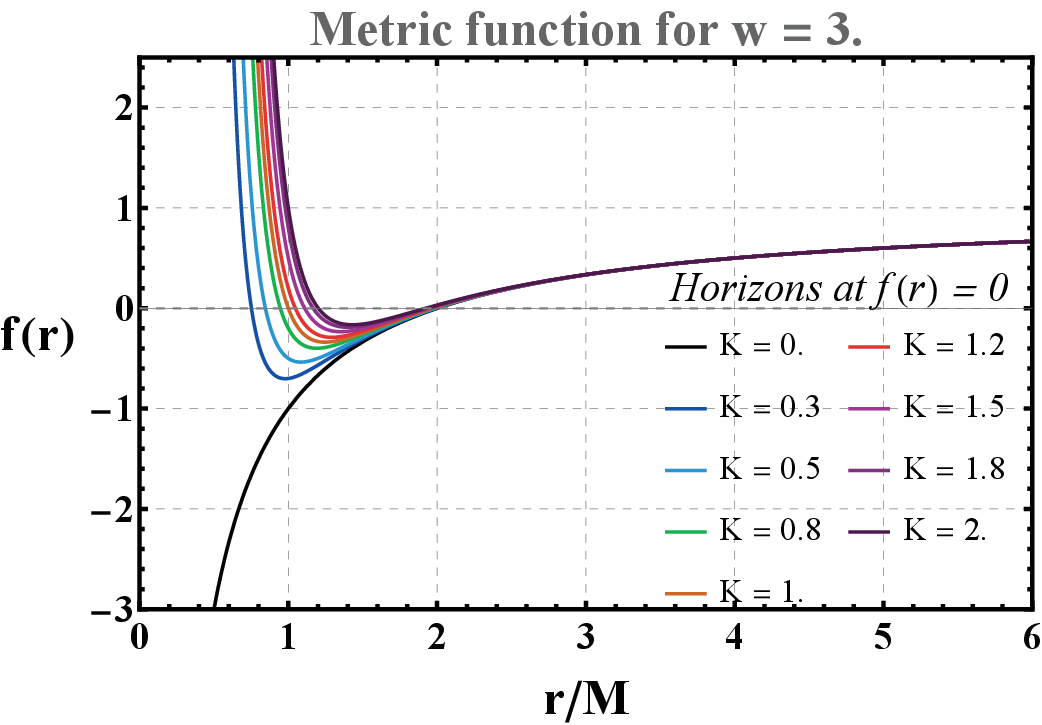}
    \includegraphics[width=7.2cm]{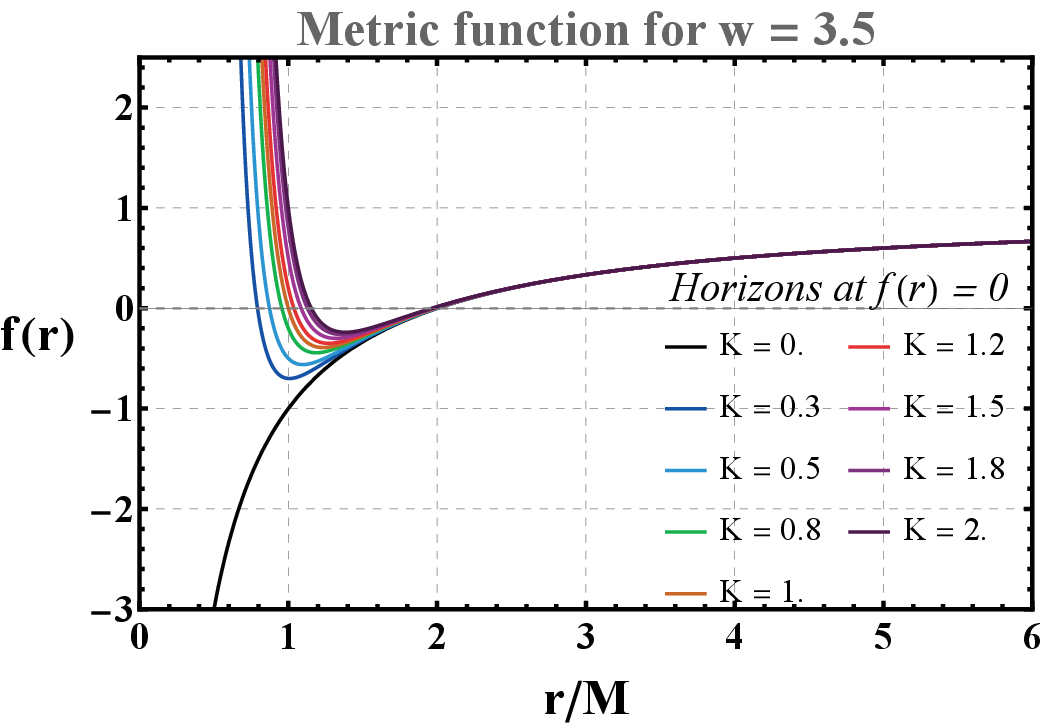}
    \caption{Metric function $f(r)$ versus $r/M$ for eight values of $\omega$ (panel labels) and different values of $K$ (colored curves). Horizons occur at $f=0$ (dashed line). Parameters: $M=1$.}
    \label{fig:omega_metric}
\end{figure}
In Fig.~\ref{fig:omega_metric}, we have plotted the metric function $f(r)$ for eight representative values of the equation-of-state parameter $w \in \{ 1.0,\, 1.5,\, 2.0,\, 2.5,\, 3.0,\, 3.5\}$, each panel containing multiple curves for $K = 0, 0.3, 0.5, 0.8, 1.0, 1.2, 1.5, 1.8, 2.0$. The event horizon is located at the zero-crossing of $f(r)$. The case where $K = 0$, corresponding to the black curve, represents the Schwarzschild black hole. Thus, this case serves as a reference for comparison. For each value of $w$, if $K$ is not equal to zero, two horizons are formed. As the value of $K$ increases, the number of horizons reduces to one.
It becomes an extreme black hole, and if it increases any more, a naked singularity is formed.
\begin{figure}[H]
    \centering
    \includegraphics[width=8.1cm]{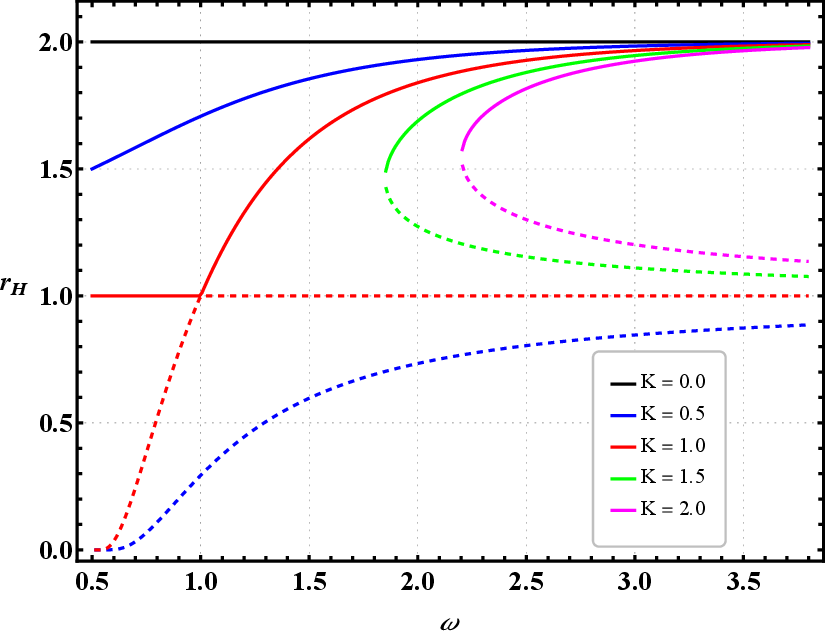}
    \includegraphics[width=8.1cm]{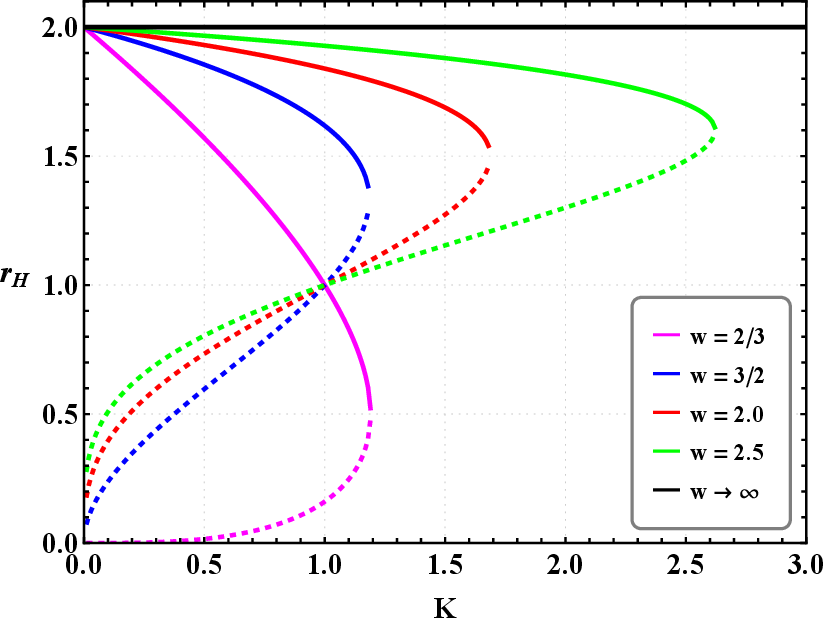}
    \caption{Event horizon radius $r_H/M$ versus $\omega$ (left) and $K$ (right).}
    \label{fig:horizon}
\end{figure}
Figure~\ref{fig:horizon} shows how $r_H$ varies with $w$ and $K$ parameters of the black hole, with $M=1$, $B_0=0.1$.
For a Schwarzschild black hole with $K=0$, the size of the event horizon is the largest. It is shown with a black solid line.
The left panel shows $r_H/M$ versus $w$ for $K = 0, 0.5, 1.01, 1.5, 2.0$.
The right panel shows $r_H/M$ versus $K$ for $w = 2/3, 3/2, 2, 2.5$.
This work looks at the case where $w > 1/2$.
To observe the general trend, imagine a line perpendicular to the $y$-axis.
Where this line intersects the color curves twice, the upper value indicates the position of the outer horizon $r_H$,
In cases where the lower value indicates the position of the inner horizon.
In cases where the lines meet at a single point on the color curves, this indicates the location of the horizon for an extremal black hole.
Regions where the lines never meet indicate a naked singularity with no horizon.
In the left panel, the value $K$ is fixed for each curve, and as the value $w$ increases,
the location of the inner horizon on the lower left approaches the value $1$.

\begin{figure}[H]	
    \centering
    \includegraphics[width=8.1cm]{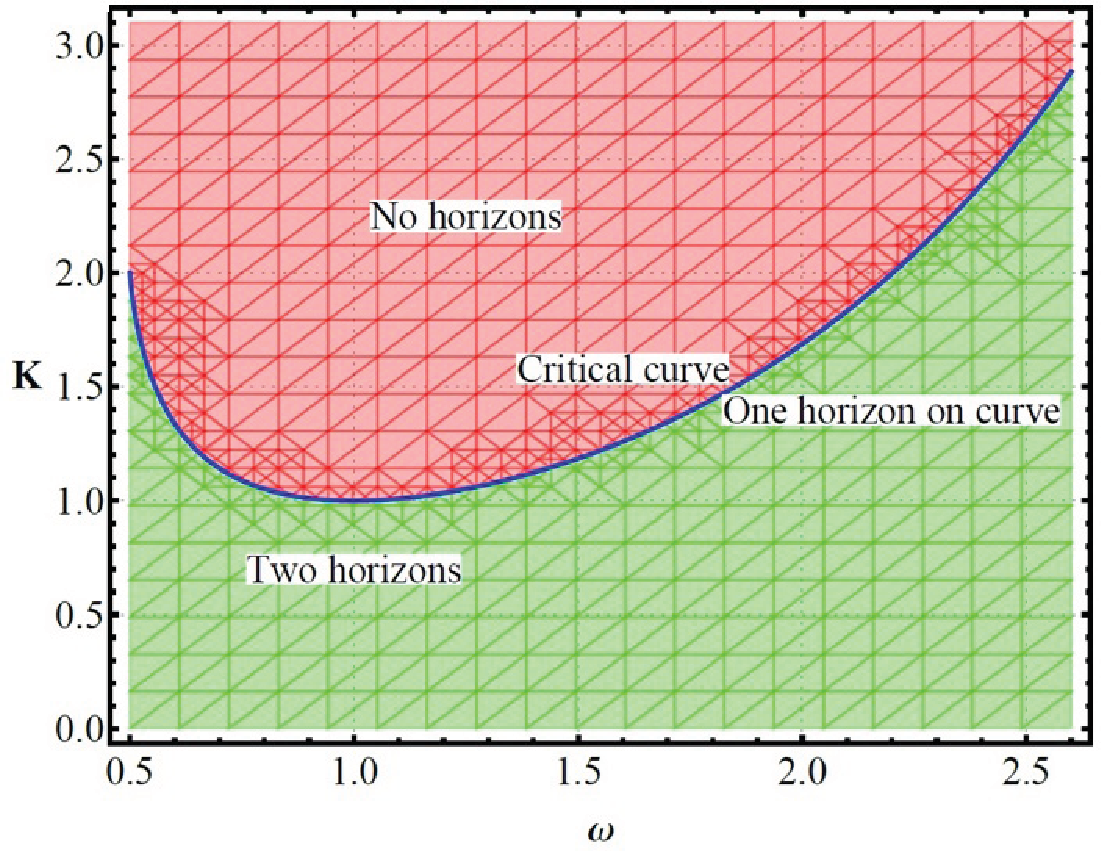}
    \includegraphics[width=8.1cm]{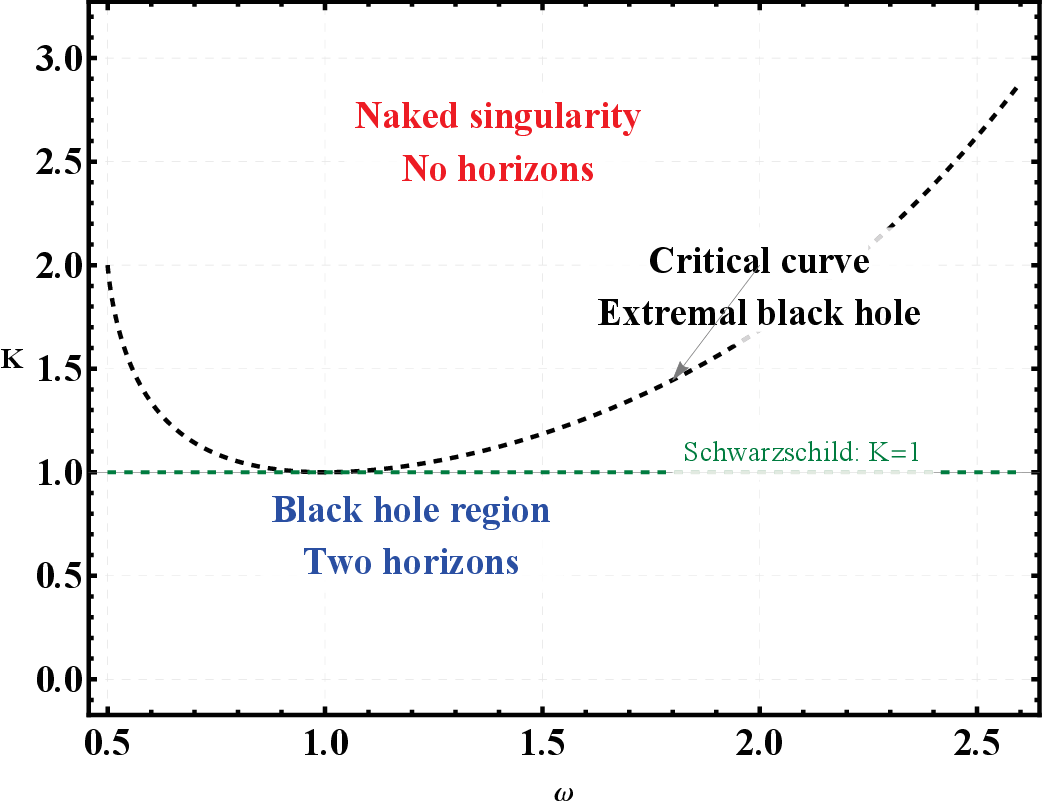}
    \caption{Parameter space $(w, K)$ showing horizon structure regions. Parameters: $M=1$, $B_o=0.1$.}
    \label{fig:param_space}
\end{figure}
Figure~\ref{fig:param_space} provides a more detailed analysis of the cases where the black hole has two horizons, one horizon, or a naked singularity, depending on the values of $w$ and $K$.
The analytical extremal boundary $K_{\rm crit}(\omega)$ is obtained from $f=0$ and $\partial_r f=0$ simultaneously
\begin{equation}
    r_{\rm ext}(\omega) = \frac{M(2\omega-1)}{\omega}, \qquad K_{\rm crit}(\omega) = \frac{M - r_{\rm ext}}{(1-\omega)\,r_{\rm ext}^{1-2\omega}},
    \label{eq:Kcrit}
\end{equation}
with the special case $K_{\rm crit}(\omega=1) = M^2 = 1$. Below this line, the black hole has two horizons; above it the region is the naked-singularity regime.
It is shown that for small $K$ spacetime always has two horizons (or one in the Schwarzschild limit). As $K$ increases, a critical curve is reached along which the two horizons merge into an extremal black hole; above this curve no horizon exists and the spacetime contains a naked singularity. The critical curve has a global minimum at $w = 1$ (where $K_{\rm crit} = M^2$) and rises on both sides.
The right panel makes the physical regions clear: the lower-left region (two horizons) is the physically relevant domain for our orbital analysis.

\begin{figure}[H]
\begin{center}
\subfigure[For the fixed value of $w=3/2$]
{\includegraphics[width=6.5in]{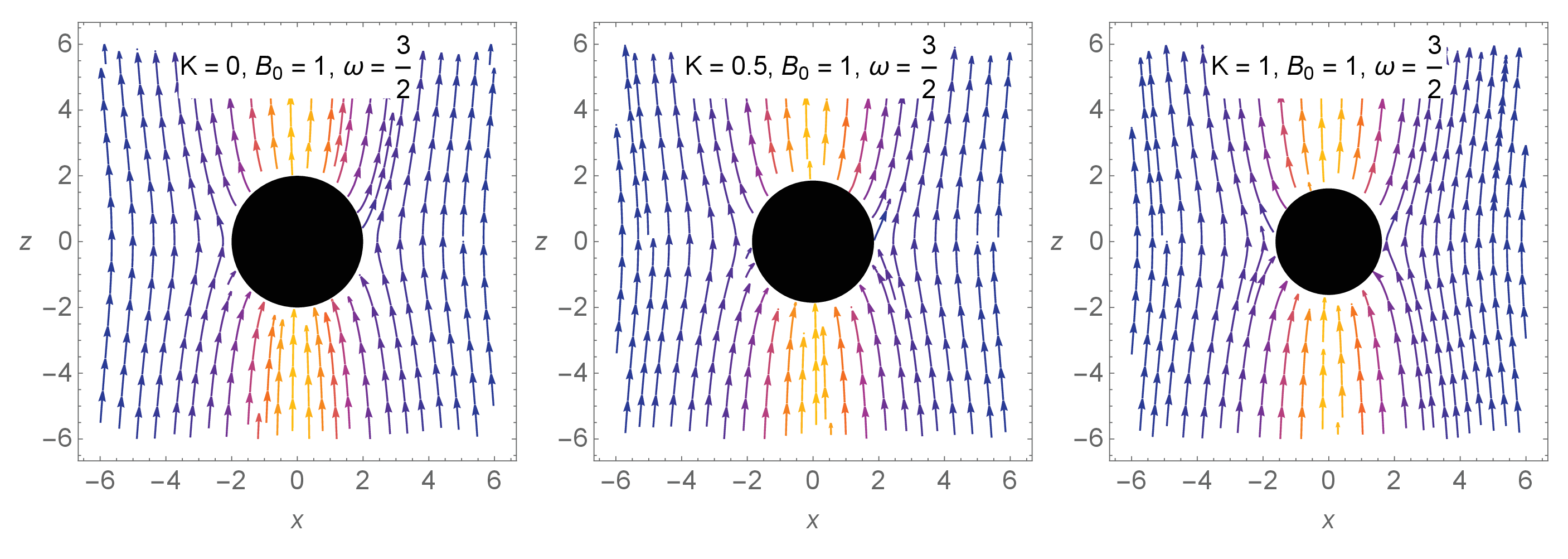}} \\
\subfigure[For the fixed value of $K=0.5$]
{\includegraphics[width=6.5in]{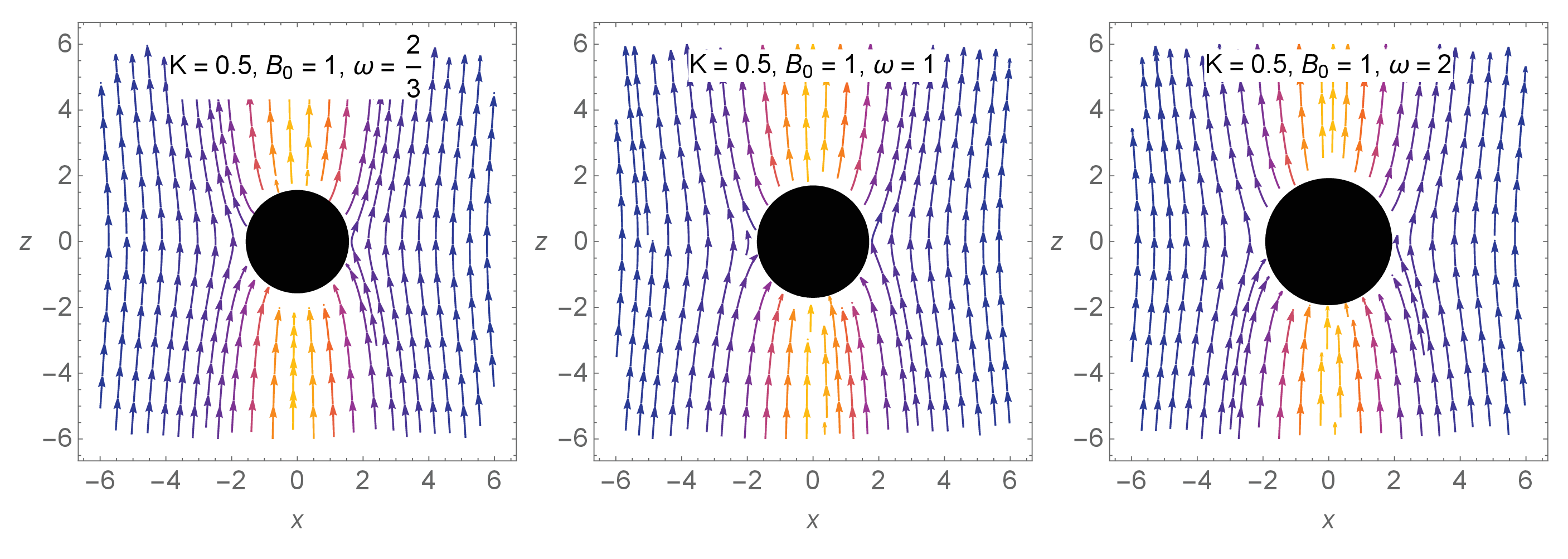}}
\end{center}
\caption{\footnotesize{(color online).
Magnetic field lines around a black hole immersed in an external magnetic field in the presence of an anisotropic matter field }}
\label{fig:magneticfieldw}
\end{figure}
In Fig.~\ref{fig:magneticfieldw}, we present the magnetic field configuration (\ref{bfieldortho}) for a black hole surrounded by an anisotropic matter field and immersed in an external magnetic field. At large radial distances the field lines remain nearly uniform, consistent with the asymptotic Melvin geometry, whereas in the near-horizon region they undergo significant bending due to the combined action of the gravitational field and the external magnetic flux. The presence of the anisotropic matter parameter $K$ enhances this effect (Fig.~\ref{fig:magneticfieldw}): larger values of $K$ lead to a stronger focusing of magnetic flux tubes toward the black hole horizon. The same effect can be seen in Fig.~\ref{fig:magneticfieldw} with the parameter $w$. Correspondingly, the radius of the event horizon exhibits a nontrivial dependence on these parameters: increasing K decreases the radius of the horizon, while larger values of $w$ produce an expansion of the horizon. This distortion modifies the structure of the effective potential experienced by both neutral and charged test particles. In particular, neutral geodesics are shifted due to the change in the background metric, while charged particles are directly guided along the distorted field lines, leading to enhanced confinement and possible acceleration in the near-horizon region. These features highlight the role of anisotropic matter in amplifying  gravitational-magnetic interaction and may have direct implications for plasma dynamics and jet collimation processes around a black hole.

\subsection{Separability structure of the spacetime and the  Hamilton-Jacobi equation \label{sec2-2} }

\quad

In this section, we analyze the separability structure and related separation of variables for the equations of motion~\cite{Carter:1968rr, Walker:1970un, Benenti:1979erw, Demianski:1980mgt}. This involves identifying the number of Killing vectors and tensors in the given spacetime~\cite{Frolov:2017kze}. These Killing vectors and tensors are then contracted with the momentum of geodesic particles to form constants of motion. In four-dimensional spacetime geometry, four constants of motion can be formed with two Killing vectors and two Killing tensors. Consequently, the equations of motion for geodesic particles turn out to be four first-order differential equations that are integrable~\cite{Jeong:2023hom, Lee:2021sws}. If one Killing field among four is not allowed, we cannot form four first-order differential equations.
This will cause coupled equations to arise. Now, let us examine the structure of our spacetime geometry.

The two conserved quantities are related to the Killing vectors
\begin{equation}
\label{killingve}
     \xi^{\mu}_{(t)} \pi_{\mu}  = -E \,,~~ \xi^{\mu}_{(\phi)} \pi_{\mu} = L   \ ,
\end{equation}
where we adopted the generalized canonical momentum $\pi_{\mu}=p_{\mu}+ q A_{\mu}$  for the particle with charge $q$ and
$p_{\mu}$ corresponds to the mechanical momentum. $E$ and $L$ which correspond to the energy of the test particle and its angular momentum at infinity.
The quantity is related to the normalization of the four-momentum of a particle
\begin{equation}
\label{killingte}
     g^{\mu\nu} p_{\mu} p_{\nu} =-m^2 \,,
\end{equation}
which would be treated as the Killing tensor.
This spacetime geometry does not allow for a Killing tensor associated with Carter's constant~\cite{Carter:1968rr}.

From Eq.~(\ref{killingve}) we obtain two first-order differential equations
\begin{eqnarray}
\label{ptpphi}
p^t= \frac{dt}{d\sigma}= \frac{ r^2 E }{\Delta \Lambda^2}  \,, \quad
p^{\phi}=\frac{d\phi}{d\sigma} = \frac{L \Lambda^2}{r^2 \sin^2\theta} - \frac{q B_o \Lambda}{2}\,,
\end{eqnarray}
where $m\sigma=\tau$.

We now check the Hamilton-Jacobi equation, which is given by
\begin{eqnarray}
\label{hajaeq}
\frac{\partial S}{\partial \sigma} =\frac{m^2}{2} =-\frac{1}{2} g^{\mu\nu}\left(\partial_{\mu}S-q A_{\mu} \right) \left(\partial_{\nu}S- q A_{\nu} \right)\,.
\end{eqnarray}
Thanks to cyclic coordinates for $t$ and $\phi$, $S$ can be represented as follows:
\begin{eqnarray}
S= \frac{\sigma}{2} m^2 -E t + L \phi + S_{r\theta}(r, \theta)\,,
\end{eqnarray}
where the coupled $S_{r\theta}(r, \theta)$ is due to the lack of the Killing tensor associated with Carter's constant.
Thus, the coupled equation is given by
\begin{eqnarray}
\label{mpart}
\frac{m}{2}\left( \frac{dr}{d\tau} \right)^2 + \frac{m}{2}\Delta \left( \frac{d \theta}{d\tau} \right)^2 + {\cal V}_{\rm effp} (r, \theta)  =0 \,,
\end{eqnarray}
where ${\cal V}_{\rm effp} (r, \theta) =\frac{\Delta}{2m \Lambda^2 r^2} \left[  -\frac{r^2 E^2}{\Delta \Lambda^2} +  \frac{\Lambda^2}{r^2\sin^2\theta} \left(L - \frac{q B_o r^2 \sin^2\theta}{2\Lambda} \right)^2   + m^2 \right ]$, $p^r =dr/d\sigma$, $p^{\theta}=d\theta/d\sigma$. The Hamiltonian is given by
\begin{eqnarray}
\label{hamilton}
{\cal H} &=& \frac{1}{2}g^{\mu\nu}(\pi_{\mu} -q A_{\mu})(\pi_{\nu} -q A_{\nu}) = - \frac{m^2}{2} \nonumber \\
    &=& \frac{1}{2} \left[  -\frac{r^2 E^2}{\Delta \Lambda^2} + \frac{\Delta \pi^2_r}{\Lambda^2 r^2}  + \frac{\pi^2_{\theta}}{\Lambda^2 r^2}   +  \frac{\Lambda^2}{r^2\sin^2\theta} \left(L - \frac{q B_o r^2 \sin^2\theta}{2\Lambda} \right)^2  \right ]  \,.
\end{eqnarray}

\section{Particle trajectories and chaotic behaviors \label{sec3}}
\quad

In this section, we aim to examine the trajectories of particles.
By investigating the geodesic behavior of particles, we hope to understand the present black hole geometry.
In particular, Einstein's theory of gravitation encompasses the Keplerian orbits of Newtonian theory,
as well as two additional orbits caused by curvature effects that do not exist in Newtonian theory.
These are the innermost stable circular orbit (ISCO)~\cite{Bardeen:1972fi, Ferrari:2020nzo} and the homoclinic orbit~\cite{Levin:2008yp, Perez-Giz:2008ajn, Li:2023bgn}.
Furthermore, two indicators are employed to detect chaotic behavior in particle motion:
the Lyapunov exponent and the Poincar\'e section in phase space.
In this section, the focus is on the homoclinic orbit and the presence of chaotic behavior.

\subsection{Homoclinic orbits \label{sec3-1} }

\quad

In the geometry of a black hole, a homoclinic orbit is characterized as an unstable circular orbit that serves as a separatrix between inspiral and plunging orbits. This can generally be identified by examining the shape of the effective potential in the geodesic equation of the particle.

In the present geometry, we look for it in the equatorial plane, where $\theta=\pi/2$.
From Eq.~(\ref{killingve}), we obtain the radial equation
\begin{equation}
\label{mpartthe}
\frac{m}{2}\left( \frac{dr}{d\tau} \right)^2  + {\cal V}_{\rm effp} (r)  =0 \,,
\end{equation}
where ${\cal V}_{\rm effp} (r) = \frac{\Delta}{2m\Lambda^2(r) r^2} \left[  -\frac{r^2 E^2}{\Delta \Lambda^2(r)} +  \frac{\Lambda^2(r)}{r^2} \left(L - \frac{q B_o r^2}{2\Lambda(r)} \right)^2   + m^2 \right ]$ with $\Lambda(r) = 1+ \frac{1}{4} B^2_o r^2$.
The location of the homoclinic orbit corresponds to finding the local maximum of the potential; therefore, it locates the point where $V_{\rm effp}|_{r_{\rm ho}}=0$, $dV_{\rm effp}/dr|_{r_{\rm ho}}=0$, and $d^2V_{\rm effp}/dr^2|_{r_{\rm ho}} < 0$.
\begin{figure}[H]
\begin{center}
\subfigure[$w=0.7$ and $K=0.01$]
{\includegraphics[width=6.8cm]{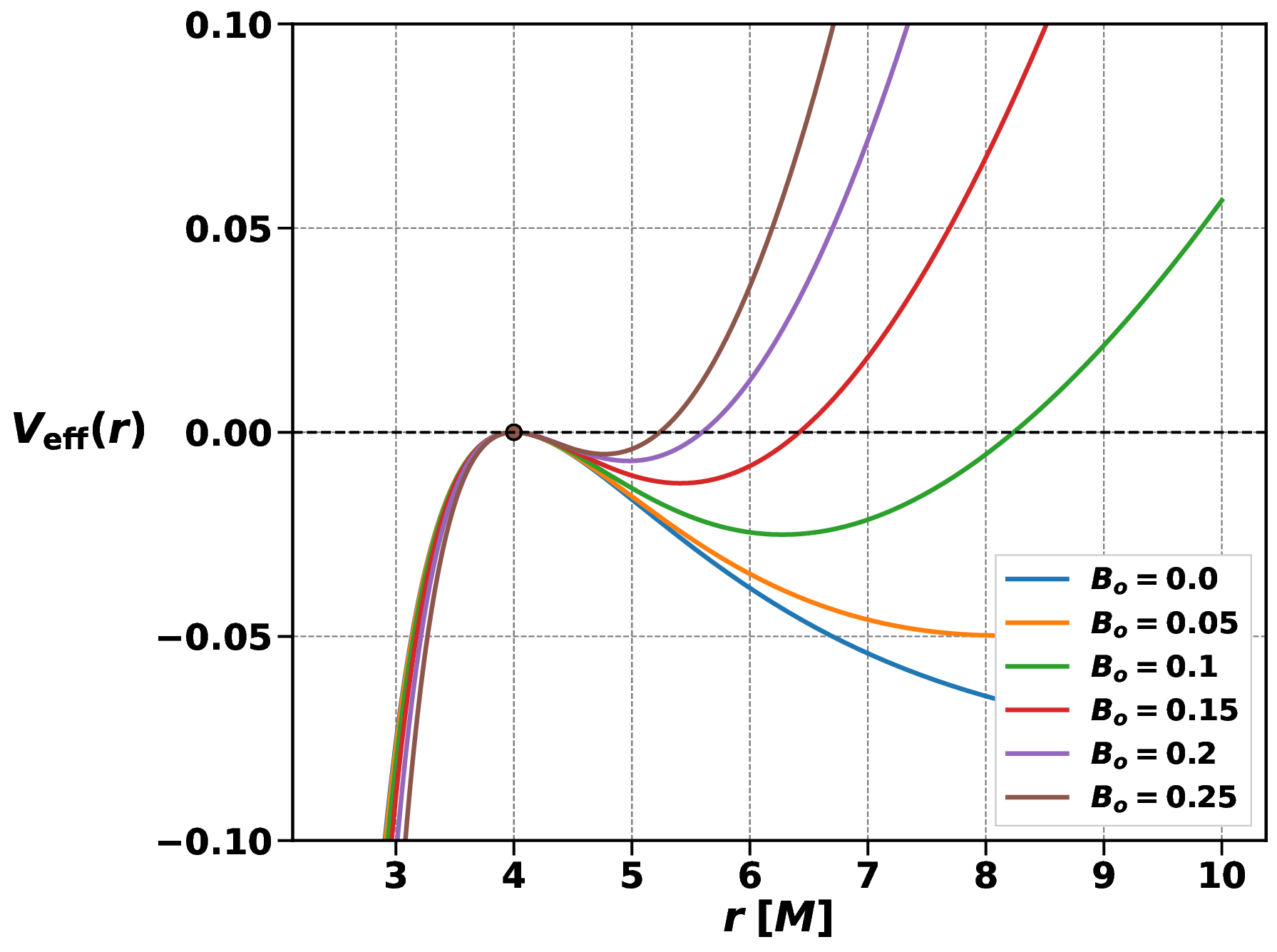}}
\subfigure[$w=0.7$ and $B_o=0.1$]
{\includegraphics[width=6.8cm]{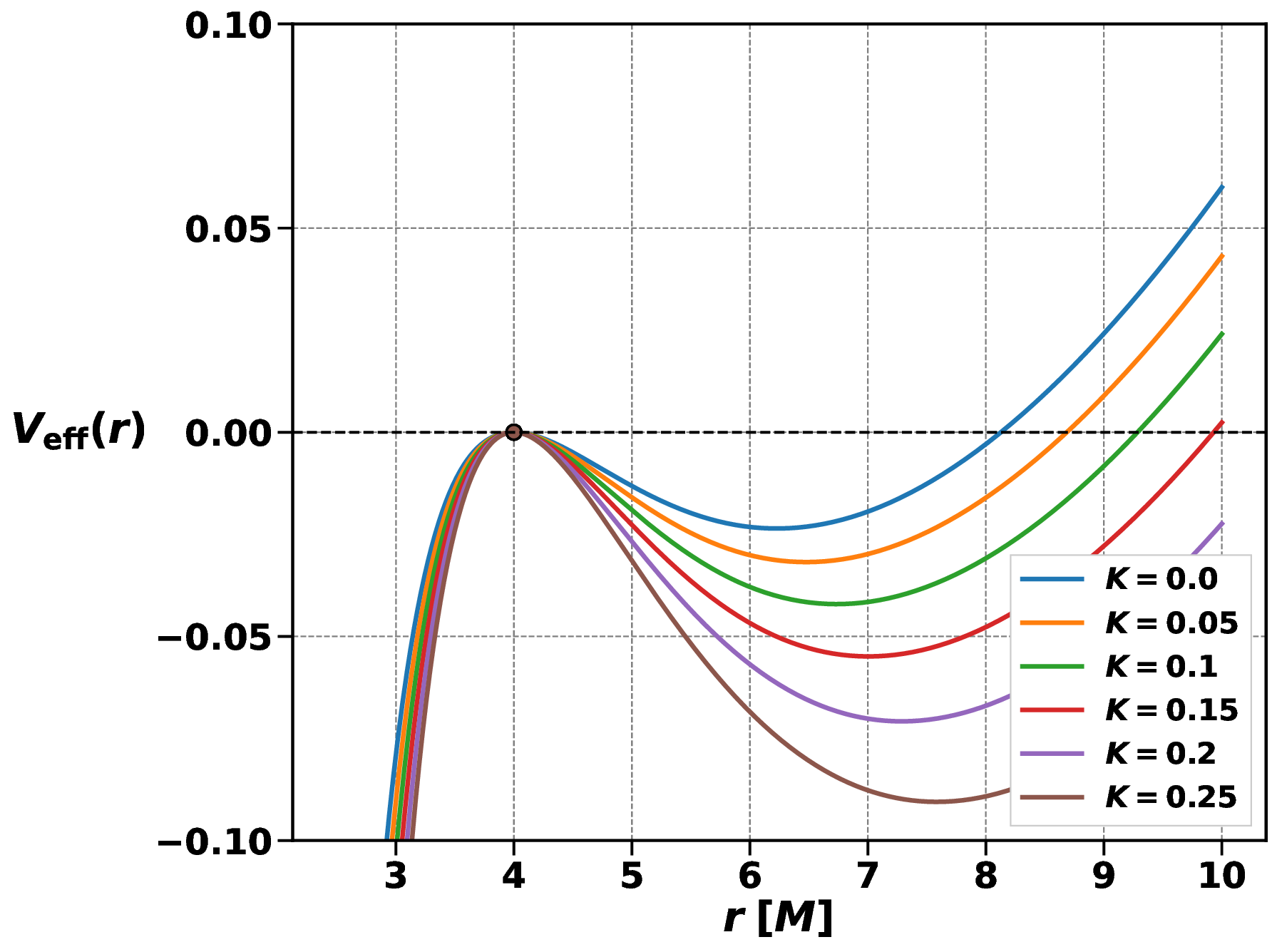}}
\subfigure[$w=1.5$ and $K=0.01$]
{\includegraphics[width=6.8cm]{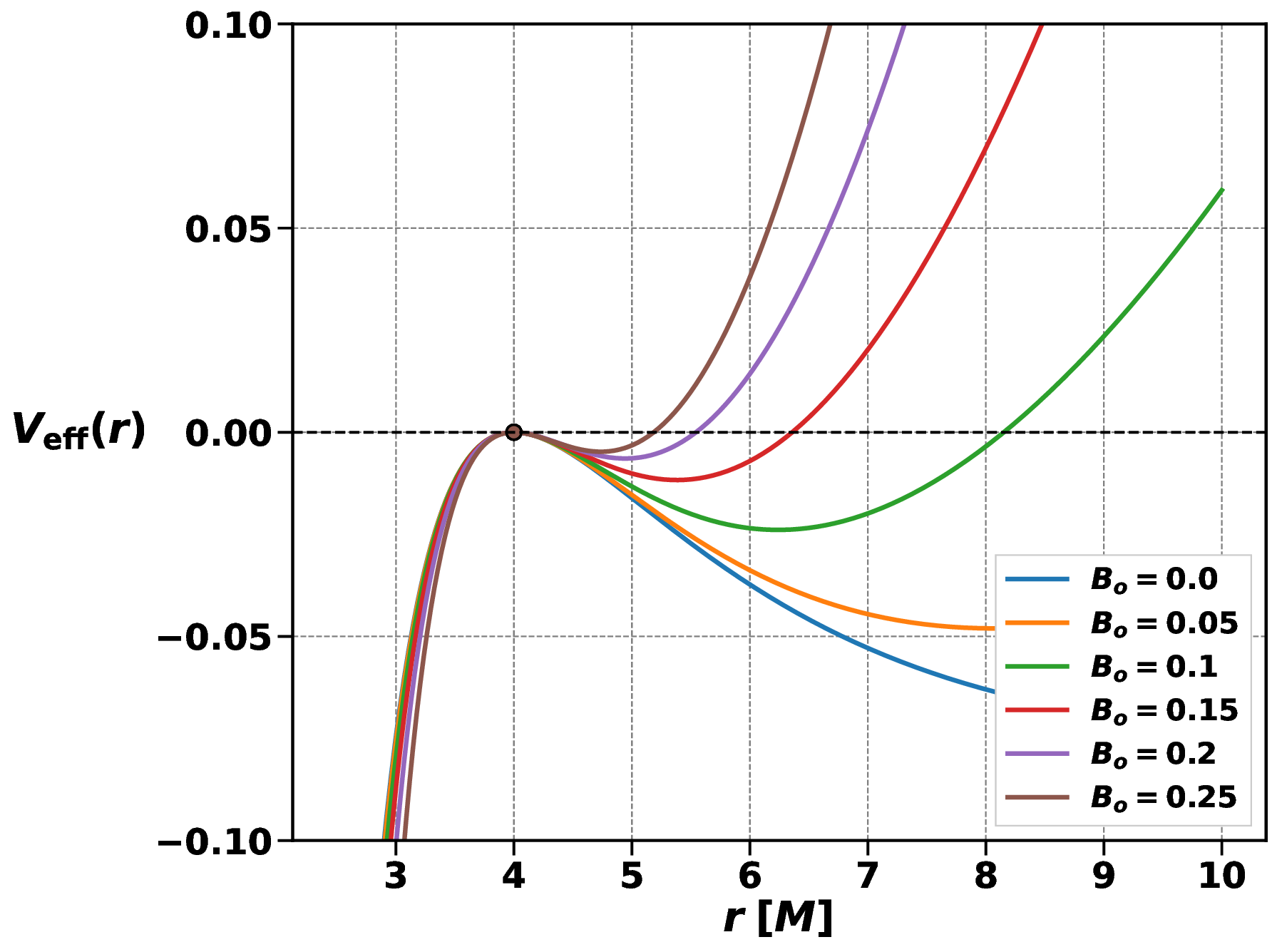}}
\subfigure[$w=1.5$ and $B_o=0.1$]
{\includegraphics[width=6.8cm]{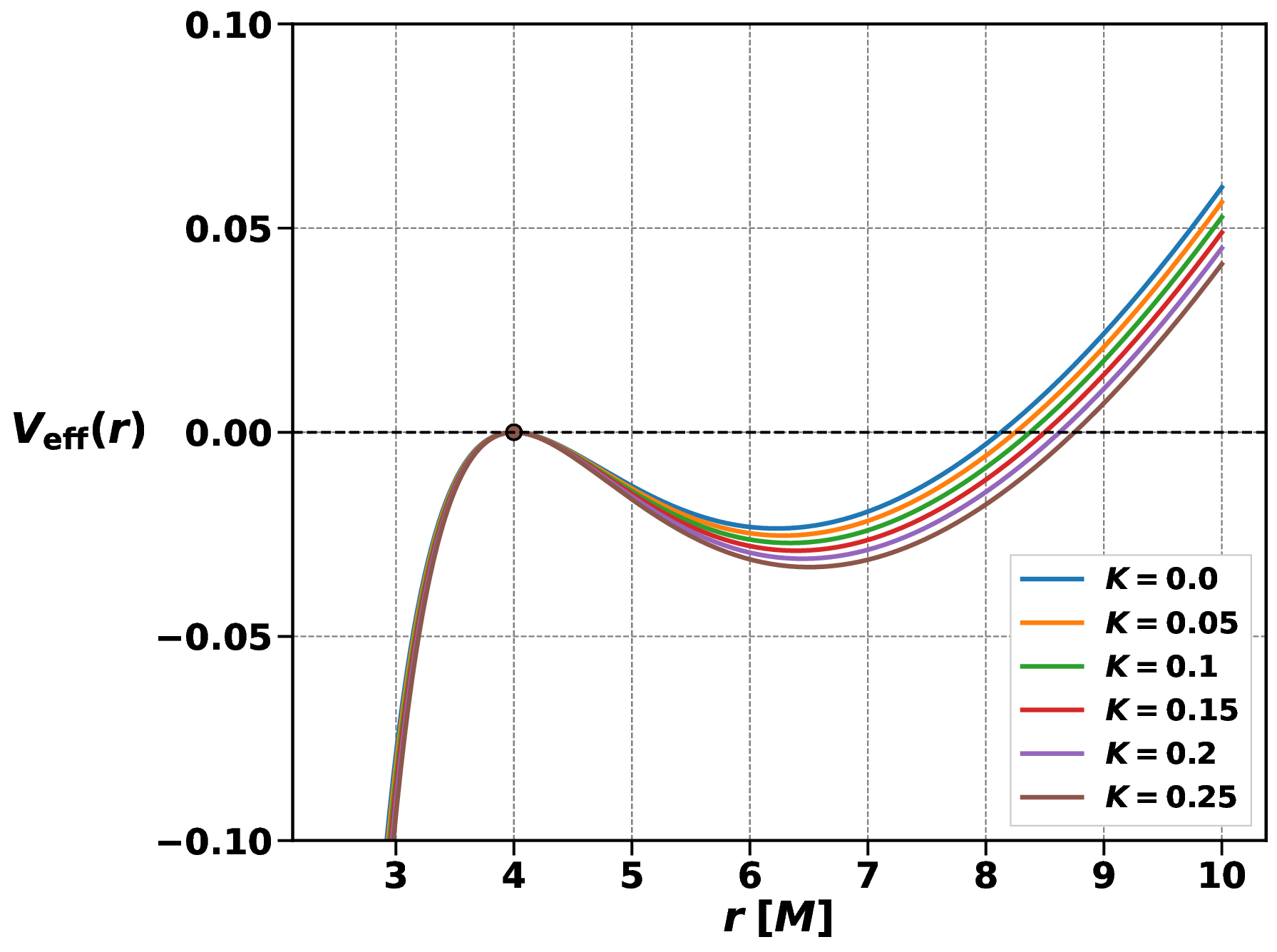}}
\subfigure[$w=2.0$ and $K=0.01$]
{\includegraphics[width=6.8cm]{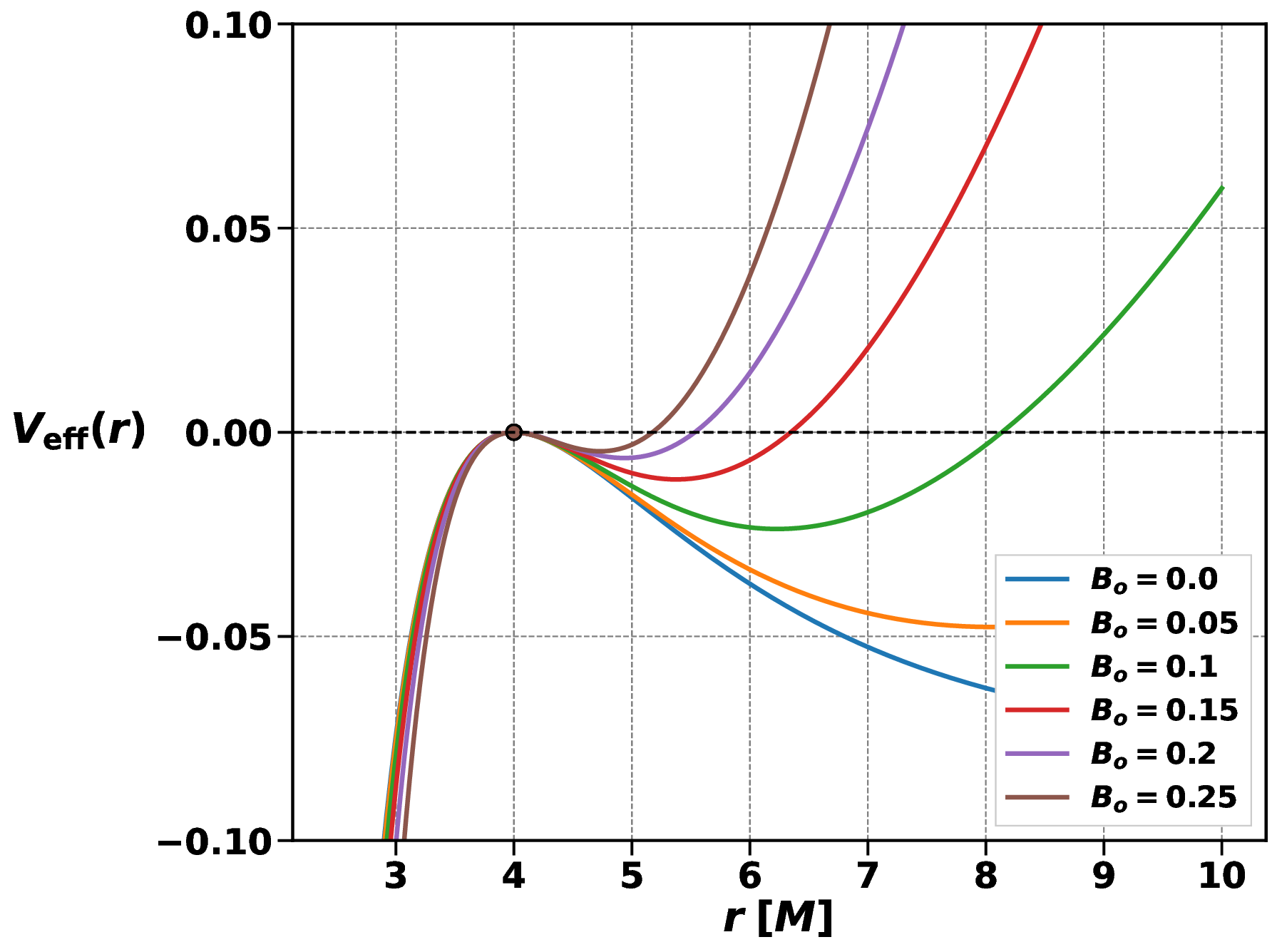}}
\subfigure[$w=2.0$ and $B_o=0.1$]
{\includegraphics[width=6.8cm]{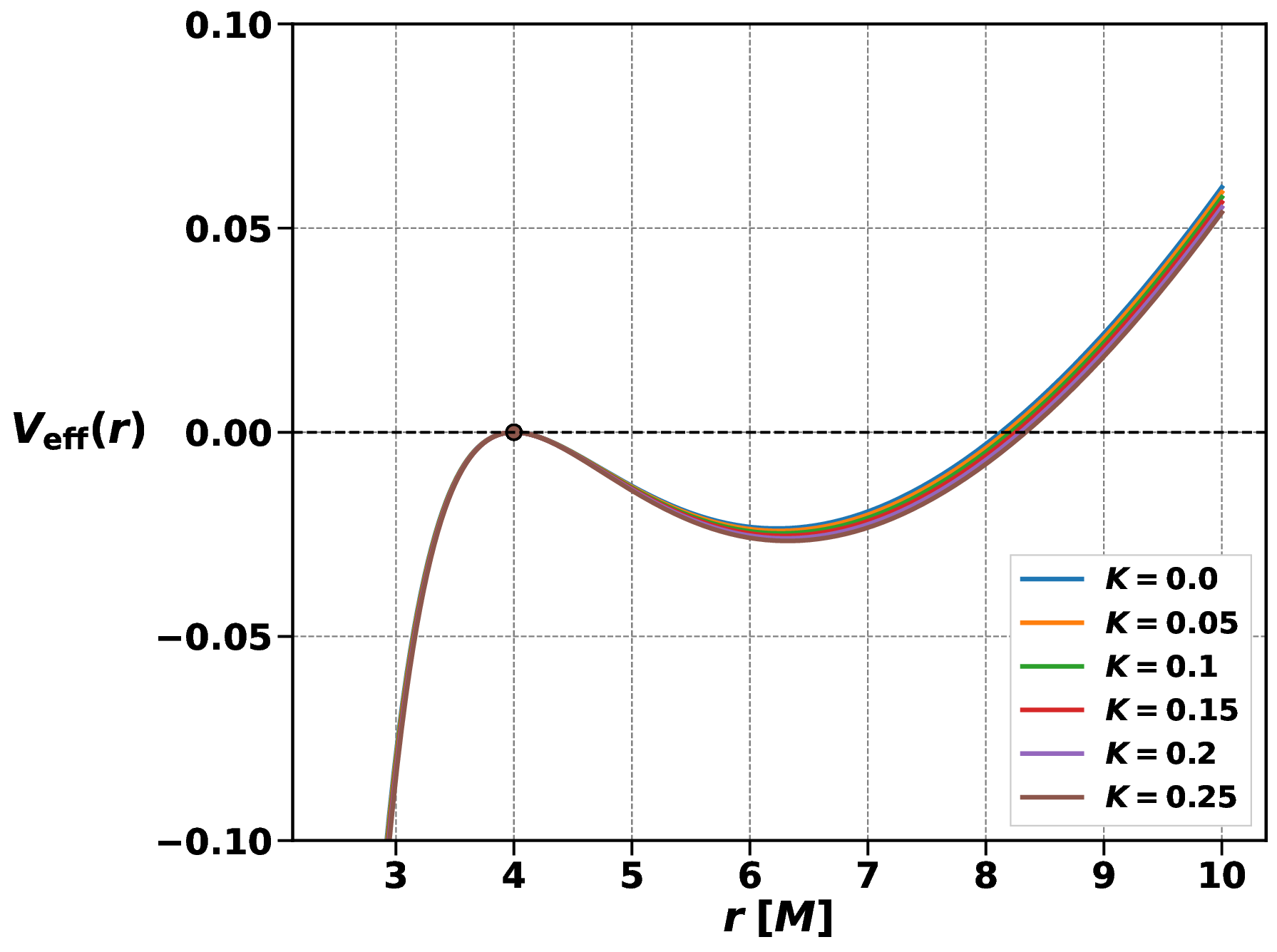}}
\end{center}
\caption{\footnotesize{(color online).
${\cal V}_{\rm effp}(r)-E^2$ for varying $B_o \in [0, 0.25]$ at $K=0.01$ for the left panel and for varying $K\in[0,\,0.025]$ at $B_o=0.1$ for the right panel. Each row corresponds to $w=0.7$, $w=1.5$, and $w=2.0$, respectively. In the left panel of the figure, the right side is clearly separated based on the $B_o$ value. }}
\label{fig:effB_varyB}
\end{figure}
In Fig.~\ref{fig:effB_varyB}, we plot the effective potential ${\cal V}_{\rm effp}(r)$ in the equatorial plane for three values of $w = 0.7, 1.5, 2.0$, varying the magnetic field strength $B_o \in \{0.0, 0.05, 0.1, 0.15, 0.2, 0.25\}$ at fixed $K = 0.01$ for the left panel, while varying the anisotropic matter parameter $K \in \{0.0, 0.005, 0.01, 0.015, 0.02, 0.025\}$ at fixed $B_o=0.1$ for the right panel.
Each figure shows how the magnitude of the magnetic field modifies the shape of the effective potential at the corresponding parameter values.
The black dot corresponds to the location of the homoclinic orbits.
In the left panel, the blue curve corresponds to $B_o = 0.00$ and represents the effective potential shown in Fig,~$1$ of Ref.~\cite{Jeong:2023hom}.
The orange curve corresponds to $B_o = 0.05$, the green curve to $B_o = 0.10$, the red curve to $B_o = 0.15$, the purple curve to $B_o = 0.20$,
and the brown curve to $B_o = 0.25$, respectively.
As the magnetic field becomes stronger, a wall of potential energy forms on the right side of each figure, causing the curve to rise. In other words, the region of the allowed bound orbit that is designated as the $r$-direction region undergoes a decrease.
We find that as $B_o$ increases, the local maximum (which marks the unstable circular orbit or the homoclinic orbit, shown by the filled black circle) shifts to smaller radii. This tells us that a stronger magnetic field brings the unstable orbit closer to the black hole. The inset in each panel magnifies the region around the maximum so that small differences between $B_o$ values are visible. We also figure out that the overall shape of ${\cal V}_{\rm effp}$ becomes more sensitive to the magnetic field at smaller $w$, while the three panels look increasingly similar as $w$ grows, confirming that the matter effect weakens at high $w$.
In the right panel, the blue curve corresponds to $K = 0.00$ and represents the effective potential of the Schwarzschild black hole immersed in a magnetic field.
The orange curve corresponds to $K = 0.005$, the green curve to $K = 0.010$, the red curve to $K = 0.015$, the purple curve to $K = 0.020$,
and the brown curve to $K = 0.025$, respectively.
We find that increasing $K$ causes the local maximum (black dot) to shift and its height to change, in a way that is opposite to the effect of the magnetic field: while $B_o$ generally enhances the barrier, $K$ acts to lower and modify it.
The suppression is strongest with $w = 0.7$ (panel b) because the long-range matter profile reaches further into the orbital region, whereas with $w = 2.0$ (panel f) the different $K$ curves nearly coincide, showing that the matter effect becomes negligible.

\subsection{Lyapunov exponent \label{sec3-2} }
\quad

Our aim here is to examine the Lyapunov exponent as an indicator of the sensitivity of particles to initial conditions. We focus on two particles: one is situated on an unstable circular orbit, specifically a homoclinic orbit described by $r=r_{\rm ho}$, while the other experiences slight variations in its parameters due to a small perturbation, represented by $r=r_{\rm ho} + \varepsilon$. By analyzing the Lyapunov exponent within this framework, we aim to explore the local chaotic behavior of the particles. For this analysis, we utilize the coordinate time $t$ as the independent variable instead of the proper time $\tau$, as the particles have differing proper times~\cite{Wu:2003pe}. This approach is essential for the numerical integration of the equations of motion.

For two initial points $r_{\rm ho}$ and $r_{\rm ho} + \varepsilon$, the Lyapunov exponent $\lambda$ corresponds to the measure
that shows the average exponential growth per unit of time between two trajectories:
\begin{equation}
\label{eq:lyapunov_definition}
d(t) \sim \epsilon e^{\lambda t}\,,
\end{equation}
where the negative $\lambda$ indicates that the orbits converge, while a positive value signifies divergence, which can lead to chaotic behavior.

It has been conjectured that there is a universal upper bound on the Lyapunov exponent in thermal quantum systems~\cite{Maldacena:2015waa}.
This concept has also been applied to black hole systems~\cite{Hashimoto:2016dfz} as
\begin{equation}
\lambda \leq \frac{2\pi T_H}{\hbar} = \kappa,
\label{eq:chaos_bound_BH}
\end{equation}
where $T_H$ is  the black hole  temperature and $\kappa$ is the surface gravity.
The radial equation in coordinate time is given by
\begin{equation}
\label{eq:radial_coordinate_time}
\frac{1}{2}m\left(\frac{dr}{dt}\right)^2 + {\cal V}_{\rm effc} = 0\,,
\end{equation}
where the effective potential is
\begin{equation}
\label{eq:Veff_coordinate}
{\cal V}_{\rm effc} = \frac{m \Delta^3 \Lambda^2(r)}{2 r^6 E^2} \left[  -\frac{r^2 E^2}{\Delta \Lambda^2(r)} +  \frac{\Lambda^2(r)}{r^2} \left(L - \frac{q B_o r^2}{2\Lambda(r)} \right)^2   + m^2 \right ]\,.
\end{equation}
At $r \sim r_{\rm ho} + \varepsilon$, the equation becomes
\begin{equation}
\label{eq:perturbation_eq}
0 \simeq \frac{1}{2}m\left(\frac{d\epsilon}{dt}\right)^2 + \left[{\cal V}_{\rm effc}(r_{\rm ho}) + \frac{1}{2}{\cal V}''_{\rm effc}(r_{\rm ho})\epsilon^2\right] \simeq \frac{1}{2}m\left[\left(\frac{d\epsilon}{dt}\right)^2 - \lambda^2\epsilon^2\right]\,,
\end{equation}
where the coefficient of $ \varepsilon^2$ gives the Lyapunov exponent
\begin{equation}
\label{eq:lyapunov_formula}
\lambda^2 = -\frac{{\cal V}''_{\rm effc}(r)|_{r=r_{\rm ho}}}{m}\,.
\end{equation}

\begin{figure}[H]
\begin{center}
\subfigure[$w=0.7$]
{\includegraphics[width=8.1cm]{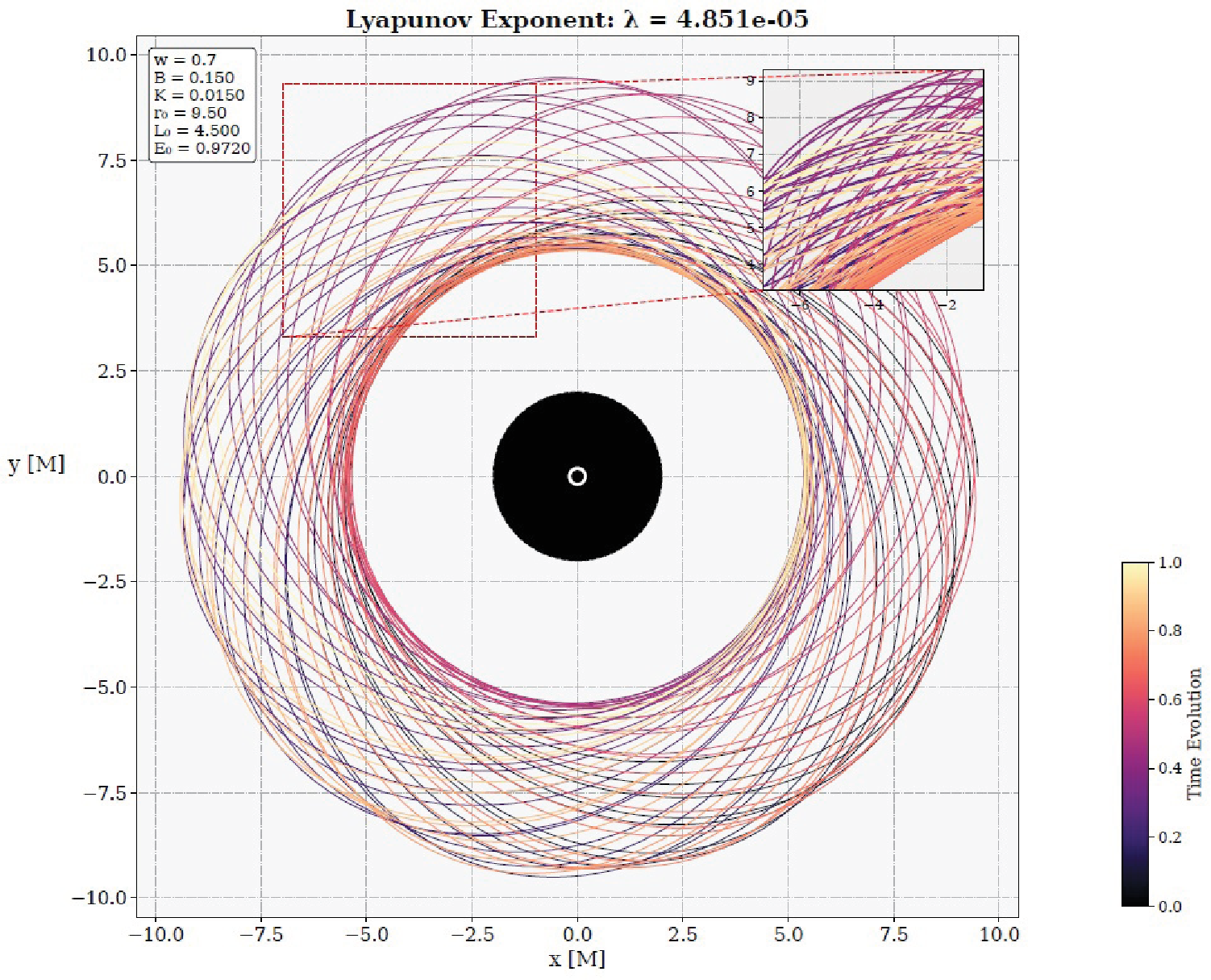}}
\subfigure[$w=1.5$]
{\includegraphics[width=8.1cm]{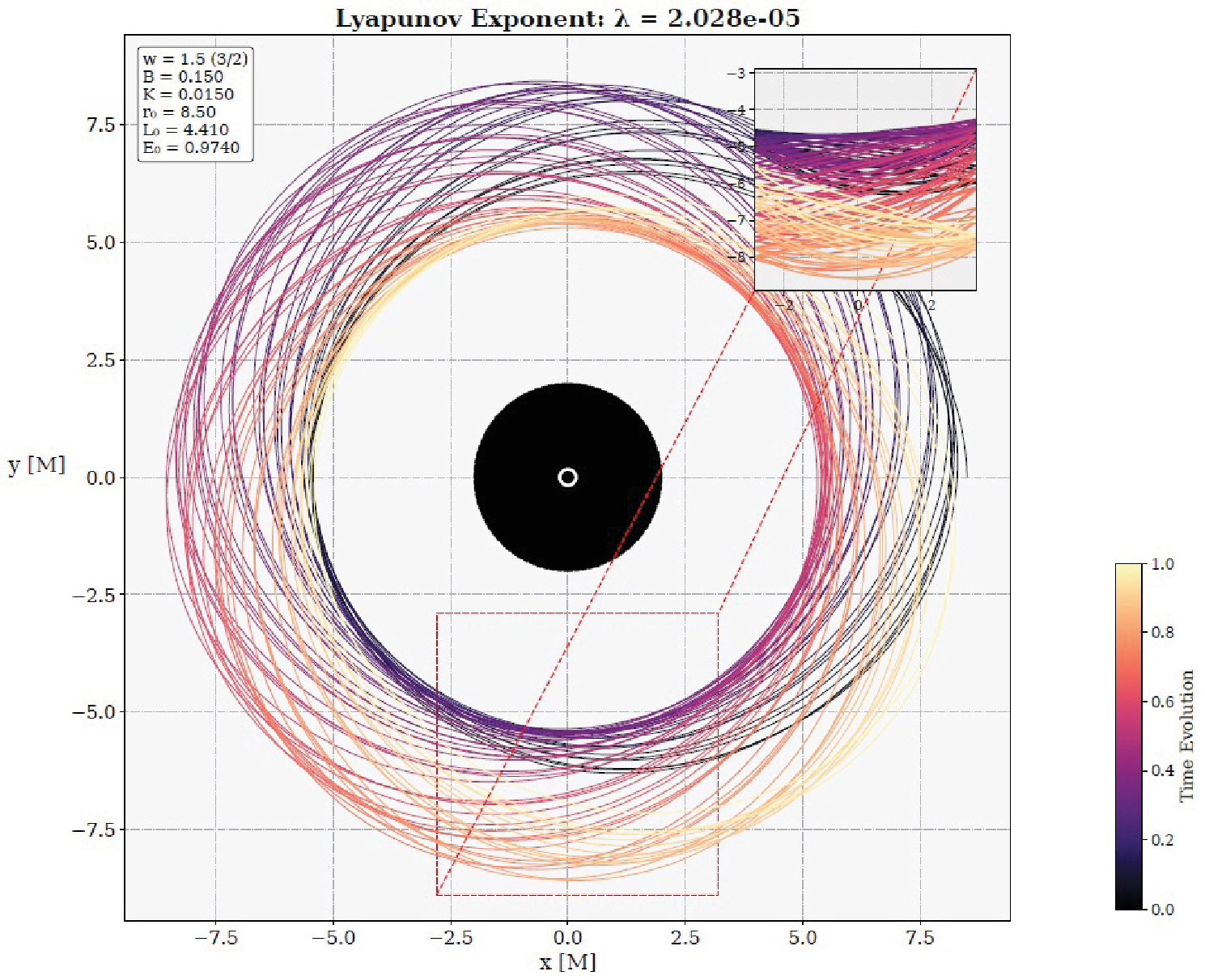}}
\subfigure[$w=1.5$]
{\includegraphics[width=8.1cm]{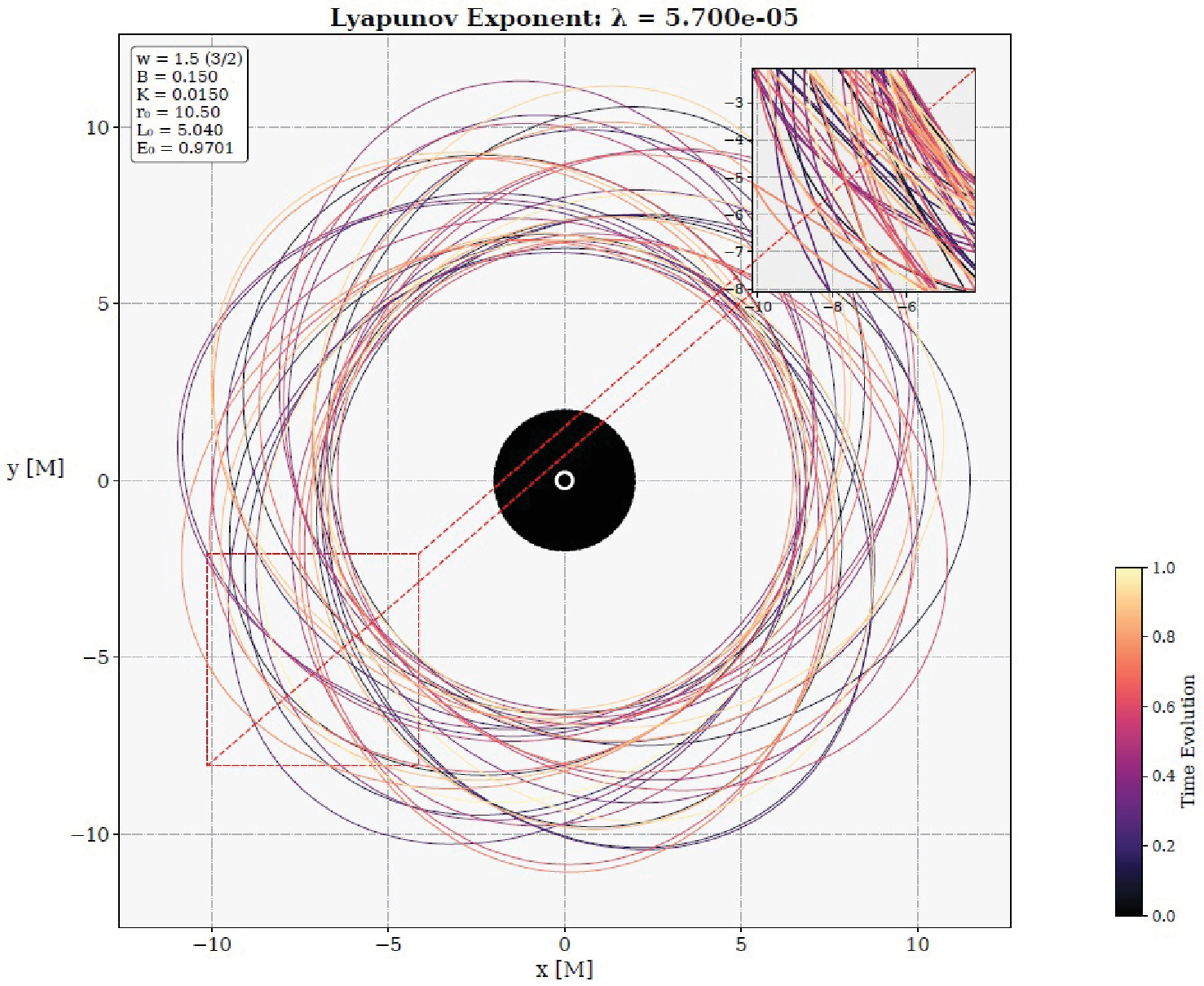}}
\subfigure[$w=2.0$]
{\includegraphics[width=8.1cm]{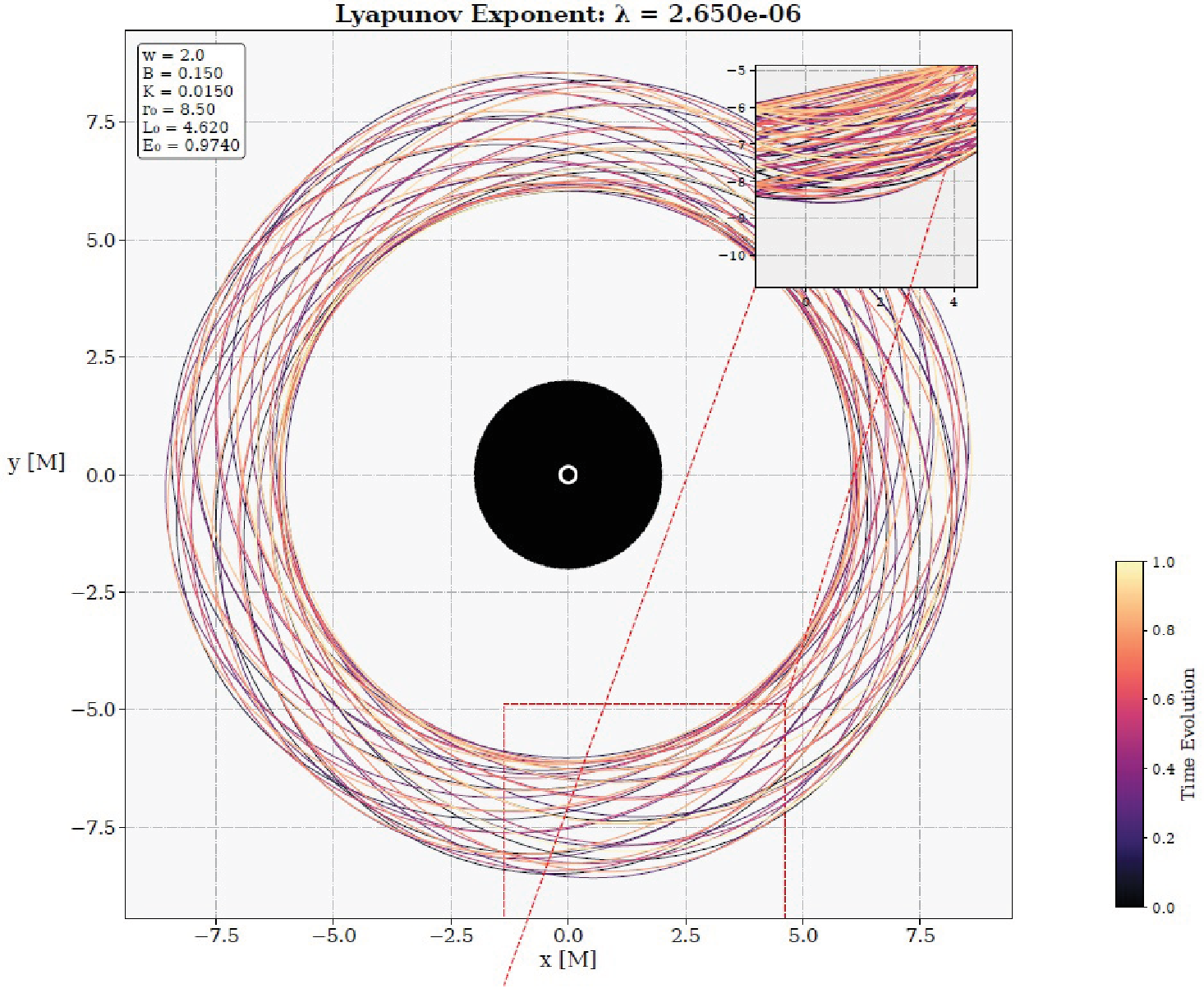}}
\end{center}
\caption{Particle trajectories for $w=0.7$ (top-left), $w=1.5$ (top-right, middle-left), and $w=2.0$ (middle-right). Color: time evolution.  }
\label{fig:partrajec}
\end{figure}
Figure~\ref{fig:partrajec} illustrates the local chaotic behavior of particles near the homoclinic orbits shown in Fig.~\ref{fig:effB_varyB}.
The four panels correspond to $(a) w = 0.7$, $(b) w = 1.5$, and $(c) w = 2.0 $, respectively.
The color bar indicates the evolution of time, showing that the particle does not return to the same position after each cycle.
The trajectory of one particle located at a homoclinic orbit, along with the Lyapunov exponent, is depicted, while the initial conditions of another particle differ only slightly. Consequently, even a minor variation in the initial conditions can result in the other particle entering a plunge orbit into the black hole or a scattering orbit away from it. It may also be on a bounded orbit, revolving in the vicinity of the black hole. In the context of the black hole geometry under consideration, the magnetic field is shown to generate a potential wall, the consequence of which is that particles are confined to a bounded orbit.  As demonstrated in Eq.~\eqref{eq:lyapunov_definition}, a negative $\lambda$ signifies the convergence of orbits, while a positive $\lambda$ indicates divergence and eventually chaotic behavior.
Figure $(d)$ shows a negative lambda value.

\subsubsection{Lyapunov exponent versus angular momentum: parameter dependence  \label{sec3-2-1} }

\quad

In this subsection, we conduct a detailed analysis of how the Lyapunov exponent depends on the angular momentum $L$, as well as on each of the four physical parameters $B_o$, $K$, $w$, and $q$. Our findings indicate that $\lambda$ increases monotonically with $L$ in all combinations of field parameters examined. A larger orbit is found to couple more intensively with the curvature and modifications of the field, resulting in a faster divergence of geodesics. This outcome confirms that $L$ serves as the main driver of the chaotic behavior of the particle trajectories within our system.

\begin{figure}[H]
\begin{center}
\subfigure[$w=0.7$ and $K=0.01$]
{\includegraphics[width=7.0cm]{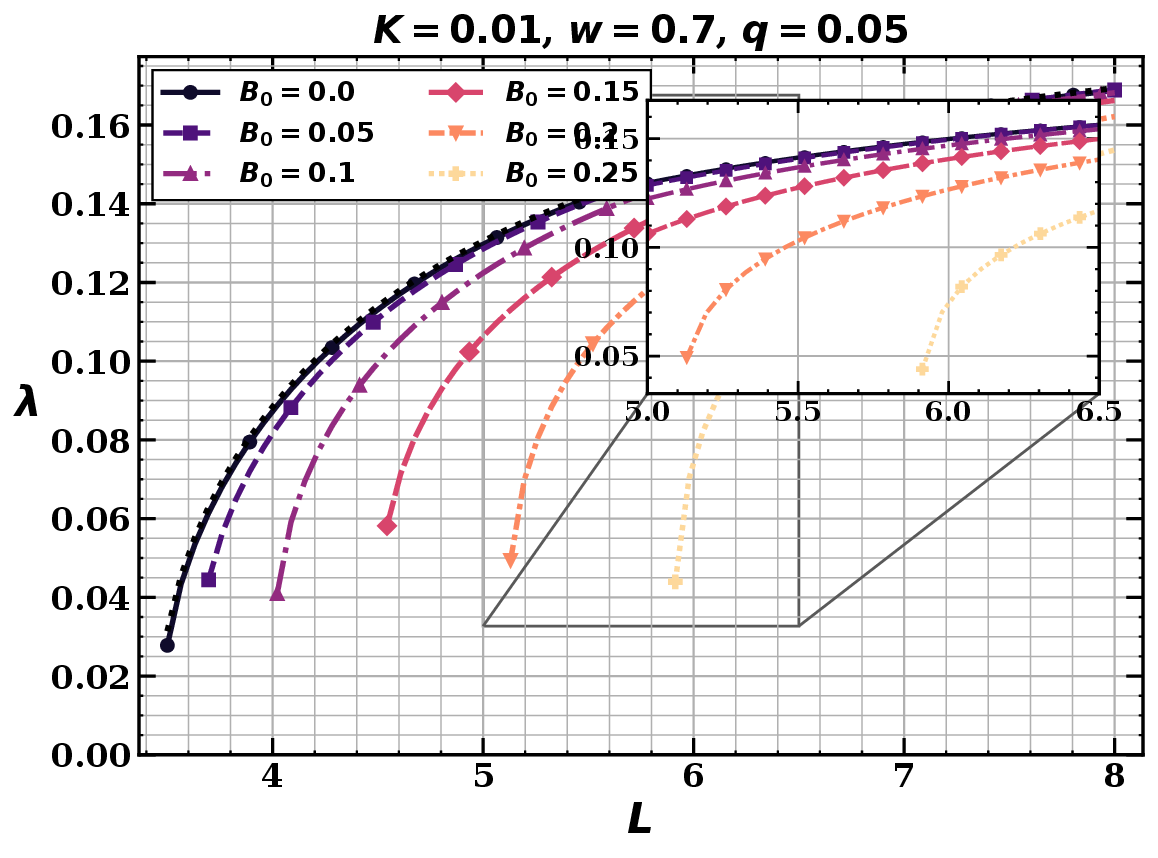}}
\subfigure[$w=0.7$ and $B_o=0.1$]
{\includegraphics[width=7.0cm]{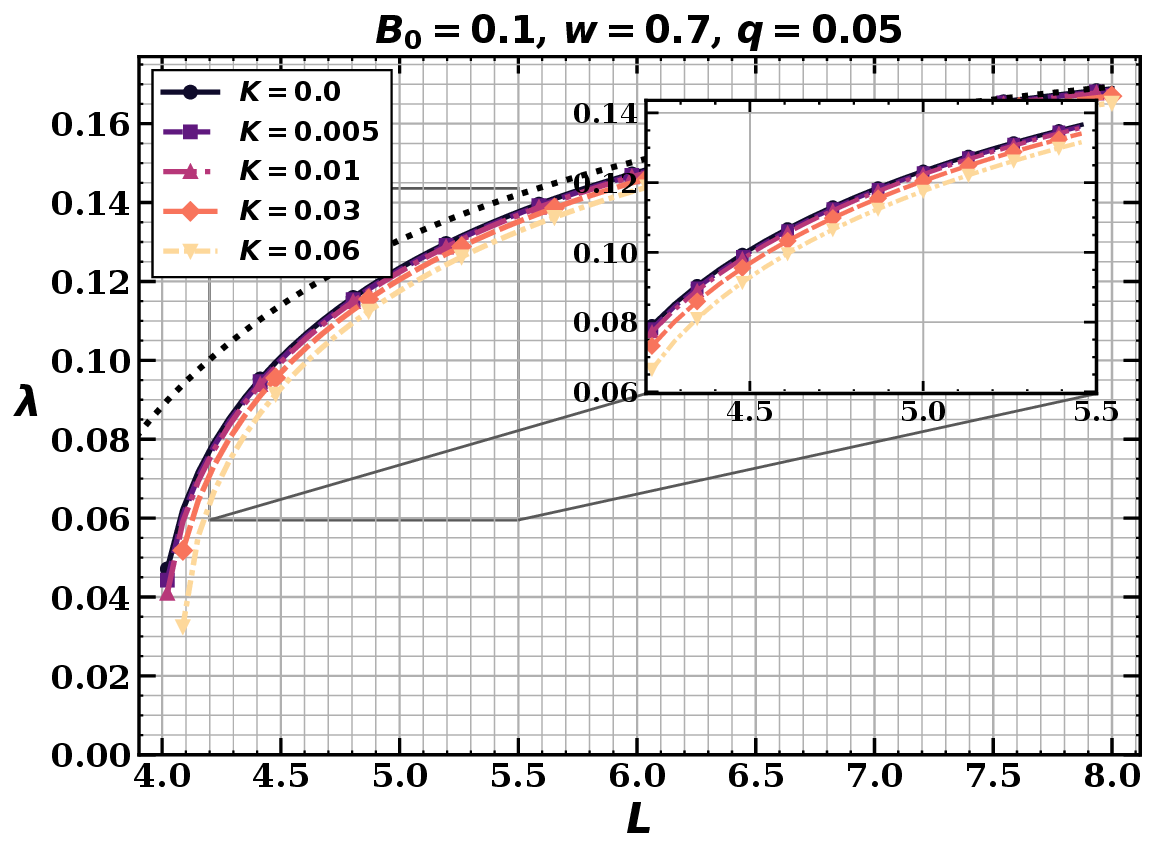}}
\subfigure[$w=1.5$ and $K=0.01$]
{\includegraphics[width=7.0cm]{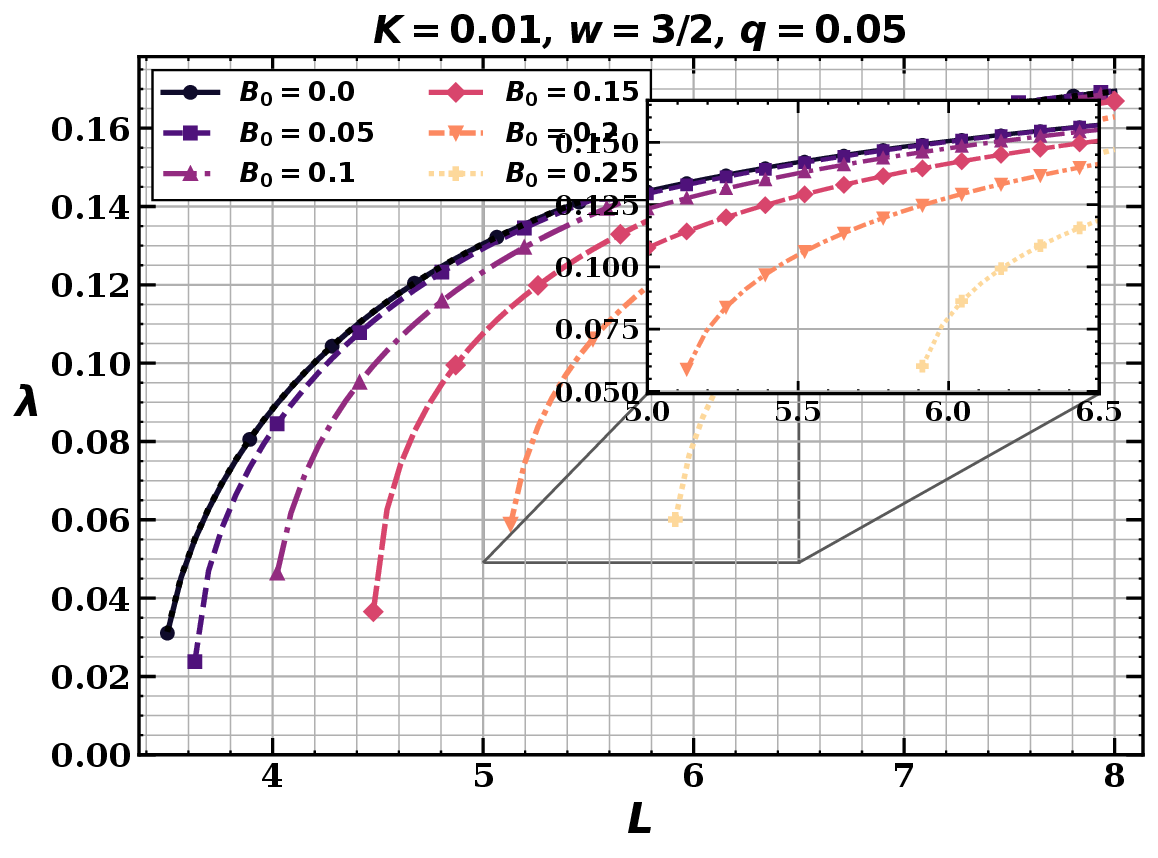}}
\subfigure[$w=1.5$ and $B_o=0.1$]
{\includegraphics[width=7.0cm]{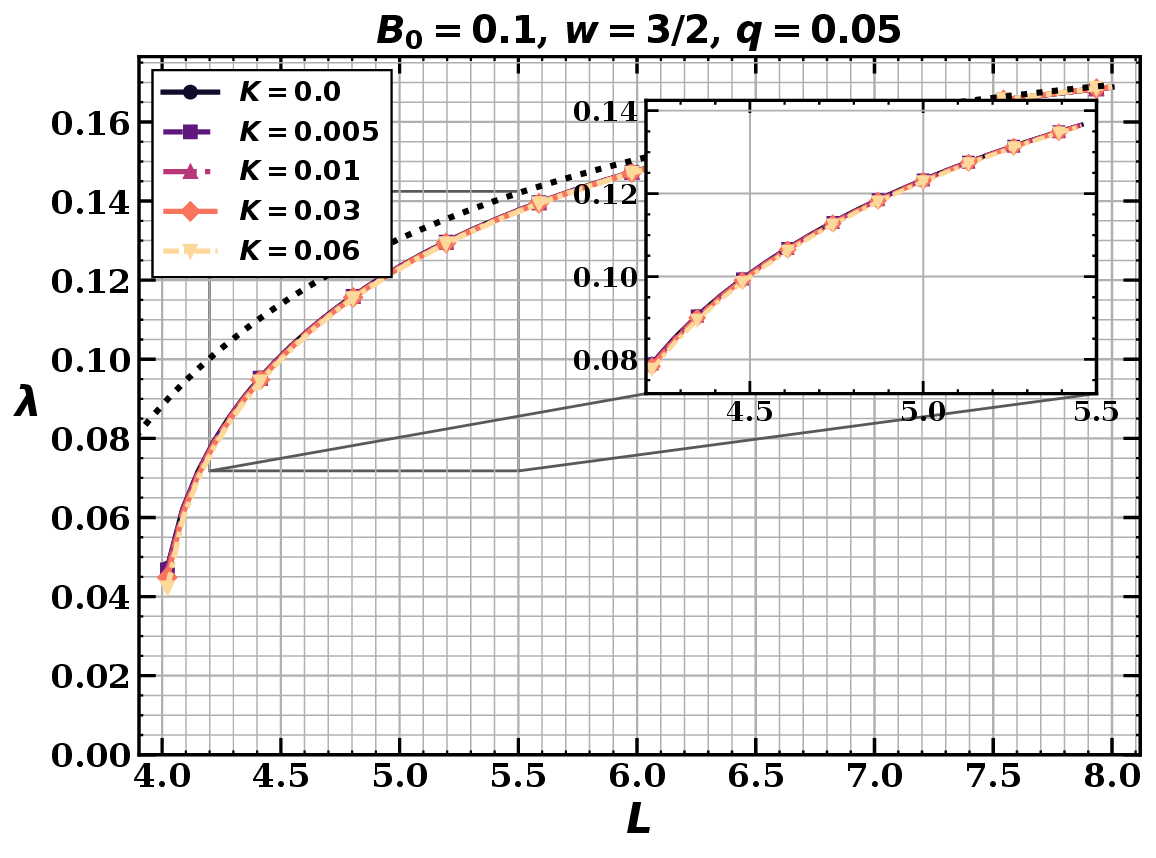}}
\subfigure[$w=2.0$ and $K=0.01$]
{\includegraphics[width=7.0cm]{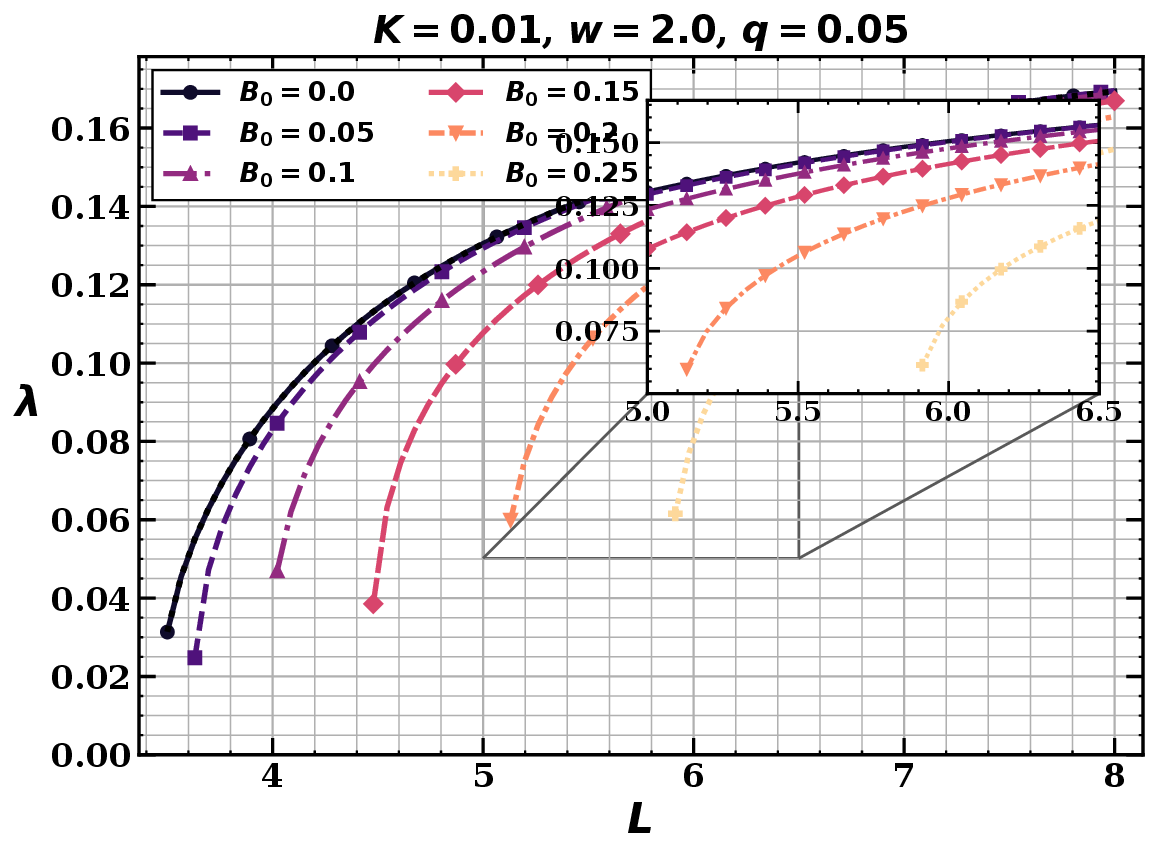}}
\subfigure[$w=2.0$ and $B_o=0.1$]
{\includegraphics[width=7.0cm]{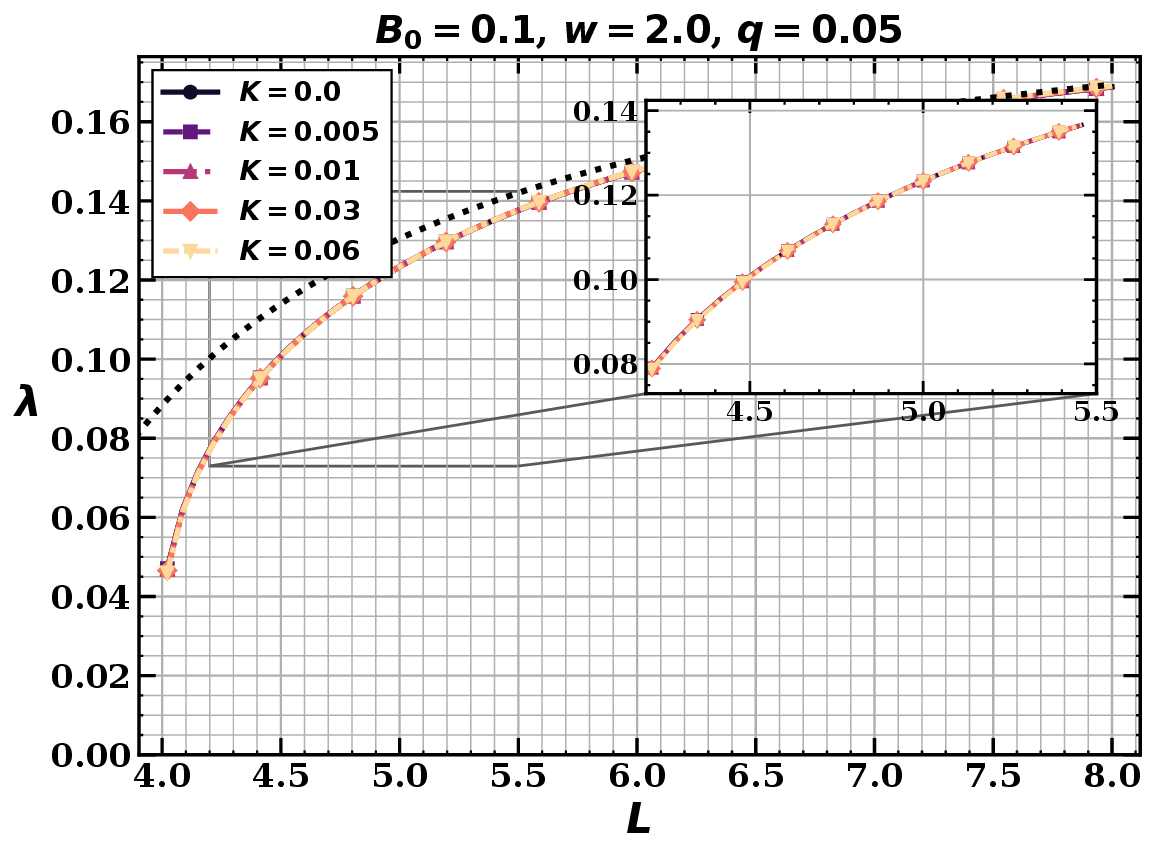}}
\end{center}
\caption{\footnotesize{(color online).
$\lambda$ vs.\ $L$ for $B_o\in[0,\,0.25]$ ($K=0.01$, $q=0.05$) for left panels, while $\lambda$ vs.\ $L$ for $K\in[0,\,0.06]$ ($B_o=0.1$, $q=0.05$) for right panels. Dotted: Schwarzschild BH.   }}
\label{fig:lam_L_B}
\end{figure}
Figure~\ref{fig:lam_L_B} shows the effect of the angular momentum $L$ on $\lambda$, with $K$, $q$ and $w$ being constant.
The figures in the left-hand column illustrate how the value of $\lambda$ changes with different values of $B_o$.
The values of $B_o$ were plotted in the range $[0, 0.25]$, covering six different values.
Figures (a), (c), and (e) show the results for different $w$ values.
Increasing $B_o$ increases $\lambda(L)$ and widens the gap between the curves as $L$ decreases.
This is because a stronger magnetic field reduces the effective angular momentum $L_{\rm eff} = L - qA_{\phi}$,
shrinking the potential barrier and allowing the unstable circular orbit to sit closer to the black hole,
where $|V''_{\rm eff}(r_c)|$ is larger.
The figures also show that magnetic amplification is nearly independent of the $w$ values when comparing the three figures.

The figures in the right-hand column show how $\lambda$ changes depending on $K$, while $B_o$, $q$, and $w$ are kept constant.
The range of values $K$ is $[0, 0.06]$, with five values plotted.
Figures (b), (d), and (f) show the results for different $w$ values.

\subsubsection{Two-dimensional parameter maps \label{sec3-2-2} }

\quad

We move beyond one-dimensional parameter sweeps and produce two-dimensional color maps of $\lambda$ over several parameter planes. These maps reveal how the competing effects of the magnetic field and the quintessence correction combine to determine the degree of chaos.

{\bf (A) Chaotic and nonchaotic regions in the $(B_o, K)$ plane }

\begin{figure}[H]
  \centering
  \includegraphics[width=13cm]{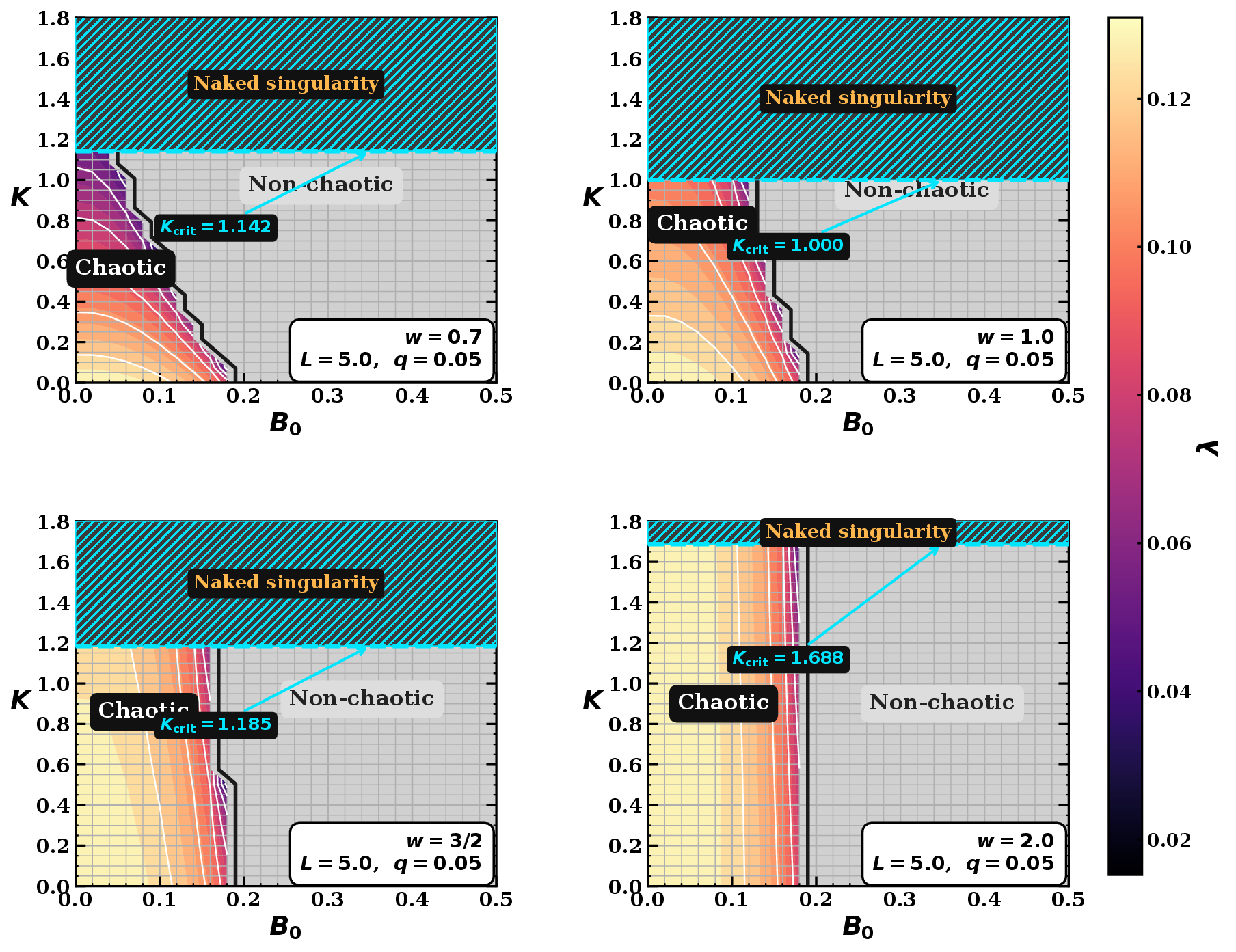}
  \caption{Chaotic-region map in the $(B_o,K)$ plane for $w\in\{0.7,1.0,3/2,2.0\}$ ($L=5$, $q=0.05$). Colored: $\lambda>0$. Gray hatched: no unstable orbit. Cyan dashed: $K_{\rm crit}(w)$.   }
\label{fig:chaotic_regionsw}
\end{figure}
We map the full $(B_o,K)$ plane in Fig.~\ref{fig:chaotic_regionsw} to show which parameter combinations support chaotic orbits (colored pixels, where an unstable circular orbit exists and $\lambda > 0$) versus nonchaotic ones (gray-hatched, where no unstable circular orbit exists). We have analyzed these maps for all four values $w \in \{0.7, 1.0, 3/2, 2.0\}$. We find three key results: (i) the chaotic region grows with $B_0$, confirming the magnetic field as the primary mechanism for creating unstable orbits; (ii) the chaotic region is noticeably smaller at $w=0.7$ than at $w=2.0$, because long-range quintessence suppresses instability over a larger portion of the plane; (iii) the extremal-black-hole threshold $K_{\rm crit}$ varies markedly between panels, reaching its global minimum $K_{\rm crit}=1$ at $w=1$, so the narrow valid-black-hole strip at $w=1$ leaves the least room for chaotic orbits.

{\bf (B) Chaotic and nonchaotic regions in the $(B_o, q)$ plane }

\begin{figure}[H]
    \centering
    \includegraphics[width=13cm]{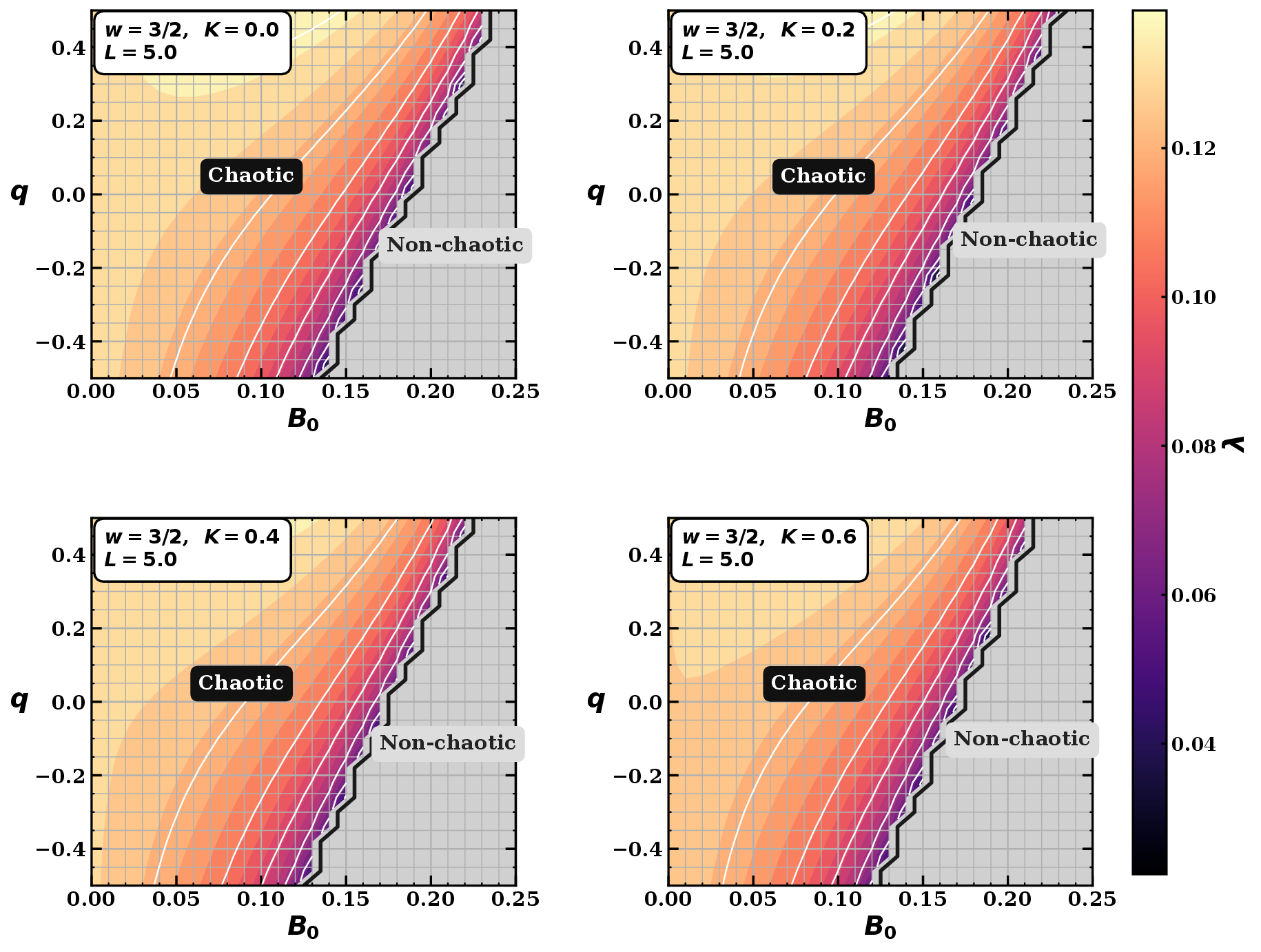}
    \caption{  Chaotic-region map in the $(B,q)$ plane for $K\in\{0.0,0.2,0.4,0.6\}$. Colored: $\lambda>0$.    }
\label{fig:chaotic_regionsk}
\end{figure}
We map the full $(B,q)$ plane in Fig.~\ref{fig:chaotic_regionsk} to show which parameter combinations support chaotic orbits (colored pixels, where an unstable circular orbit exists, and $\lambda > 0$) versus nonchaotic ones (gray-hatched, where no unstable circular orbit exists). We have analyzed these maps for the fixed value $w = 3/2$ and for four  $K \in \{0.0, 0.2, 0.4, 0.6\}$. We have found two key results: (i) for each fixed $K$, the chaotic region expands monotonically with the magnetic-field parameter $B$; (ii) increasing $K$ systematically shrinks the chaotic area.

\begin{figure}[H]
\begin{center}
\subfigure[the joint dependence of $(w, K)$ and $\lambda$]
{\includegraphics[width=8.1cm]{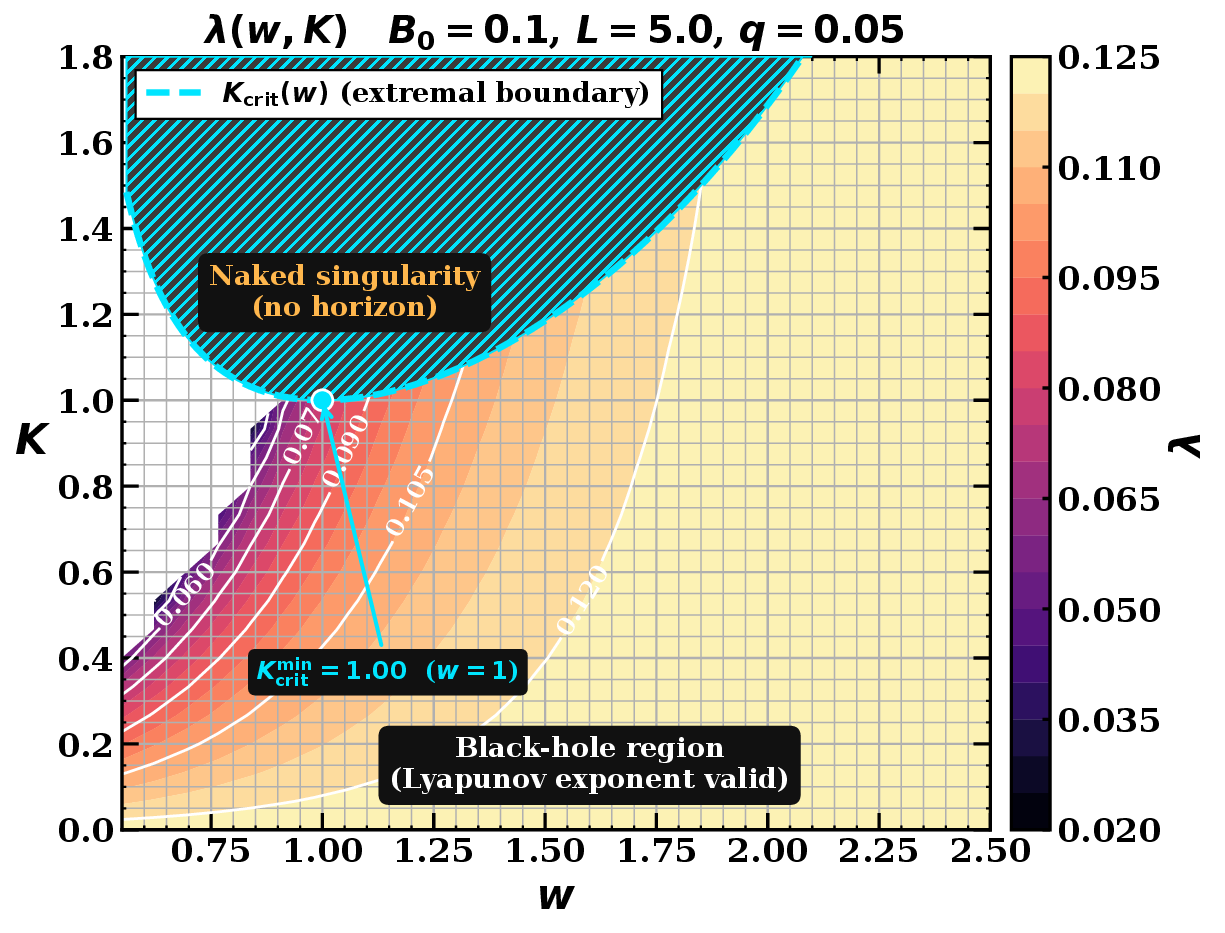}}
\subfigure[the joint dependence of $(w, B_o)$ and $\lambda$]
{\includegraphics[width=8.1cm]{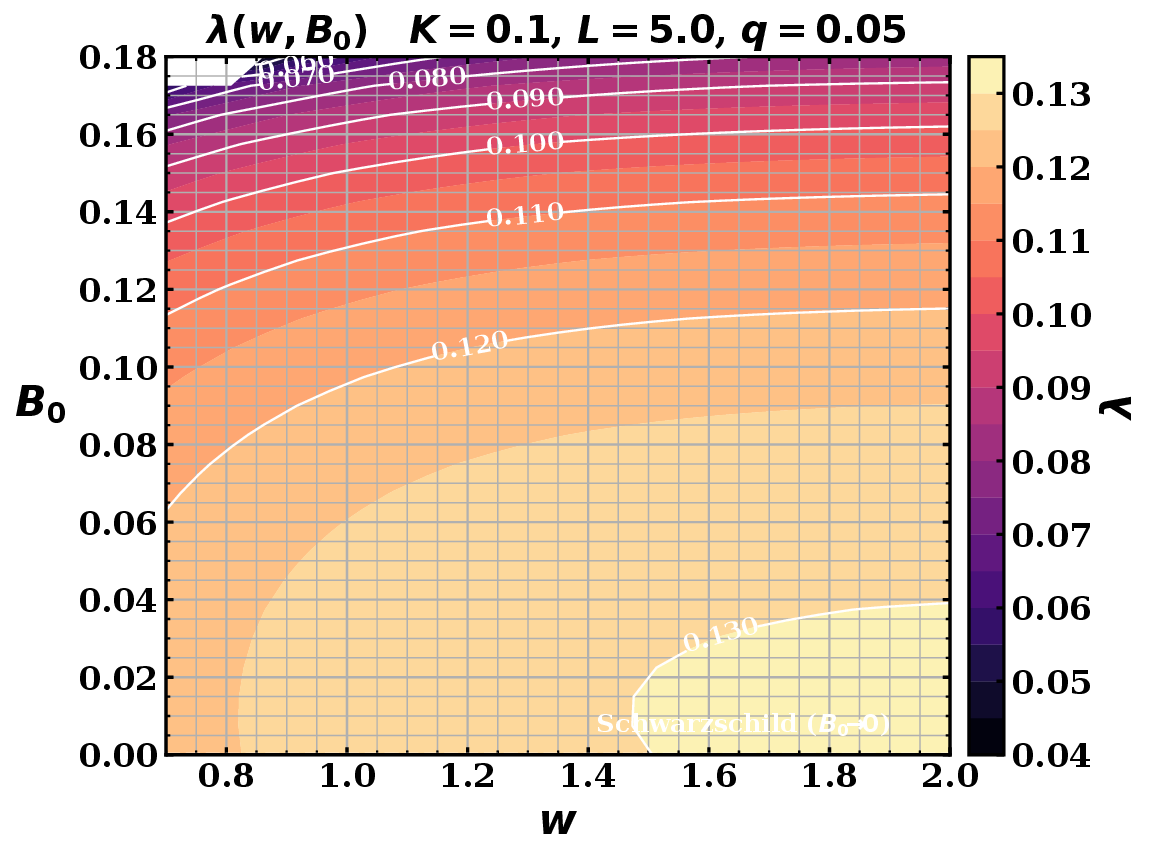}}
\end{center}
    \caption{The examination of the joint dependence of $(w, K)$ and $\lambda$, and that of $(w, B_o)$ and $\lambda$.     }
\label{fig:chaotic_regions}
\end{figure}
Figure~\ref{fig:chaotic_regions} shows the examination of the joint dependence of $(w, K)$ and $\lambda$, as well as that of $(w, B_o)$ and $\lambda$.
In Fig.~\ref{fig:chaotic_regions}(a), $B_o = 0.1$ and $L=5$.
As shown in Fig.~\ref{fig:param_space}, the cyan dashed line on the top left represents the analytical extremal boundary $K_{\rm crit}(w)$.
We are interested in this boundary and the region below it. When $w$ is large and $K$ is small, as shown in the bottom right of the figure,
$\lambda$ assumes small values. In contrast, when $w$ is small and $K$ is large, $\lambda$ has large values.
The white region on the top left is the nonchaotic region where $\lambda$ is zero or negative.
In Fig.~\ref{fig:chaotic_regions}(b), $K=0.1$.
The figure shows that the main feature is the steep rise with $B_o$.
The magnetic field is stronger than the dependence on $w$ for $B_o \gtrsim 0.08$.
When $w$ is large and $B_o$ is small (see bottom right of the figure), $\lambda$ assumes small values.
On the other hand, when $w$ is small and $B_o$ is large, $\lambda$ has large values. The white part of the top left is the nonchaotic region.

\subsubsection{Initial-condition Space: The \texorpdfstring{$(L,E)$}{(L,E)} plane \label{sec3-2-4} }

\begin{figure}[H]
    \centering
    \includegraphics[width=5.2cm]{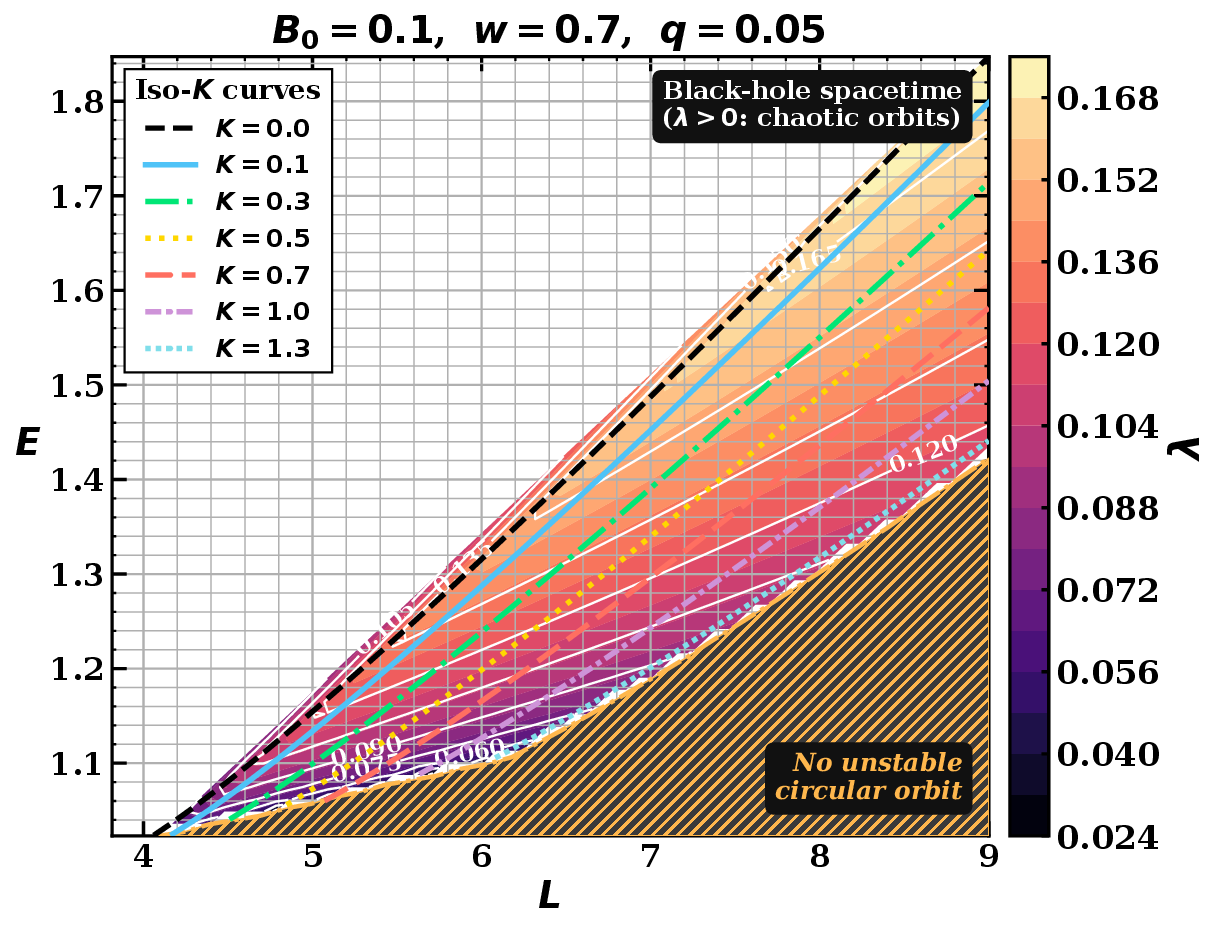}
    \includegraphics[width=5.2cm]{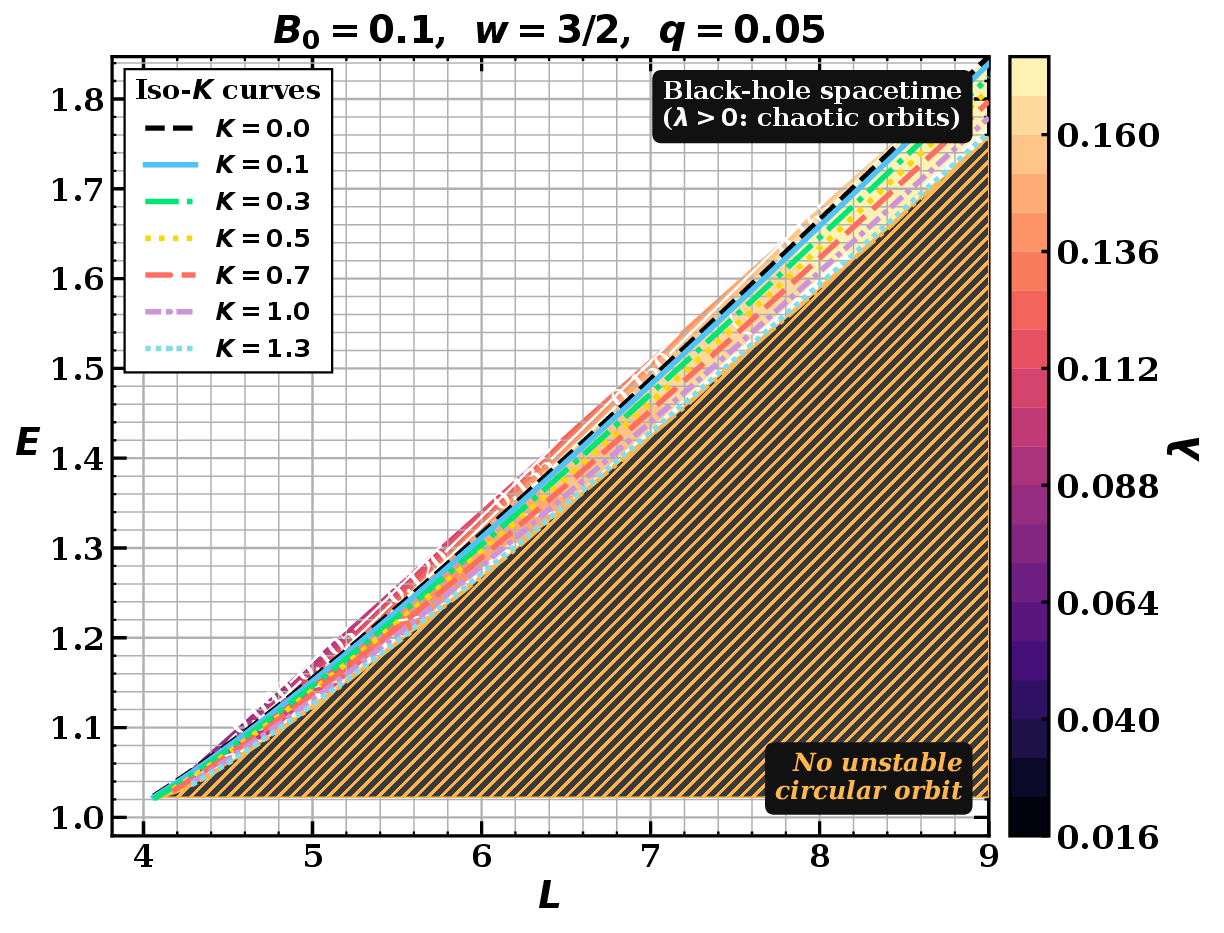}
    \includegraphics[width=5.2cm]{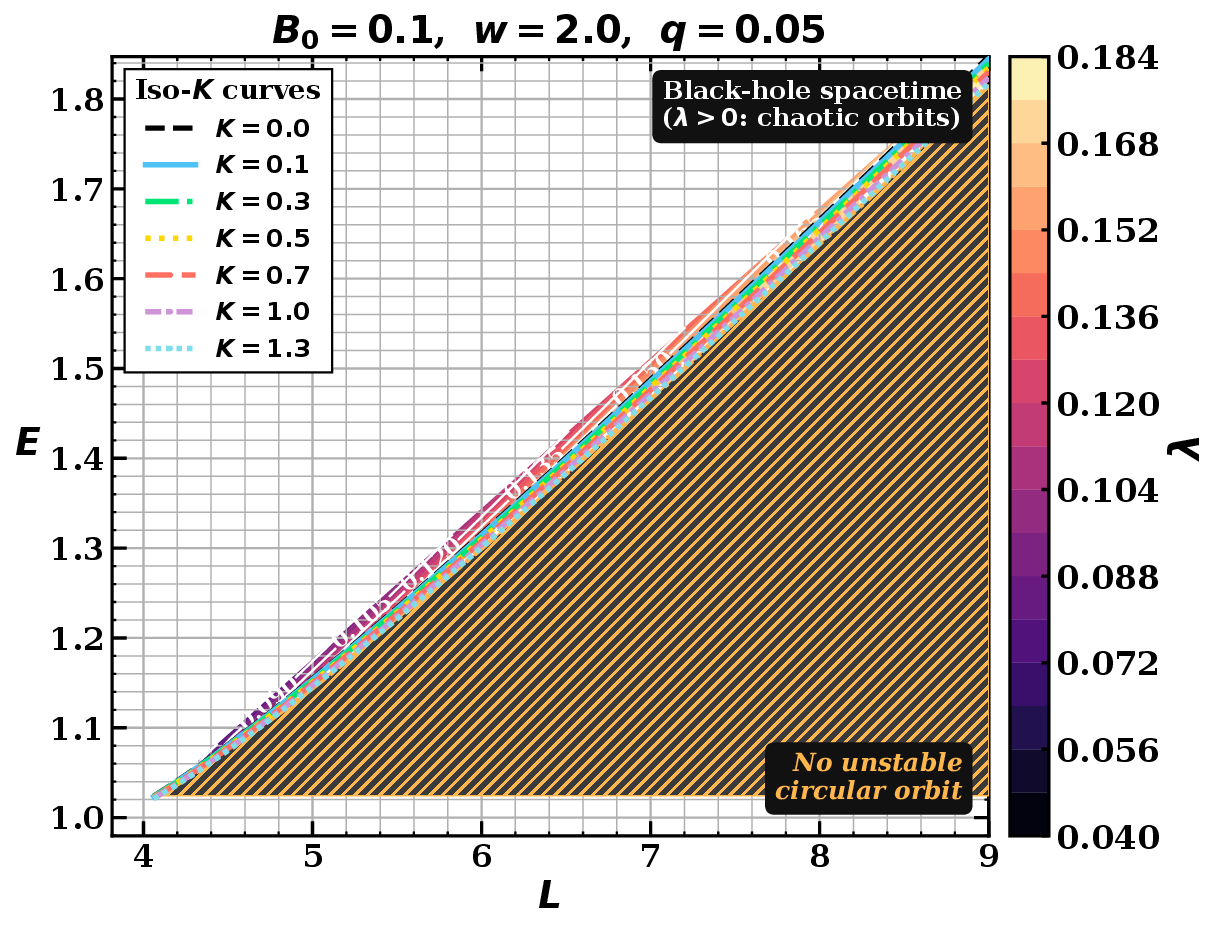}
    \caption{Color map of $\lambda(L,E)$, $B_o=0.1$, $q=0.05$; $K\in[0,1.4]$. Iso-$K$ curves overlaid (legend). Gold dash-dot: $E_{\rm min}(L)$. (a)~$w=0.7$, (b)~$w=3/2$, (c)~$w=2.0$.     }
  \label{fig:density_LE}
\end{figure}
We investigate the Lyapunov exponent in the physical initial-condition space spanned by the angular momentum $L$ and the orbital energy $E$.
For each spacetime specified by fixed values of $B_o$, $K$, $w$, and $q$, the energy of the unstable circular orbit is obtained numerically from the circular orbit conditions as a function of the angular momentum $L$. Figure~\ref{fig:density_LE} presents a color map of the Lyapunov exponent in the $(L,E)$ plane for $K \in [0, 1.4]$. In the background are seven iso-$K$ curves corresponding to $K = 0, 0.1, 0.3, 0.5, 0.7, 1.0, 1.3$, each representing the locus of unstable circular orbits in a spacetime with the corresponding value of $K$.
As the anisotropic matter parameter increases, the iso-$K$ curves systematically shift toward lower values of both $L$ and $E$,
while simultaneously crossing progressively darker regions of the color map, indicating smaller Lyapunov exponents.
This behavior shows that increasing $K$ not only reduces the energy required for unstable circular orbits but also suppresses their local dynamical instability. The gold dashed-dotted curve denotes the lower boundary $E_{\rm min}(L)$, below which unstable circular orbits do not exist, whereas the gray hatched region is therefore physically inaccessible.

\subsection{Poincar\'e section in phase space \label{sec3-3} }

\quad

The objective of this study is to illustrate the motion of a particle within the geometry of a black hole. In this section, we analyze the Poincar\'e section. We consider a particle of mass $m$ and focus on its trajectory near the horizon of a black hole, described by the metric in~\eqref{metric}. To derive Hamilton's equations, we utilize the equations presented in Sec.~\ref{sec2-2}.

The trajectory of a particle demonstrates global chaotic behavior when the points do not align to form a few distinct locations or a smooth line in the Poincar\'e section of the phase space. Instead, they fill a particular region with a complex, nonperiodic pattern that is often fractal in nature. This indicates that the particle's history is unpredictable, yet remains bounded.

To derive Hamilton's equations, we use the equations presented in Sec.~\ref{sec2-2}. From Eq.~\eqref{hamilton}, Hamilton's equations can be expressed as
\begin{eqnarray}
\label{hamiltonseq}
\frac{d x^{\mu}}{d \lambda} = \frac{\partial \cal H }{\partial \pi_{\mu}} \,,~~   \frac{d \pi_{\mu}}{d \lambda} = - \frac{\partial \cal H }{\partial x^{\mu}} \,,
\end{eqnarray}
where $p^{\mu}=\frac{d x^{\mu}}{d \lambda}$ and $m \lambda=\tau$. This yields the following four equations:
\begin{eqnarray}
\label{hamiltonseq02}
&&{\dot r}  = \frac{f \pi_r}{m \Lambda^2}  \,, \\
&&{\dot \pi_r}  =\frac{1}{2 m r^3 \Lambda^3 f^2} \left[ r^3 f^2 \left( 2 f \partial_r \Lambda - \Lambda f' \right) \pi_r^2 + 2 f^2 \left( r \partial_r \Lambda + \Lambda \right) \pi_\theta^2 - r^3 E^2 \left( 2 f \partial_r \Lambda + \Lambda f' \right) \right]   \nonumber \\
   && \quad\qquad + \frac{\left( L - q A_\phi \right) \Lambda^2}{m r^2 \sin^2\theta} \left[ \left( L- q A_\phi \right) \left( \frac{1}{r} - \frac{\partial_r \Lambda}{\Lambda} \right) + q \partial_r A_\phi \right]  \,, \\
&&{\dot \theta} = \frac{\pi_{\theta}}{m r^2 \Lambda^2}  \,, \\
&&{\dot \pi_{\theta}} =\frac{1}{m r^2 \Lambda^3 f} \left[ r^2 f^2 \left( \partial_\theta \Lambda \right) \pi_r^2 + f \left( \partial_\theta \Lambda \right) \pi_{\theta}^2 - r^2 E^2 \left( \partial_\theta \Lambda \right) \right] \nonumber    \\
   && \quad\qquad + \frac{\left( L - q A_{\phi} \right) \Lambda^2}{m r^2 \sin^2\theta} \left[ \left( L - q A_{\phi} \right) \left( \cot\theta - \frac{\partial_{\theta} \Lambda}{\Lambda} \right) + q \partial_{\theta} A_\phi  \right] \,,
\end{eqnarray}
where $f= \frac{\Delta}{r^2}$, $\pi_r = p_r$ and $\pi_{\theta}=p_{\theta}$, $\cdot$ denotes the derivative with respect to $\tau$ and $\prime$ denotes the derivative with respect to $r$.

\begin{figure}[H]
    \centering
    \subfigure[$K = 0.2$, $B_o = 0.004$]{\includegraphics[width=8.1cm]{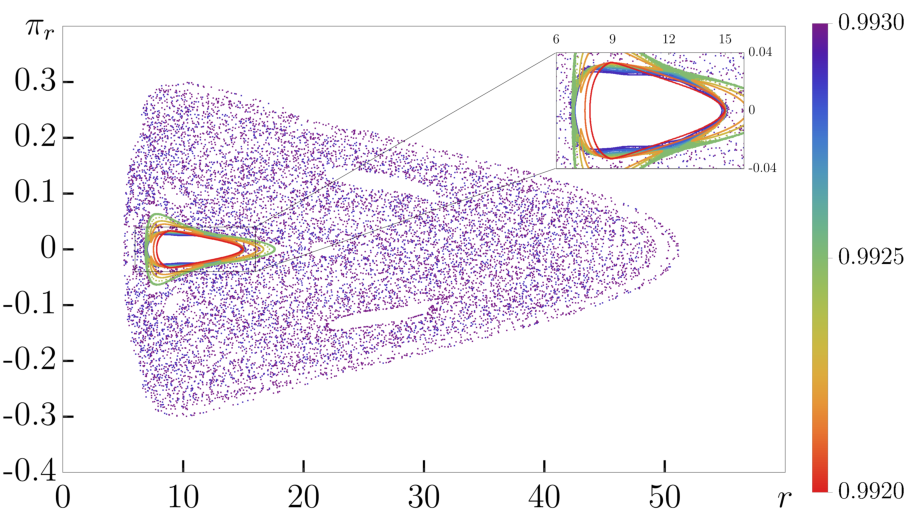}}
    \subfigure[$K = 0.2$, $B_o = 0.005$]{\includegraphics[width=8.1cm]{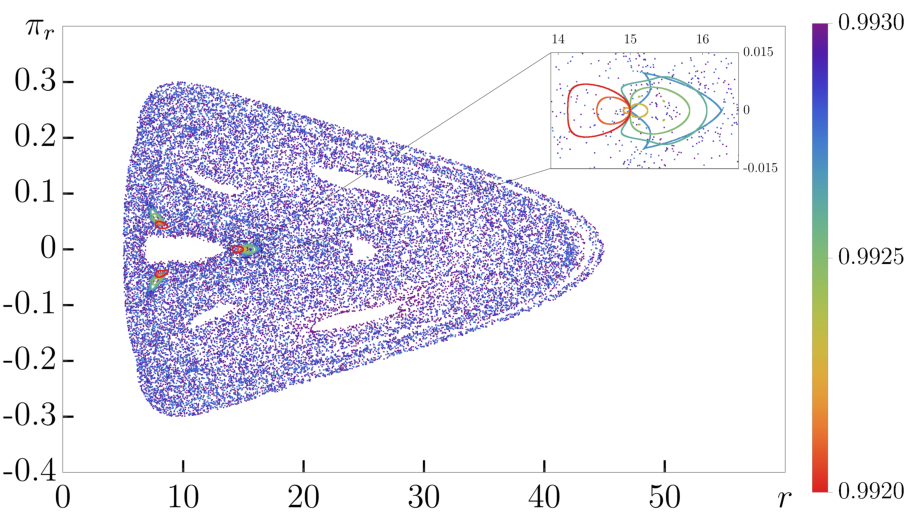}}
    \caption{Poincar\'e section on the equatorial plane with the particle energy varying from $0.992$ to $0.993$ (color scale).}
    \label{fig:poincare_section}
\end{figure}
Figure~\ref{fig:poincare_section} presents the Poincar\'e sections in the equatorial plane $( \theta = \pi/2 )$ for the parameters $M = m = 1$, $q = 0.1$, $K = 0.2$, $L = 4$, and $\omega = 3/2$, with initial conditions $r = 15$, $\theta = \pi / 2$ and $\pi_r = 0$. The left and right panels correspond to $B_o = 0.004$ and $B_o = 0.005$, respectively. Points are recorded at each crossing of the equatorial plane with positive polar momentum $( \pi_\theta > 0 )$. The sections are shown for $11$ energy values ranging from $E = 0.992$ to $0.993$, indicated by the color scale from red to purple. As the particle energy increases, the figure shows a qualitative change. For lower particle energies, the trajectories are located in narrow regions of phase space, with details shown in the insets. At higher energies, the points spread over a wider region in the $( r, \, \pi_r )$ plane, indicating that the trajectories explore a larger portion of phase space.
\begin{figure}[H]
    \centering
    \subfigure[$K = 0.1$, $B_o = 0.004$]{\includegraphics[width=5.4cm]{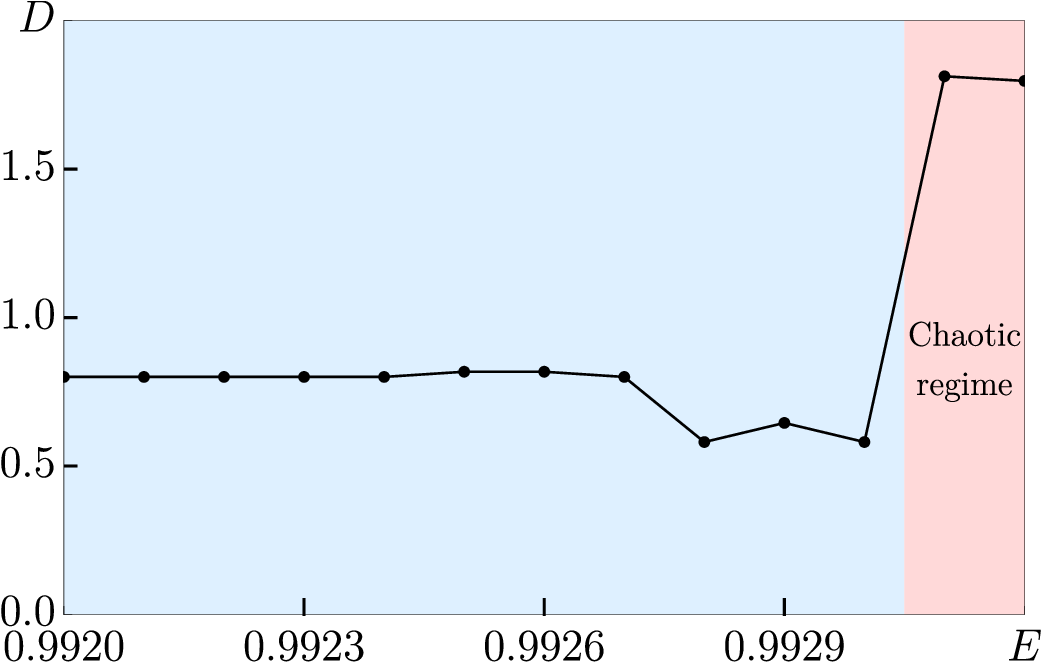}}
    \subfigure[$K = 0.2$, $B_o = 0.004$]{\includegraphics[width=5.4cm]{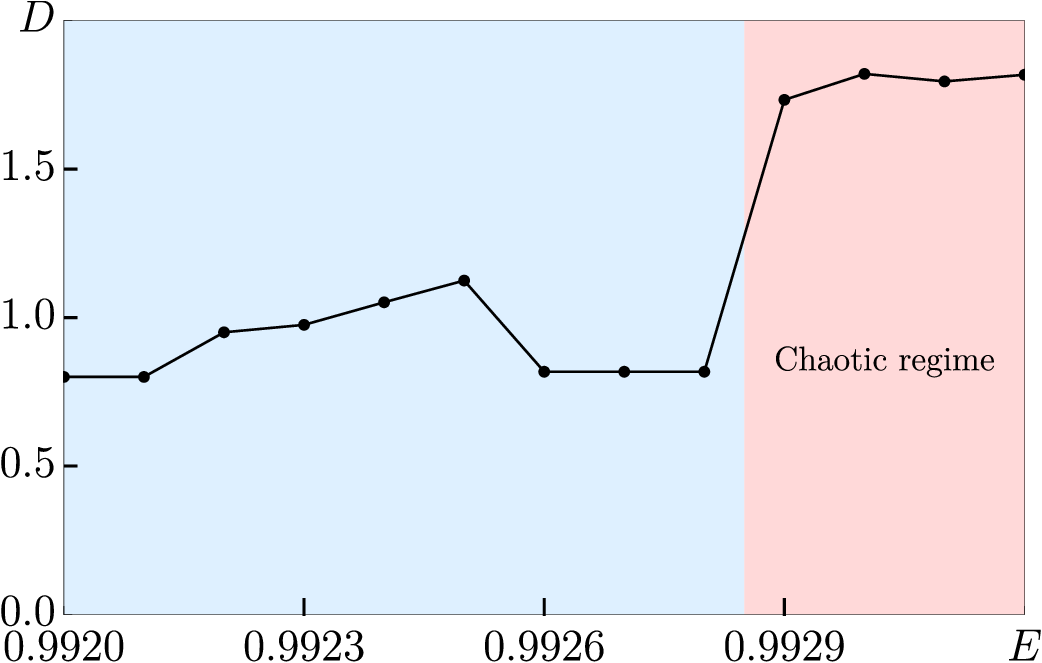}}
    \subfigure[$K = 0.3$, $B_o = 0.004$]{\includegraphics[width=5.4cm]{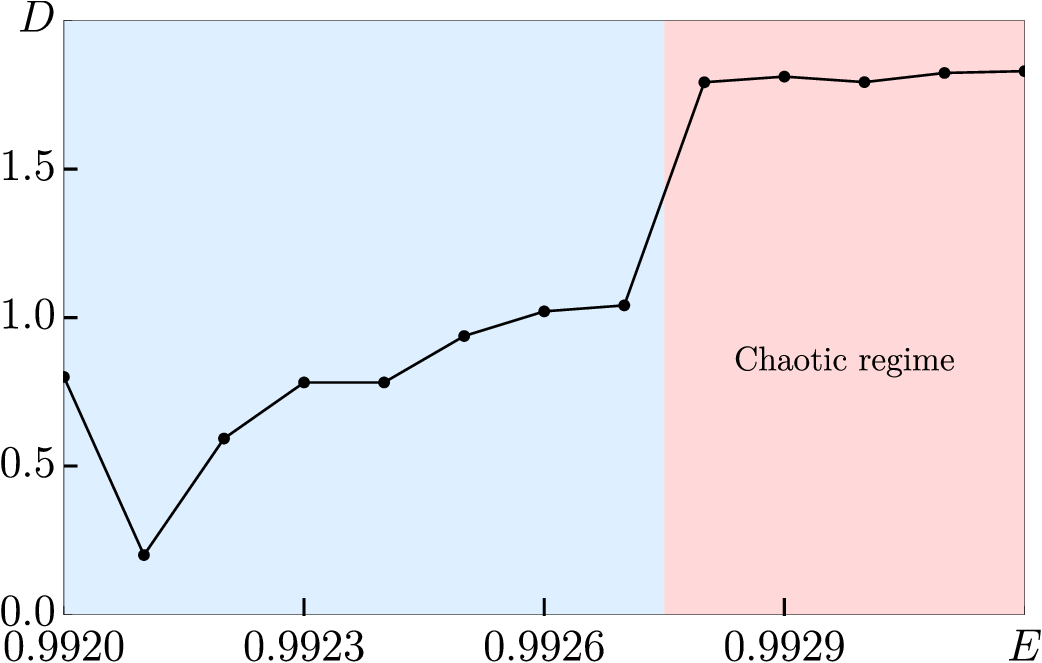}}
    \subfigure[$K = 0.1$, $B_o = 0.005$]{\includegraphics[width=5.4cm]{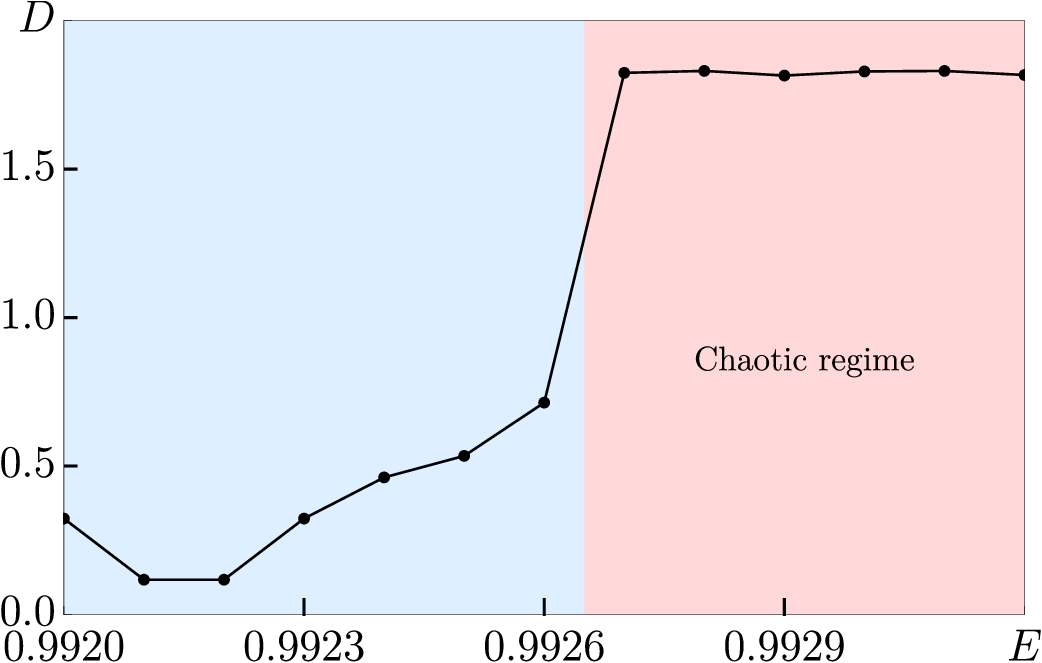}}
    \subfigure[$K = 0.2$, $B_o = 0.005$]{\includegraphics[width=5.4cm]{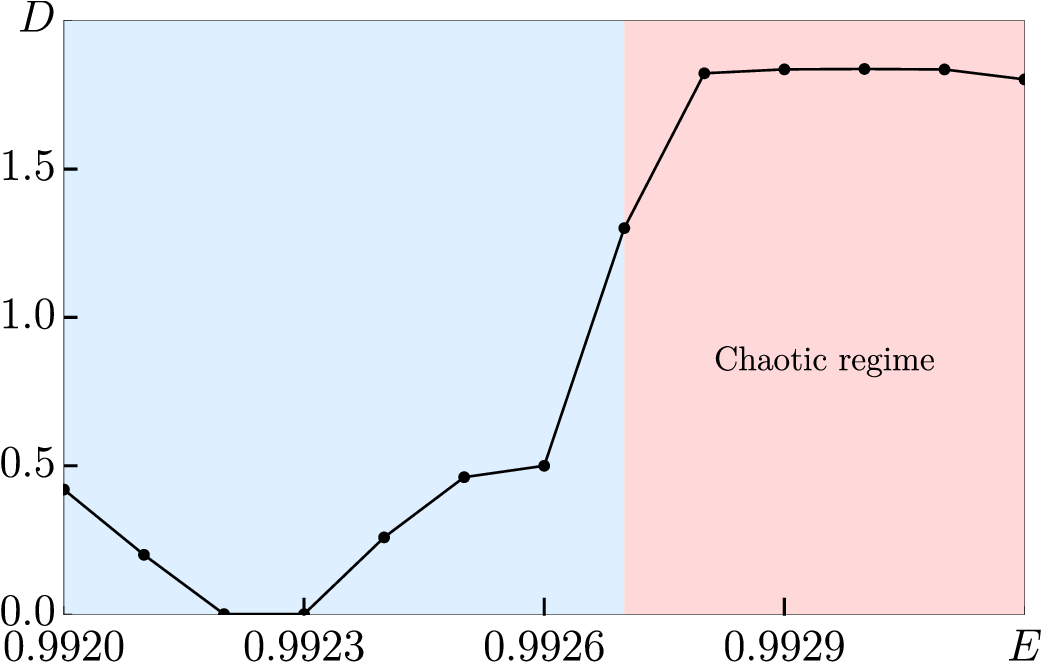}}
    \subfigure[$K = 0.3$, $B_o = 0.005$]{\includegraphics[width=5.4cm]{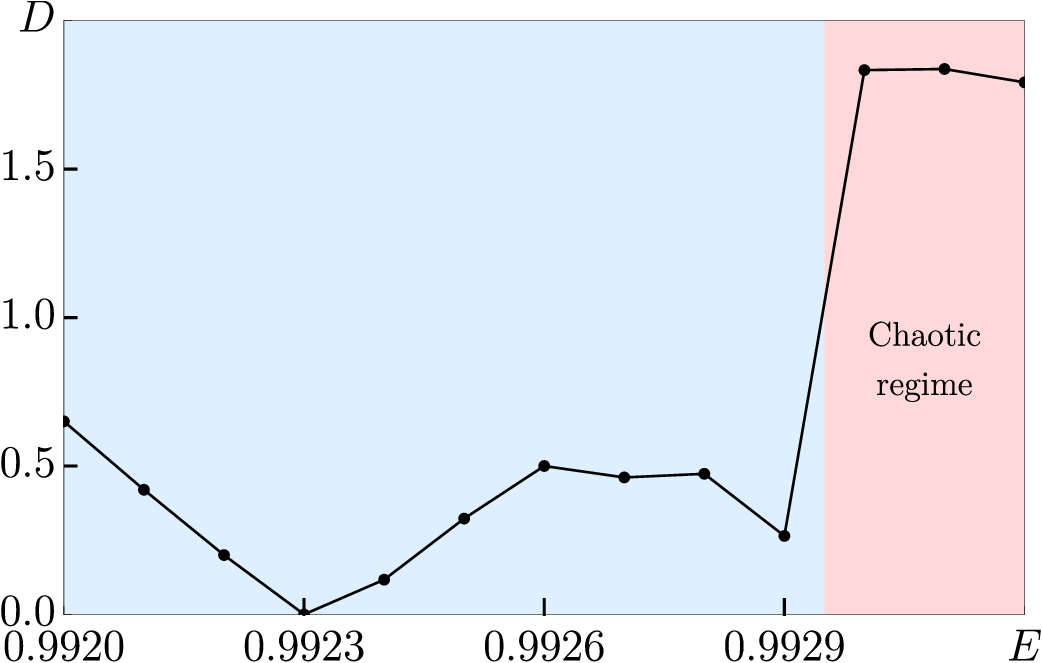}}
    \caption{Minkowski--Bouligand (box-counting) dimension}
    \label{fig:dimension}
\end{figure}
Figure~\ref{fig:dimension} illustrates the Minkowski--Bouligand (box-counting) dimension~\cite{Falconer:2003} for the Poincar\'e sections shown in Fig.~\ref{fig:poincare_section}, along with additional configurations. The first and second rows correspond to $B_o = 0.004$ and $B_o = 0.005$, respectively. In each column, the three panels correspond to $K = 0.1$, $0.2$, and $0.3$, from left to right. The particle energy range from $0.9920$ to $0.9932$ is adopted, and all other initial conditions are identical to those in Fig.~\ref{fig:poincare_section}. The red regions indicate chaotic motion, while the blue regions correspond to regular motion, as quantified by the Minkowski--Bouligand dimension. The Minkowski--Bouligand dimension is defined as
\begin{equation}
    D = \lim_{\epsilon \to 0} \frac{\log N(\epsilon)}{\log (1 / \epsilon)}.
\end{equation}
For each particle energy, the dimension is estimated using $10^4$ coordinate pairs $(r, \pi_r)$ extracted from the corresponding Poincar\'e section. The phase space coordinates are uniformly scaled to the interval $[0, 1]$ using the global maximum absolute values across all configurations. To estimate the dimension $D$, a linear regression of $\log N(\epsilon)$ against $\log(1/\epsilon)$ is performed over the scaling region $\epsilon = 2^{-k}$ for $k = 2$, $3$, $\cdots$, $6$, where the slope of the fit yields the dimension. A lower value of $D$ indicates that the trajectory occupies a relatively limited region of phase space, with $D$ approaching zero when the Poincar\'e section is composed of a small number of localized points. As the trajectory explores a larger portion of phase space, the dimension increases, with $D \sim 2$ corresponding to a fully chaotic motion. For $B_o = 0.004$, the critical energy at which chaotic regions first emerge decreases as $K$ increases, indicating that higher values of $K$ facilitate the onset of chaotic motion. In contrast, for $B_o = 0.005$, the critical energy increases with increasing $K$, which implies that stronger values of $K$ tend to suppress the emergence of chaos within the energy range considered. For $B_o = 0.004$, the critical energy, which corresponds to the energy where chaotic behavior first emerges within the energy range of Fig.~\ref{fig:dimension}, decreases as $K$ increases. In contrast, for $B_o = 0.005$, the critical energy increases with $K$, exhibiting a dependence on $K$ that is opposite to that observed for $B_o = 0.004$. This opposite dependence suggests that the highly nonlinear interactions within the nonintegrable system alter the phase space structure qualitatively under small variations of the magnetic field.

\section{Summary and discussions \label{sec4}}
\quad

Let us consider a supermassive black hole located in the center of a galaxy~\cite{Ghez:2008ms, Gillessen:2008qv, Kormendy:1995er}.
Such a black hole would coexist with a nonzero, albeit extremely small, density of dark matter~\cite{Rubin:1970zza}.
For dark matter to remain outside the horizon without being drawn into the black hole, it ideally should have a density greater than zero coupled with a radial pressure that is less than zero.
If we consider multiple potential dark matter candidates~\cite{Sin:1992bg, Lee:1995af, Hu:2000ke, Lee:2008jp, Hui:2016ltb, Kang:2023gef}, a black hole coexisting with an anisotropic fluid would be a promising model for describing the supermassive black holes found in galactic centers. Additionally, some of these black holes may interact closely with magnetic fields~\cite{Marrone:2006vu, Eatough:2013nva, EventHorizonTelescope:2025vum}. Treating such a black hole as embedded within its astrophysical environment, it would be fascinating to investigate the motion of light or particles in its vicinity~\cite{Bardeen:1973tla, Dymnikova1986, Virbhadra:1998dy, Virbhadra:2002ju}.

In this work, we have constructed a new exact solution of the Einstein-Maxwell equations describing a static black hole coexisting with anisotropic matter and immersed in an external magnetic field through the Harrison transformation~\cite{Harrison:1968wue}.
The resulting spacetime interpolates between the anisotropic matter black hole and Ernst geometry~\cite{Ernst:1976mzr}, thereby providing a unified framework for investigating the combined influence of matter fields and magnetic fields on particle trajectories.

We first investigated the geometrical and physical properties of the solution. The event horizon structure was analyzed as a function of the anisotropic matter parameter $K$ and the equation-of-state parameter $w$, showing the parameter regions corresponding to regular black holes, extremal black holes, and naked singularities. The magnetic field configuration was also obtained explicitly. We showed that anisotropic matter modifies the near horizon magnetic field structure while preserving the asymptotic Melvin behavior~\cite{Melvin:1963qx}, indicating that the gravitational field and external magnetic field interact nontrivially in the vicinity of the black hole.

A central result of this work is the analysis of the spacetime symmetry responsible for particle motion in the vicinity of the black hole, coexisting with anisotropic matter immersed in the magnetic field, and the effect of anisotropic matter fields.
We demonstrated that spacetime admits only the Killing symmetries associated with stationarity and axisymmetry and does not possess the second-rank Killing tensor responsible for Carter's constant.
Consequently, the Hamilton-Jacobi equation cannot be completely separated, and the radial and angular equations remain coupled.
This establishes that geodesic motion in the present spacetime is intrinsically nonintegrable. We wanted to explicitly show the geometric reason for nonintegrability and the particle's chaotic behavior through the loss of hidden spacetime symmetry.

The consequences of breaking spacetime symmetry were explored using complementary indicators of chaos. By analyzing the Lyapunov exponent associated with the homoclinic orbit as an unstable circular orbit, we assessed the local chaotic behavior.
Our findings indicate that increasing the anisotropic matter parameter systematically reduces the Lyapunov exponent,
suggesting that anisotropic matter acts to suppress local chaotic behavior.

The global chaotic behavior was explored through Poincar\'e sections~\footnote{It is said that the term `chaos' was coined by James A. Yorke~\cite{yorke:1975}.}. In contrast to the monotonic influence of anisotropic matter on the Lyapunov exponent, the external magnetic field produces qualitative changes in the global chaotic behavior by analyzing Poincar\'e sections. We demonstrated that, depending on the magnitude of the magnetic field, the highly nonlinear interactions within the nonintegrable system qualitatively alter the phase space structure under small variations of the magnetic field. These results demonstrate that anisotropic matter and magnetic fields play distinct but complementary roles in determining particle trajectories: anisotropic matter primarily controls the degree of local instability, whereas the magnetic field governs the global organization of phase space through nonlinear gravitational-electromagnetic interaction.

The present work establishes a direct connection between spacetime geometry, hidden symmetry, and nonlinear particle dynamics in magnetized black hole spacetimes.
The results suggest that environmental matter distributions and external magnetic fields can substantially modify particle trajectories around astrophysical black holes, potentially affecting particle transport, plasma confinement, and acceleration processes in strong-field environments such as a supermassive black hole that coexists with dark matter
at the galactic center and interacts with a magnetic field.
More broadly, our analysis demonstrates how variations in spacetime symmetry are manifested in the observable dynamical behavior of particles, providing a useful framework for understanding nonintegrable particle motion in an exact solution of the Einstein-Maxwell system.

In the context of the equations of motion for a particle, the geodesic equations cannot be separated into distinct variables. Instead, they lead to coupled equations with coupled potentials, rendering the system nonintegrable and potentially chaotic. Notably, the coupling of the geodesic equations may induce global chaotic behavior in the test particle,  irrespective of the symmetry of the underlying geometry, due to various other effects. The scenarios in which a test particle may exhibit chaotic behavior are outlined below:
\begin{itemize}
\item  Geometries that do not allow a separable structure: black holes immersed in a magnetic field~\cite{Ernst:1976mzr, Ernst:1976bsr, Bicak:1985, Aliev:1989wx, Karas:1992, Wang:2016wcj, Li:2018wtz}.
\item  Geometries that do not permit a separable structure: nonhomogeneous vacuum pp-wave backgrounds~\cite{Podolsky:1998ez}.
\item  Geometries that do not permit a separable structure: Weyl's cylindrically symmetric geometry, describing a Schwarzschild black hole surrounded by thin disks or rings~\cite{Semerak:2010lzj, Semerak:2012dx}.
\item  Geometries that allow for a separable structure: scenarios where a magnetic field is introduced as a probe limit, such as in the geometry of a weakly magnetized black hole~\cite{Wald:1974np, Stuchlik:2015nlt, Tursunov:2018erf}.
\item  Geometries that allow for a separable structure: the motion of a spinning particle in black hole geometry~\cite{Suzuki:1996gm, Kao:2004qs, Jefremov:2015gza, Yang:2026ekp, Dalui:2026oaa}.
\item  Geometries that allow for a separable structure: the motion of a string in black hole geometry~\cite{Frolov:1999pj, Bai:2014wpa, Basu:2016zkr}.
\item  Geometries that allow for a separable structure: cases in which a coupled potential is introduced as an assumption within a black hole geometry~\cite{Hashimoto:2016dfz, Dalui:2018qqv, Das:2024iuf, Das:2025vja, Das:2025eiv}. Properly elucidating the conditions under which such a potential could be introduced in astrophysical environments would be advantageous.
\end{itemize}

Several compelling extensions arise naturally from the current study. Since the geometry of the black hole immersed in a magnetic field we examined is not asymptotically flat, the analysis of weakly magnetized black holes with anisotropic matter becomes relevant to astrophysical scenarios. In this paper, we have introduced a methodology that can be effectively applied to such weakly magnetized black holes, with expectations of direct applicability. The framework could be expanded to include rotating black holes. Additionally, it would be valuable to explore how these effects impact quasinormal modes and black hole shadows~\cite{EventHorizonTelescope:2022wkp}.
Lastly, incorporating plasma dynamics~\cite{Atamurotov:2015nra} and general-relativistic magnetohydrodynamic environments~\cite{Ruffini:1975ne}
would yield a more accurate portrayal of particle behavior around astrophysical black holes and create stronger ties to future observations.

\section*{Acknowledgments}
This research was supported by the Grants F-FA-2021-510 and MRB-2021-527 from the Uzbekistan Ministry for Innovative Development. K. Alibekov was supported by Mayflower Scholarship Funding from the University of Southampton.
H. Lee (NRF-2022R1I1A2063176 and Dongguk University Research Fund of 2026), B.-H. Lee (RS-2026-25473640 and RS-2020-NR049598), and W. Lee (RS-2026-25484780 and RS-2020-NR049598) were supported by Basic Science Research Program through the National Research Foundation of Korea. W. Lee is grateful to Sang-Pyo Kim for his helpful comments on black holes with the magnetic field.


\newpage

\begin{thebibliography}{99}

\bibitem{Ghez:2008ms}
  A.~M.~Ghez, S.~Salim, N.~N.~Weinberg, J.~R.~Lu, T.~Do, J.~K.~Dunn, K.~Matthews, M.~Morris, S.~Yelda and E.~E.~Becklin, \textit{et al.}
  Astrophys.\ J.\ \textbf{689}, 1044-1062 (2008)
  [arXiv:0808.2870 [astro-ph]].

\bibitem{Gillessen:2008qv}
  S.~Gillessen, F.~Eisenhauer, S.~Trippe, T.~Alexander, R.~Genzel, F.~Martins and T.~Ott,
  Astrophys.\ J.\ \textbf{692}, 1075-1109 (2009)
  [arXiv:0810.4674 [astro-ph]].

\bibitem{Blandford:1977ds}
  R.~D.~Blandford and R.~L.~Znajek,
  Mon.\ Not.\ Roy.\ Astron.\ Soc.\ \textbf{179}, 433-456 (1977).

\bibitem{Johnson:2015iwg}
  M.~D.~Johnson, V.~L.~Fish, S.~S.~Doeleman, D.~P.~Marrone, R.~L.~Plambeck, J.~F.~C.~Wardle, K.~Akiyama, K.~Asada, C.~Beaudoin and L.~Blackburn, \textit{et al.}
  Science \textbf{350}, no.6265, 1242-1245 (2015)
  [arXiv:1512.01220 [astro-ph.HE]].

\bibitem{Blandford:2018iot}
  R.~Blandford, D.~Meier and A.~Readhead,
  Ann.\ Rev.\ Astron.\ Astrophys.\ \textbf{57}, 467-509 (2019)
  [arXiv:1812.06025 [astro-ph.HE]].

\bibitem{LIGOScientific:2020iuh}
  R.~Abbott \textit{et al.} [LIGO Scientific and Virgo],
  Phys.\ Rev.\ Lett.\ \textbf{125}, no.10, 101102 (2020)
  [arXiv:2009.01075 [gr-qc]].

\bibitem{EventHorizonTelescope:2022wkp}
  K.~Akiyama \textit{et al.} [Event Horizon Telescope],
  Astrophys.\ J.\ Lett.\ \textbf{930}, no.2, L12 (2022)
  [arXiv:2311.08680 [astro-ph.HE]].

\bibitem{EventHorizonTelescope:2025vum}
  K.~Akiyama \textit{et al.} [Event Horizon Telescope],
  Astron.\ Astrophys.\ \textbf{704}, A91 (2025)
  [arXiv:2509.24593 [astro-ph.HE]].

\bibitem{Fernandes:2025osu}
  P.~G.~S.~Fernandes and V.~Cardoso,
  Phys.\ Rev.\ Lett.\ \textbf{135}, no.21, 211403 (2025)
  [arXiv:2507.04389 [gr-qc]].

\bibitem{Schwarzschild:1916uq}
  K.~Schwarzschild,
  Sitzungsber.\ Preuss.\ Akad.\ Wiss.\ Berlin (Math. Phys. ) \textbf{1916}, 189-196 (1916)

\bibitem{Kerr:1963ud}
  R.~P.~Kerr,
  Phys.\ Rev.\ Lett.\ \textbf{11}, 237-238 (1963).

\bibitem{Kerr:2007dk}
  R.~P.~Kerr,
  [arXiv:0706.1109 [gr-qc]].

\bibitem{Carter:1968rr}
  B.~Carter,
  Phys.\ Rev.\ \textbf{174}, 1559-1571 (1968).

\bibitem{Carter:1968ks}
  B.~Carter,
  Commun.\ Math.\ Phys.\ \textbf{10}, no.4, 280-310 (1968).

\bibitem{Walker:1970un}
  M.~Walker and R.~Penrose,
  Commun.\ Math.\ Phys.\  {\bf 18}, 265 (1970).

\bibitem{Benenti:1979erw}
  S.~Benenti and M.~Francaviglia,
  Gen.\ Rel.\ Grav.\ \textbf{10}, no.1, 79-92 (1979).

\bibitem{Bardeen:1973tla}
  J.~M.~Bardeen, 1973, in Black Holes, ed.\ C.~DeWitt $\&$ B.~S.~DeWitt (New York: Gordon $\&$ Breach), 215-240.

\bibitem{Dymnikova1986}
  I.~G.~Dymnikova,
  Sov.\ Phys.\ Usp.\ {\bf 29}, 215 (1986).

\bibitem{Levin:2008yp}
  J.~Levin and G.~Perez-Giz,
  Phys.\ Rev.\ D \textbf{79}, 124013 (2009)
  [arXiv:0811.3814 [gr-qc]].

\bibitem{Pugliese:2010ps}
  D.~Pugliese, H.~Quevedo and R.~Ruffini,
  Phys.\ Rev.\ D \textbf{83}, 024021 (2011)
  [arXiv:1012.5411 [astro-ph.HE]].

\bibitem{Li:2023bgn}
  Y.~T.~Li, C.~Y.~Wang, D.~S.~Lee and C.~Y.~Lin,
  Phys.\ Rev.\ D \textbf{108}, no.4, 044010 (2023)
  [arXiv:2302.09471 [gr-qc]].

\bibitem{Cardoso:2008bp}
  V.~Cardoso, A.~S.~Miranda, E.~Berti, H.~Witek and V.~T.~Zanchin,
  Phys.\ Rev.\ D \textbf{79}, no.6, 064016 (2009)
  [arXiv:0812.1806 [hep-th]].

\bibitem{Gwak:2022xje}
  B.~Gwak, N.~Kan, B.~H.~Lee and H.~Lee,
  JHEP \textbf{09}, 026 (2022)
  [arXiv:2203.07298 [gr-qc]].

\bibitem{Jeong:2023hom}
  S.~Jeong, B.~H.~Lee, H.~Lee and W.~Lee,
  Phys.\ Rev.\ D \textbf{107}, no.10, 104037 (2023)
  [arXiv:2301.12198 [gr-qc]].

\bibitem{Lei:2023jqv}
  Y.~Q.~Lei and X.~H.~Ge,
  Phys.\ Rev.\ D \textbf{107}, no.10, 10 (2023)
  [arXiv:2302.12812 [hep-th]].

\bibitem{Lee:2025vih}
  H.~Lee and B.~Gwak,
  Phys.\ Rev.\ D \textbf{112}, no.4, 046018 (2025)
  [arXiv:2506.00833 [gr-qc]].

\bibitem{Targema:2026anu}
  T.~V.~Targema, K.~Bamba and U.~Zafar,
  [arXiv:2605.26829 [hep-th]].

\bibitem{Karas:1992}
  V.~Karas and D.~Vokroulflicky,
  Gen.\ Relativ.\ Gravit.\ \textbf{24}, 729-743 (1992).

\bibitem{Wang:2016wcj}
  M.~Wang, S.~Chen and J.~Jing,
  Eur.\ Phys.\ J.\ C \textbf{77}, no.4, 208 (2017)
  [arXiv:1605.09506 [gr-qc]].

\bibitem{Li:2018wtz}
  D.~Li and X.~Wu,
  Eur.\ Phys.\ J.\ Plus \textbf{134}, no.3, 96 (2019)
  [arXiv:1803.02119 [gr-qc]].


\bibitem{Melvin:1963qx}
  M.~A.~Melvin,
  Phys.\ Lett.\ \textbf{8}, 65-70 (1964).

\bibitem{Ernst:1976mzr}
  F.~J.~Ernst,
  J.\ Math.\ Phys.\ \textbf{17}, no.1, 54-56 (1976).

\bibitem{Ernst:1976bsr}
  F.~J.~Ernst and W.~J.~Wild,
  J.\ Math.\ Phys.\ \textbf{17}, no.2, 182 (1976).

\bibitem{Harrison:1968wue}
  B.~K.~Harrison,
  J.\ Math.\ Phys.\ \textbf{9}, no.11, 1744 (1968).

\bibitem{Wald:1974np}
  R.~M.~Wald,
  Phys.\ Rev.\ D \textbf{10}, 1680-1685 (1974).

\bibitem{Hiscock:1980zf}
  W.~A.~Hiscock,
  J.\ Math.\ Phys.\ \textbf{22}, 1828 (1981).

\bibitem{Thorne:1986iy}
  K.~S.~Thorne, R.~H.~Price and D.~A.~Macdonald,
  {\it Black Holes: The Membrane Paradigm },
  (Yale University Press, New Haven and London, 1986).

\bibitem{Rezzolla:2000dk}
  L.~Rezzolla, B.~J.~Ahmedov and J.~C.~Miller,
  Mon.\ Not.\ Roy.\ Astron.\ Soc.\ \textbf{322}, 723 (2001)
  [arXiv:astro-ph/0011316 [astro-ph]].

\bibitem{Podolsky:2025tle}
  J.~Podolsky and H.~Ovcharenko,
  Phys.\ Rev.\ Lett.\ \textbf{135}, no.18, 181401 (2025)
  [arXiv:2507.05199 [gr-qc]].

\bibitem{Astorino:2025lih}
  M.~Astorino,
  Phys.\ Rev.\ D \textbf{112}, no.10, 104077 (2025)
  [arXiv:2508.12908 [gr-qc]].

\bibitem{Ma:2026ima}
  L.~Ma and H.~Lu,
  [arXiv:2606.23782 [gr-qc]].

\bibitem{Kiselev:2002dx}
  V.~V.~Kiselev,
  Class.\ Quant.\ Grav.\ \textbf{20}, 1187-1198 (2003)
  [arXiv:gr-qc/0210040 [gr-qc]].

\bibitem{Toshmatov:2015npp}
  B.~Toshmatov, Z.~Stuchl{\'\i}k and B.~Ahmedov,
  Eur.\ Phys.\ J.\ Plus \textbf{132}, no.2, 98 (2017)
  [arXiv:1512.01498 [gr-qc]].

\bibitem{Cho:2017nhx}
  I.~Cho and H.~C.~Kim,
  Chin.\ Phys.\ C \textbf{43}, no.2, 025101 (2019)
  [arXiv:1703.01103 [gr-qc]].

\bibitem{Visser:2019brz}
  M.~Visser,
  Class.\ Quant.\ Grav.\ \textbf{37}, no.4, 045001 (2020)
  [arXiv:1908.11058 [gr-qc]].

\bibitem{Kim:2019hfp}
  H.~C.~Kim, B.~H.~Lee, W.~Lee and Y.~Lee,
  Phys.\ Rev.\ D \textbf{101}, no.6, 064067 (2020)
  [arXiv:1912.09709 [gr-qc]].

\bibitem{Kim:2021vlk}
  H.~C.~Kim, B.~H.~Lee, W.~Lee and Y.~Lee,
  AIP Conf.\ Proc.\ \textbf{2874}, no.1, 020008 (2024)
  [arXiv:2112.04131 [gr-qc]].

\bibitem{Kim:2025sdj}
  H.~C.~Kim and W.~Lee,
  Eur.\ Phys.\ J.\ C \textbf{85}, no.11, 1245 (2025)
  [arXiv:2503.06961 [gr-qc]].

\bibitem{Lee:2026cit}
  W.~Lee and Y.~S.~Myung,
  Eur.\ Phys.\ J.\ C \textbf{86}, no.5, 525 (2026)
  [arXiv:2603.01507 [gr-qc]].

\bibitem{Xu:2026jjg}
  M.~Xu, J.~Lu, Y.~Liu and S.~Wu,
  [arXiv:2606.29881 [gr-qc]].

\bibitem{Gibbons:1976ue}
  G.~W.~Gibbons and S.~W.~Hawking,
  Phys.\ Rev.\ D \textbf{15}, 2752-2756 (1977).

\bibitem{Hawking:1995ap}
  S.~W.~Hawking and S.~F.~Ross,
  Phys.\ Rev.\ D \textbf{52}, 5865-5876 (1995)
  [arXiv:hep-th/9504019 [hep-th]].

\bibitem{Vigano:2022hrg}
  A.~Vigan{\`o},
  [arXiv:2211.00436 [gr-qc]].

\bibitem{Lungu:2024iob}
  V.~Lungu, M.~A.~Dariescu and C.~Stelea,
  Phys.\ Rev.\ D \textbf{111} (2025) no.6, 064014
  [arXiv:2405.14420 [gr-qc]].


\bibitem{Carter:1969zz}
  B.~Carter,
  J.\ Math.\ Phys.\ \textbf{10}, 70-81 (1969).

\bibitem{Demianski:1980mgt}
  M.~Demianski and M.~Francaviglia,
  Int.\ J.\ Theor.\ Phys.\ \textbf{19}, no.9, 675-680 (1980).

\bibitem{Frolov:2017kze}
  V.~P.~Frolov, P.~Krtous and D.~Kubiznak,
  Living Rev.\ Rel.\ \textbf{20}, no.1, 6 (2017)
  [arXiv:1705.05482 [gr-qc]].

\bibitem{Lee:2021sws}
  B.~H.~Lee, W.~Lee and Y.~S.~Myung,
  Phys.\ Rev.\ D \textbf{103}, no.6, 064026 (2021)
  [arXiv:2101.04862 [gr-qc]].


\bibitem{Bardeen:1972fi}
  J.~M.~Bardeen, W.~H.~Press and S.~A.~Teukolsky,
  Astrophys.\ J.\ \textbf{178}, 347 (1972).

\bibitem{Ferrari:2020nzo}
  V.~Ferrari, L.~Gualtieri and P.~Pani,
  {\it General Relativity and its Applications}, CRC Press, (2020).

\bibitem{Perez-Giz:2008ajn}
  G.~Perez-Giz and J.~Levin,
  Phys.\ Rev.\ D \textbf{79}, 124014 (2009)
  [arXiv:0811.3815 [gr-qc]].

\bibitem{Wu:2003pe}
  X.~Wu and T.~y.~Huang,
  Phys.\ Lett.\ A \textbf{313}, 77-81 (2003)
  [arXiv:gr-qc/0302118 [gr-qc]].

\bibitem{Maldacena:2015waa}
  J.~Maldacena, S.~H.~Shenker and D.~Stanford,
  JHEP \textbf{08}, 106 (2016)
  [arXiv:1503.01409 [hep-th]].

\bibitem{Hashimoto:2016dfz}
  K.~Hashimoto and N.~Tanahashi,
  Phys.\ Rev.\ D \textbf{95}, no.2, 024007 (2017)
  [arXiv:1610.06070 [hep-th]].

\bibitem{Falconer:2003}
  K.~Falconer,
   {\it Fractal Geometry: Mathematical Foundations and Applications},
  John Wiley \& Sons, Ltd, 2003.

\bibitem{Kormendy:1995er}
  J.~Kormendy and D.~Richstone,
  Ann.\ Rev.\ Astron.\ Astrophys.\ \textbf{33}, 581 (1995).

\bibitem{Rubin:1970zza}
  V.~C.~Rubin and W.~K.~Ford, Jr.,
  Astrophys.\ J.\ \textbf{159}, 379-403 (1970).

\bibitem{Sin:1992bg}
  S.~J.~Sin,
  Phys.\ Rev.\ D \textbf{50}, 3650-3654 (1994)
  [arXiv:hep-ph/9205208 [hep-ph]].

\bibitem{Lee:1995af}
  J.~w.~Lee and I.~g.~Koh,
  Phys.\ Rev.\ D \textbf{53}, 2236-2239 (1996)
  [arXiv:hep-ph/9507385 [hep-ph]].

\bibitem{Hu:2000ke}
  W.~Hu, R.~Barkana and A.~Gruzinov,
  Phys.\ Rev.\ Lett.\ \textbf{85}, 1158-1161 (2000)
  [arXiv:astro-ph/0003365 [astro-ph]].

\bibitem{Lee:2008jp}
  J.~W.~Lee and S.~Lim,
  JCAP \textbf{01}, 007 (2010)
  [arXiv:0812.1342 [astro-ph]].

\bibitem{Hui:2016ltb}
  L.~Hui, J.~P.~Ostriker, S.~Tremaine and E.~Witten,
  Phys.\ Rev.\ D \textbf{95}, no.4, 043541 (2017)
  [arXiv:1610.08297 [astro-ph.CO]].

\bibitem{Kang:2023gef}
  S.~Kang, A.~Kar and S.~Scopel,
  JCAP \textbf{11}, 077 (2023)
  [arXiv:2308.13203 [hep-ph]].

\bibitem{Marrone:2006vu}
  D.~P.~Marrone, J.~M.~Moran, J.~H.~Zhao and R.~Rao,
  Astrophys.\ J.\ Lett.\ \textbf{654}, L57-L60 (2006)
  [arXiv:astro-ph/0611791 [astro-ph]].

\bibitem{Eatough:2013nva}
  R.~P.~Eatough, H.~Falcke, R.~Karuppusamy, K.~J.~Lee, D.~J.~Champion, E.~F.~Keane, G.~Desvignes, D.~H.~F.~M.~Schnitzeler, L.~G.~Spitler and M.~Kramer, \textit{et al.}
  Nature \textbf{501}, 391-394 (2013)
  [arXiv:1308.3147 [astro-ph.GA]].

\bibitem{Virbhadra:1998dy}
  K.~S.~Virbhadra, D.~Narasimha and S.~M.~Chitre,
  Astron.\ Astrophys.\ \textbf{337}, 1-8 (1998)
  [arXiv:astro-ph/9801174 [astro-ph]].

\bibitem{Virbhadra:2002ju}
  K.~S.~Virbhadra and G.~F.~R.~Ellis,
  Phys.\ Rev.\ D \textbf{65}, 103004 (2002).


\bibitem{yorke:1975}
  T.-Y.~Li and J.~A.~Yorke,
  Am.\ Math.\ Mon.\ \textbf{82}(10), 985-992 (1975).

\bibitem{Bicak:1985}
  J.~Bicak and V.~Janis,
  Mon.\ Not.\ R.\ astr.\ Soc.\ 212, 899 (1985).

\bibitem{Aliev:1989wx}
  A.~N.~Aliev and D.~V.~Galtsov,
  Sov.\ Phys.\ Usp.\ \textbf{32}, 75 (1989).

\bibitem{Podolsky:1998ez}
  J.~Podolsky and K.~Vesely,
  Phys.\ Rev.\ D \textbf{58}, 081501 (1998)
  [arXiv:gr-qc/9805078 [gr-qc]].

\bibitem{Semerak:2010lzj}
  O.~Semerak and P.~Sukova,
  Mon.\ Not.\ Roy.\ Astron.\ Soc.\ \textbf{404}, 545-574 (2010)
  [arXiv:1211.4106 [gr-qc]].

\bibitem{Semerak:2012dx}
  O.~Semerak and P.~Sukova,
  Mon.\ Not.\ Roy.\ Astron.\ Soc.\ \textbf{425}, 2455-2476 (2012)
  [arXiv:1211.4107 [gr-qc]].

\bibitem{Stuchlik:2015nlt}
  Z.~Stuchl{\'\i}k and M.~Kolo{\v{s}},
  Eur.\ Phys.\ J.\ C \textbf{76}, no.1, 32 (2016)
  [arXiv:1511.02936 [gr-qc]].

\bibitem{Tursunov:2018erf}
  A.~Tursunov, M.~Kolo{\v{s}}, Z.~Stuchl{\'\i}k and D.~V.~Gal'tsov,
  Astrophys.\ J.\ \textbf{861}, no.1, 2 (2018)
  [arXiv:1803.09682 [gr-qc]].

\bibitem{Suzuki:1996gm}
  S.~Suzuki and K.~i.~Maeda,
  Phys.\ Rev.\ D \textbf{55}, 4848-4859 (1997)
  [arXiv:gr-qc/9604020 [gr-qc]].

\bibitem{Kao:2004qs}
  J.~K.~Kao and H.~T.~Cho,
  Phys.\ Lett.\ A \textbf{336}, 159-166 (2005)
  [arXiv:gr-qc/0406101 [gr-qc]].

\bibitem{Jefremov:2015gza}
  P.~I.~Jefremov, O.~Y.~Tsupko and G.~S.~Bisnovatyi-Kogan,
  Phys.\ Rev.\ D \textbf{91}, no.12, 124030 (2015)
  [arXiv:1503.07060 [gr-qc]].

\bibitem{Yang:2026ekp}
  C.~Yang, C.~Gao, D.~Chen and K.~Liu,
  Eur.\ Phys.\ J.\ C \textbf{86}, no.6, 677 (2026).

\bibitem{Dalui:2026oaa}
  S.~Dalui and X.~H.~Ge,
  [arXiv:2605.19673 [gr-qc]].

\bibitem{Frolov:1999pj}
  A.~V.~Frolov and A.~L.~Larsen,
  Class.\ Quant.\ Grav.\ \textbf{16}, 3717-3724 (1999)
  [arXiv:gr-qc/9908039 [gr-qc]].

\bibitem{Bai:2014wpa}
  X.~Bai, B.~H.~Lee, T.~Moon and J.~Chen,
  J.\ Korean Phys.\ Soc.\ \textbf{68}, no.5, 639-644 (2016)
  [arXiv:1406.5816 [hep-th]].

\bibitem{Basu:2016zkr}
  P.~Basu, P.~Chaturvedi and P.~Samantray,
  Phys.\ Rev.\ D \textbf{95}, no.6, 066014 (2017)
  [arXiv:1607.04466 [hep-th]].

\bibitem{Dalui:2018qqv}
  S.~Dalui, B.~R.~Majhi and P.~Mishra,
  Phys.\ Lett.\ B \textbf{788}, 486-493 (2019)
  [arXiv:1803.06527 [gr-qc]].

\bibitem{Das:2024iuf}
  S.~Das, S.~Dalui and R.~Samanta,
  Phys.\ Rev.\ D \textbf{110}, no.12, 124037 (2024)
  [arXiv:2405.09945 [gr-qc]].

\bibitem{Das:2025vja}
  S.~Das, S.~Dalui, B.~H.~Lee and Y.~F.~Cai,
  [arXiv:2511.03657 [gr-qc]].

\bibitem{Das:2025eiv}
  S.~Das, S.~Dalui, B.~H.~Lee and Y.~F.~Cai,
  [arXiv:2512.04848 [gr-qc]].

\bibitem{Atamurotov:2015nra}
  F.~Atamurotov and B.~Ahmedov,
  Phys.\ Rev.\ D \textbf{92}, 084005 (2015)
  [arXiv:1507.08131 [gr-qc]].

\bibitem{Ruffini:1975ne}
  R.~Ruffini and J.~R.~Wilson,
  Phys.\ Rev.\ D \textbf{12}, 2959 (1975).




\end{thebibliography}
\end{document}